\documentclass[aps,rmp,amsmath,amssymb,longbibliography]{revtex4-1}   
\usepackage{times}
\usepackage[colorlinks=true,citecolor=blue]{hyperref}
\usepackage{bm}
\usepackage{epsfig}
\usepackage{amsmath,amssymb}
\usepackage{graphicx}
\usepackage{xcolor}
\usepackage{dcolumn}
\usepackage{bm}
\usepackage{lineno}
\usepackage{inputenc}
\usepackage{bbold}
\newcommand{\Yb}{\ensuremath{^{171}\mathrm{Yb}^+}}

\newcommand{\bra}[1]{\left \langle #1 \right |}
\newcommand{\ket}[1]{\left | #1 \right \rangle}
\newcommand{\expval}[1]{\left \langle #1 \right \rangle}
\newcommand{\up}{\left | \uparrow \right \rangle}
\newcommand{\dn}{\left | \downarrow \right \rangle}
\newcommand{\upx}{\left | \uparrow_x \right \rangle}
\newcommand{\dnx}{\left | \downarrow_x \right \rangle}
\newcommand{\upy}{\left | \uparrow_y \right \rangle}
\newcommand{\dny}{\left | \downarrow_y \right \rangle}
\newcommand{\upz}{\left | \uparrow_z \right \rangle}
\newcommand{\dnz}{\left | \downarrow_z \right \rangle}

\newcommand{\ba}{\begin{eqnarray}}
\newcommand{\ea}{\end{eqnarray}}

\newcommand{\q}[1]{\color{black}{#1}}

\begin{document}

\title{Programmable Quantum Simulations of Spin Systems with Trapped Ions}


\author{C. Monroe$^{* 1}$}
\author{W. C. Campbell$^2$}
\author{L.-M. Duan$^3$}
\author{Z.-X. Gong$^4$} 
\author{A.\ V.\ Gorshkov$^1$}
\author{P.\ W.\ Hess$^5$}
\author{R. Islam$^6$}
\author{K. Kim$^3$}
\author{N.\ M.\ Linke$^{10}$}
\author{G. Pagano$^7$}
\author{P. Richerme$^8$}
\author{C. Senko$^6$}
\author{N. Y. Yao$^9$}
\affiliation{$^{1}$Joint Quantum Institute and Joint Center on Quantum Information and Computer Science, University of Maryland Department of Physics and National Institute of Standards and Technology, College Park, MD  20742}
\affiliation{$^{2}$Department of Physics and Astronomy, University of California, Los Angeles, CA  90095}
\affiliation{$^{3}$Center for Quantum Information, Institute for Interdisciplinary Information Sciences, Tsinghua University, Beijing  100084  China}
\affiliation{$^{4}$Department of Physics, Colorado School of Mines, Golden, CO 80401}
\affiliation{$^{5}$Department of Physics, Middlebury College, Middlebury, VT  05753}
\affiliation{$^{6}$Institute for Quantum Computing and Department of Physics and Astronomy, University of Waterloo, Waterloo, Ontario N2L 3G1  Canada}
\affiliation{$^{7}$Department of Physics and Astronomy, Rice University, Houston, TX  77005}
\affiliation{$^{8}$Department of Physics, Indiana University, Bloomington, IN  47405}
\affiliation{$^{9}$Department of Physics, University of California, Berkeley CA 94720}
\affiliation{$^{10}$Joint Quantum Institute, University of Maryland Department of Physics, College Park, MD  20742}


\begin{abstract}
\noindent
Laser-cooled and trapped atomic ions form an ideal standard for the simulation of interacting quantum spin models.  Effective spins are represented by appropriate internal energy levels within each ion, and the spins can be measured with near-perfect efficiency using state-dependent fluorescence techniques.  By applying optical fields that exert optical dipole forces on the ions, their Coulomb interaction can be modulated to produce long-range and tunable spin-spin interactions that can be reconfigured by shaping the spectrum and pattern of the laser fields{\q , in a prototypical example of a quantum simulator}.  Here we review the theoretical mapping of atomic ions to interacting spin systems, the preparation of complex equilibrium states, the study of dynamical processes in these many-body interacting quantum systems{\q , and the use of this platform for optimization and other tasks}.  The use of such quantum simulators for studying spin models may inform our understanding of exotic quantum materials and shed light on {\q the behavior of} interacting quantum systems that cannot be modeled with conventional computers.

\noindent
* to whom correspondence should be addressed: monroe@umd.edu

\end{abstract}
\pacs{81.05.Uw,68.37.-d,73.20-r}
\maketitle
\onecolumngrid
\vspace{-\baselineskip}
\begin{footnotesize}
\tableofcontents{}
\end{footnotesize}

\pagebreak

\section{Introduction} \label{sec:intro}

Interacting quantum systems cannot generally be efficiently simulated using classical computational techniques, due to the exponential scaling of the information encoded in a quantum state with the size of the system. Accurate modeling of quantum phenomena such as the magnetism of interacting spins, material superconductivity, electronic structure and molecules and their chemistry, Fermi-Hubbard models of electron transport in solids, and interactions within atomic nuclei, are all beyond the reach of classical computation even for just small numbers of interacting degrees of freedom.
A quantum simulator exploits the controlled manipulation of a standard quantum system in order to mimick the properties or evolution of a different--perhaps intractable--quantum system \cite{feynman1982simulating,naturephysics2012quantum}.  A quantum simulator is a restricted quantum computer \cite{nielsen2000quantum, ladd2010quantum}, with operations that may not be universal but instead be tailored to a particular quantum physical model under study.  While useful general purpose and universal quantum computers are still a future prospect, special purpose quantum simulations have already been demonstrated and may in fact become the first useful application of quantum computers. 

The equivalence of quantum bits (qubits) and their quantum gates to effective spin$-1/2$ systems and their interactions has guided one of the most important classes of quantum simulations: interacting spin systems and quantum magnetism. The most advanced physical system for quantum bits or effective spins is {\q arguably} a collection of trapped atomic ions \cite{wineland2008entangled,monroe2013scaling,brown2016codesigning, Steane1997, wineland1998experimental}.  This is evidenced by the number of controlled and interacting qubits, the quality of quantum gates and interactions, and the {\q high fidelity of quantum state initialization and measurement}.  Trapped atomic ions, held in a vacuum chamber and confined by electromagnetic fields to be distant from surfaces or solids, laser cooled to be nearly at rest and ``wired" together with external laser or microwave fields, offer a very clean quantum system to perform quantum simulations \cite{porras2004effective,deng2005effective,taylor2008wigner, blatt2012quantum}. 

This review assesses recent progress in the quantum simulation of magnetism {\q and related phenomena} using trapped atomic ion crystals.  Following an introduction of the mapping of spins to atomic ions, we first cover experimental results on the simulation of magnetic ordering, equilibrium states, and phase transitions in quantum magnetic systems.  Then we move to dynamical studies of quantum magnetism, touching on general issues of information and entanglement dynamics, quantum thermalization, inhibitors to thermalization such as many-body localization and prethermalization, ``time crystals," dynamical phase transitions{\q, and Hamiltonian engineering and sequencing techniques} that translate quantum physics models to effective magnetic spin models. {\q Recent techniques of variational quantum simulations (VQS) and quantum approximate optimization algorithms are also reviewed in the context of trapped ion systems.}
We conclude with speculations on how these types of simulations with trapped ions may scale in the future and perhaps guide the development of real magnetic material function or more general quantum simulations as special cases of quantum computations. 

{\q Trapped ion spin simulators generally exploit the collective motion of the ions to generate many-body spin interactions, with a controlled coupling through collective bosonic harmonic oscollator modes, or phonons.  While not a focus of this review, the phonon degree of freedom itself is an interesting medium to host simulations of bosonic models and are complementary to the field of photonic cavity-QED in atomic  \cite{Raimond2001manipulating,Ye2008quantum} and superconducting \cite{wallraff2004strong} systems. There have been many demonstrations of phonon control and phonon-based simulations in trapped ion systems \cite{kish2003experimental,um2016phonon, Kienzler2017quantum}, including bosonic interference \cite{leibfried2002trapped,Toyoda2015}, blockades \cite{debnath2018hopping,ohira2020}, sampling molecular vibronic spectrum \cite{shen2018quantum}, quantum walks \cite{schmitz2009quantum, zahringer2010realization}, phononic ``lasing" \cite{valhala2009phonon,ip2018phonon}, quantum thermodynamics \cite{an2015experimental}, and quantum heat engines/refrigerators \cite{rossnagel2016engine,maslennikov2019quantum}.}

We note there are many other quantum systems amenable to quantum spin simulation, including superconducting circuits \cite{GoogleSupremacy2019, Otterbach2017, Kandala2017}, neutral atoms in optical lattices \cite{BlochNeutralReview}, and solid-state arrays of effective individual spin systems \cite{choi17,choi2017observation}.  Comparisions to these other platforms is made where appropriate, although many of the simulations covered in this review rely on individual quantum state control with long range interactions that are not native to other platforms. 

\subsection{Atomic Ion Spins}\label{sec:atomic_ion_spin}

We represent a collection of spins by a crystal of electromagnetically trapped atomic ions, with two electronic energy levels within each ion behaving as an effective spin-1/2 particle.  The particular choice of electronic levels depends on the atomic element and also the desired type of control fields used to manipulate and measure the spin state.  The most important features of these spin states for executing quantum simulations are (a) the spin levels are long-lived and exhibit excellent coherence properties, (b) the spin states have appropriate strong optical transitions to auxiliary excited states, {\q with selection rules} allowing for qubit initialization through optical pumping and qubit detection through fluorescence, and (c) the spins interact through a coherent coupling that can be externally controlled and gated. These requirements restrict the atomic species to a handful of elements and spin states that are {\q encoded in} either $S_{1/2}$ hyperfine or Zeeman ground states of single outer-electron atoms (e.g., Be$^+$, Mg$^+$, Ca$^+$, Sr$^+$, Ba$^+$, Cd$^+$, Zn$^+$, Hg$^+$, Yb$^+$) with radiofrequency/microwave frequency splittings or {\q ground and $D$ or $F$} metastable electronic excited states of single or dual outer-electron atoms (e.g., Ca$^+$, Sr$^+$, Ba$^+$, Yb$^+$, B$^+$, Al$^+$, Ga$^+$, In$^+$, Hg$^+$, Tl$^+$, Lu$^+$) with optical frequency splittings. {\q Some species (e.g., Ba$^+$, Lu$^+$, Yb$^+$) have sufficiently long $D$ or $F$ metastable excited state lifetimes to host spins in their hyperfine or Zeeman levels with radiofrequency/microwave splittings.}  

In any of these systems, we label the two relevant spin states as $\dn \equiv \dnz$ and $\up \equiv \upz$, eigenstates of the Pauli operator $\sigma_z$ separated by energy $E_{\uparrow}-E_{\downarrow}=\hbar \omega_0$.  In the transverse bases of the Bloch sphere, we define by convention the eigenstates of $\sigma_x$ as $\dnx \equiv (\dn - \up)/\sqrt{2}$ and $\upx \equiv (\dn + \up)/\sqrt{2}$, and the eigenstates of $\sigma_y$ as $\dny \equiv (\dn + i\up)/\sqrt{2}$ and $\upy \equiv (i\dn + \up)/\sqrt{2}$. {\q We note that in this review, the spin-direction is often not explicitly labeled, in which cases the notation adopts the context of the material.}

\begin{figure}
\begin{center}
\includegraphics[width=0.5\linewidth]{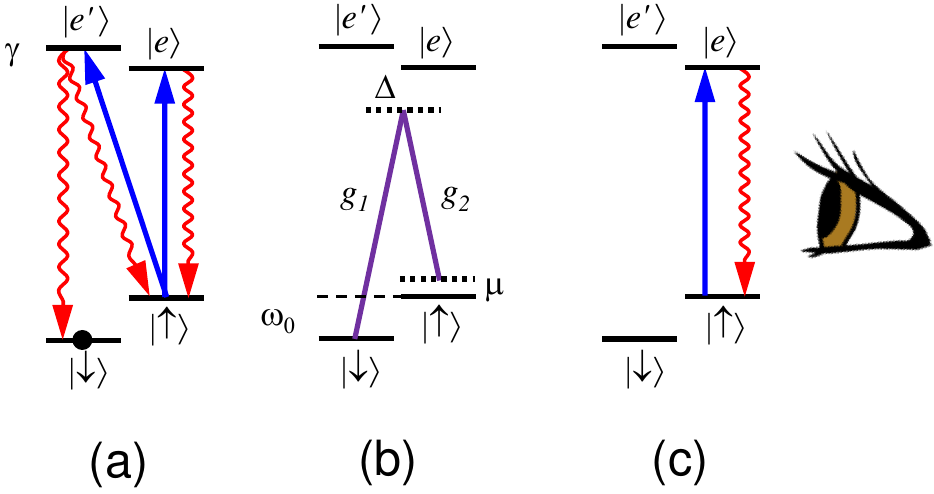}
\end{center}
\caption{\label{fig:InitMeas} Reduced energy level diagram of a single atomic ion.  Effective spin-1/2 systems are encoded within each atomic ion as stable electronic energy levels $\dn$ and $\up$.  A typical quantum simulation is comprised of three steps:
(a) Resonant radiation (blue lines) connects one of the two spin states to a pair of excited state levels (linewidth $\gamma$) and optically pumps each spin to the $\dn$ state through spontaneous  emission (red lines).  Here we assume that the excited state $\ket{e}$ couples only to $\up$ while the other excited state $\ket{e'}$ couples to both spin states.  
(Other sets of selection rules are also possible.) 
(b) In the case of ground-state (e.g., Zeeman or hyperfine) defined spins separated by frequency $\omega_0$, two tones of off-resonant radiation (purple lines) can drive stimulated Raman transitions between the spin states. The two beams have resonant Rabi frequencies $g_{1,2}$ connecting respective spin states to excited states and are detuned by $\Delta \gg \gamma$, and have a difference frequency (beatnote) detuned from the spin resonance by $\mu$. This coherently couples the spin states to create superpositions ($\mu=0$) and for non-copropagating beams also couples to the motion of the ion crystal ($\mu \ne 0$).  These processes can also be driven directly by radiofrequency or microwave signals, or for optically-defined spin states, a single laser tone.
(c) Resonant radiation (blue line) drives the $\up \leftrightarrow \ket{e}$ cycling transition, causes the $\up$ state to fluoresce strongly (red line), while the $\dn$ state is far from resonance and therefore dark.  This allows the near-perfect detection of the spin state by the collection of this state-dependent fluorescence {\q depicted by the eyeball}.}
\end{figure}

A typical quantum simulation in the ion trap system is comprised of three sequential steps: initialization, interaction, and measurement, as depicted in Fig. \ref{fig:InitMeas}.
The spins are initialized through an optical pumping process that prepares each spin in a nearly pure quantum state \cite{happer1972optical}. By applying resonant laser radiation that couples the spin states to short-lived excited states, each spin can be initialized with $>99.9\%$ state purity in a few microseconds. This relies on appropriate selection rules for the excited states as well as sufficiently split energy levels of the spin states themselves (Fig. \ref{fig:InitMeas}a). Laser cooling can prepare the motional states of the ions to near the ground state of harmonic motion \cite{leibfried2003quantum}, which is important for the control of the spin-spin interactions as detailed below.
11

Each spin can be coherently manipulated by driving the atomic ions with external fields that couple the spin states.  This can be accomplished by resonantly driving the spin levels with appropriate radiation at frequency $\omega_0$, or in Fig. \ref{fig:InitMeas}b, this is depicted as a two-field Raman process, with a beatnote of two optical fields at $\omega_0$ driving the spin (this will be assumed throughout unless otherwise indicated). This coherent coupling can also drive motional sideband transitions \cite{leibfried2003quantum} that couple the spin to the motion of the ion.  For multiple ions, this can be used to generate spin-spin couplings mediated by the Coulomb interaction, described in more detail below \cite{wineland2008entangled}. These external fields provide exquisite control over the effective spin-spin interaction, with the ability to gate the interaction, program different forms of the interaction strength and range, and even reconfigure the interaction graph topology.

At the end of the quantum simulation, the spins are measured by applying resonant laser radiation that couples one of the two spin states to a short-lived excited state through a cycling transition and detecting the resulting fluorescence \cite{nagourney1986shelved,sauter1986observation,bergquist1986observation}.  This is depicted in Fig. \ref{fig:InitMeas}c, where we take the $\up$ or ``bright" state as fluorescing and the $\dn$ or ``dark" state as not fluorescing. Even though the photon collection efficiency may be small (typically less than $1\%$), the effective spin detection efficiency can be well above $99\%$ owing to the low probability of leaving the fluorescence cycle or having the other (dark) state entering the cycle \cite{acton2007manipulation, hume2007high,InnsbruckFT2008, myerson2008high,noek2013high}. In order to detect the spins in other bases in the Bloch sphere ($\sigma_x$ or $\sigma_y$), previous to fluorescence measurement the spins are coherently rotated by polar angle $\pi/2$ along the $y$ or $x$ axis of the Bloch sphere, respectively.

\subsection{Coulomb-Collective Motion of Trapped Ion Crystals}

\begin{figure}
\includegraphics[width=0.75\linewidth]{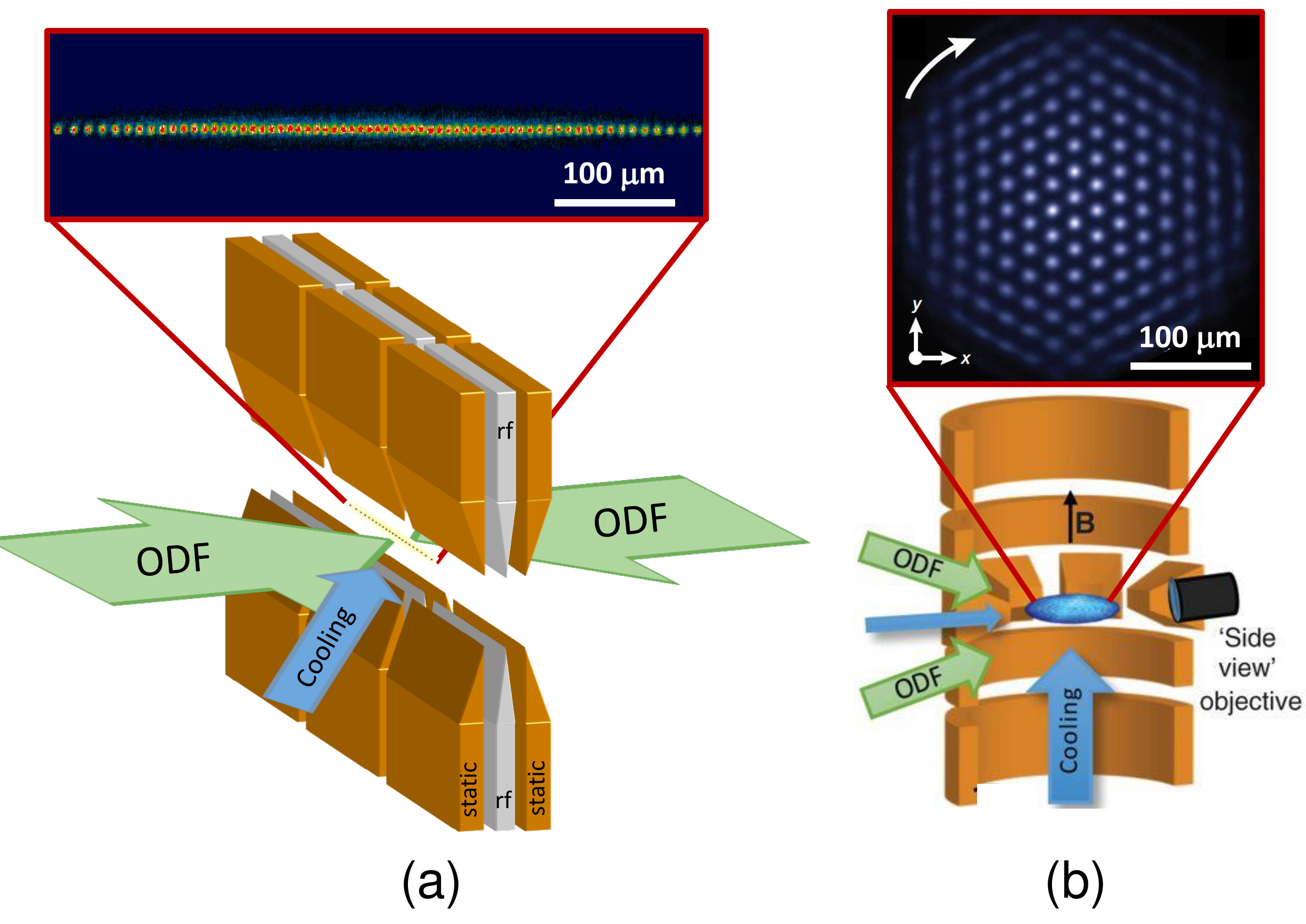}
\caption{(a) Radiofrequency (rf) linear trap used to prepare a 1D crystal of atomic ions.  The geometry in this trap has three layers of electrodes, with the central layer (gray) carrying rf potentials to generate a 2D quadrupole along the axis of the trap.  Static electrodes (gold) confine the ions along the axis of the trap.  For sufficiently strong transverse confinement, the ions form a linear crystal along {\q the trap axis}, with an image of 64 ions shown above with characteristic spacing $5$ $\mu$m for $\Yb$ ions (from \citealp{islam2011onset}). (b) Penning trap used to prepare a 2D crystal of atomic ions.  The gold electrodes provide a static quadrupole field that confines the ions along the vertical axis, and the vertical magnetic field stabilizes their orbits in the transverse plane.  For sufficiently strong axial confinement, the lowest energy configuration of the ions is a single plane triangular lattice that undergoes rigid body rotation, with an image of $\sim 200$ $^9$Be$^+$ ions shown above with a characteristic spacing of $20$ $\mu$m (from \citealp{bohnet2016quantum}). {\q For both traps, green arrows indicate non-copropagating optical dipole force (ODF) laser beams that provide spin-dependent forces along their wavevector difference, giving rise to Ising couplings.  Blue arrows indicate cooling and spin detection beams (imaging objectives not shown).}}
\label{fig:traps}
\end{figure}

Atomic ions can be confined in free space with electromagnetic fields supplied by nearby electrodes.  Two types of ion traps are used for quantum simulation experiments: the linear radiofrequency (rf) trap and the Penning trap.  The linear rf trap (Fig \ref{fig:traps}a) \cite{raizen1992linear} juxtaposes a static axial confining potential with a two-dimensional rf quadrupole potential that provides a time-averaged or ponderomotive transverse confinement potential \cite{dehmelt1967radiofrequency,paul1990electromagnetic}.  The trap anisotropy is typically adjusted so that the static axial confinement is much weaker than the transverse confinement so that laser-cooled ions reside on the axis of the trap where the rf fields are null, resulting in a one-dimensional chain of trapped ion spins.  A harmonic axial confinement potential results in an anisotropic linear ion spacing \cite{james1998quantum}, but they can be made nearly equidistant by applying an appropriate quartic axial confining potential \cite{lin2009large}. The Penning trap (Fig \ref{fig:traps}b) employs a uniform axial magnetic field with static axial electric field confinement, where the transverse confinement is provided by the $\textbf{E} \times \textbf{B}$ drift toward the axis \cite{brown1986geonium,bohnet2016quantum}.  Here, the trap anisotropy is typically adjusted so that the ions form a two-dimensional crystal perpendicular to and rotating about the axis.  Both rf and Penning traps can be modified to support other types of crystals in any number of spatial dimensions, but the quantum simulations reviewed here are either 1-D chains in rf traps or 2-D crystals in Penning traps.  However, the dimensionality of the spin-spin interaction graph does not necessarily follow that of the spatial arrangement of spins.

Ions are typically loaded into traps by generating neutral atoms of the desired element and ionizing the atoms once in the trapping volume via electron bombardment or photoionization.  Ion trap depths are usually much larger than room  temperature, so rare collisions with background gas do not necessarily eject the ion from the trap, but they can temporarily break up the crystal and scramble the atomic ions in space.  Under typical ultra-high-vacuum conditions, these collisional interruptions occur roughly once per hour per ion \cite{wineland1998experimental}, but cryogenic vacuum chambers can reduce the collision rate by orders of magnitude, where the trapped ions can be undisturbed for weeks or longer between collisions.

When atomic ions are laser-cooled {\q and localized well below their mean spacing}, they form a stable crystal, with the Coulomb repulsion balancing the external confinement force.  Typical spacings between adjacent ions in trapped ion crystals are $\sim 3-20$ $\mu$m, depending on the ion mass, number of ions in the crystal, the characteristic dimensions of the electrodes, and the applied potential values.  The equilibrium positions of ions in the crystal can be calculated numerically \cite{james1998quantum,sawyer2012spectroscopy, Steane1997}. The motion of the ions away from their equilibrium positions is well-described by harmonic normal modes of oscillation (phonon modes), with frequencies in the range $\omega_m/2\pi \sim 0.1-10$ MHz.  The thermal motion of laser-cooled ions and also the driven motion by external forces is typically at the $10-100$ nm scale. {\q This is much smaller than the inter-ion spacing, so nonlinearities to the phonon modes \cite{Marquet2003} can be safely neglected and the harmonic approximation to the phonon modes is justified.}
Calculations of the phonon mode frequencies and normal mode eigenfunctions follow straighforwardly from the calculated ion spacings \cite{james1998quantum,sawyer2012spectroscopy, Steane1997}. 
For the systems described here, we consider {\q primarily} the motion along a single {\q spatial} dimension labeled $X$. We write the $X-$component of position of the $i$th ion as $\hat{X}_i=\bar{X}_i + \hat{x}_i$, separating the mean (stationary) position $\bar{X}_i$ of the $i$th ion from the small harmonic oscillations described by the quantum position operator $\hat{x}_i$. The motion of ions in the crystal is tightly coupled by the Coulomb interaction, so it is natural to express the position operator in terms of the $m=1\ldots N$ normal (phonon) modes,
$\hat{x}_i = \sum_{m=1}^{N} b_{im}\hat{\xi}_m$, where $b_{im}$ is the normal mode transformation matrix, 
 with $\sum_i b_{im}b_{in} = \delta_{nm}$ and $\sum_m b_{im}b_{jm} = \delta_{ij}$.
Each phonon mode $\hat{\xi}_m$ oscillates at frequency $\omega_m$, and can be described as a quantum harmonic oscillator {\q with zero-point spatial spread $\xi_m^{(0)} = \sqrt{\hbar/2M\omega_m}$,} where $M$ is the mass of a single ion. In the interaction frame for each phonon mode, the position of the $i$th ion is thus written as
\begin{equation}
\hat{X}_i = \bar{X}_i + \sum_{m=1}^{N} b_{im}\xi_m^{(0)}(a_m^\dag e^{i\omega_m t} + a_m e^{-i\omega_m t}),
\end{equation}
{\q where $a_m^\dag$ and $a_m$ are bosonic raising and lowering operators for each mode, with $[a_n,a_m^\dag]=\delta_{nm}$.}

\begin{figure}[h]
\includegraphics[width=\linewidth]{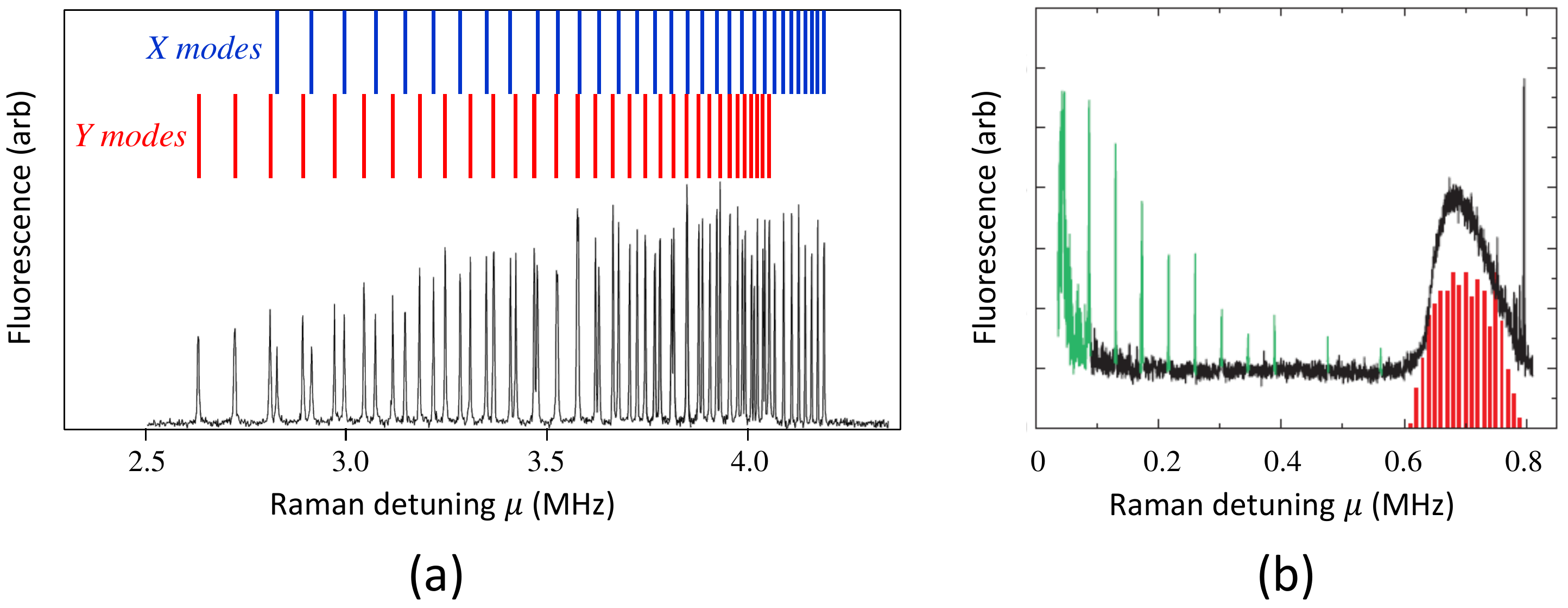}
\caption{Raman upper sideband spectrum of the transverse motion of trapped atomic ion crystals. The spectrum is measured by preparing all of the ions in (dark) state $\dn$ and driving them with global Raman laser beams with beatnote detuning $\mu$ from the spin-flip resonance and measuring the total fluorescence of the chain, which responds when the beatnote matches a sideband resonance.  (a) 32 trapped \Yb atomic ions in a linear chain (see Fig. \ref{fig:traps}a). Here, the Raman excitation is sensitive to both $X$ and $Y$ principal axes of transverse motion, and the theoretical position of both sets of 32 modes are indicated at top in blue and red. The highest frequency sidebands correspond to center-of-mass (COM) modes at $4.19$ MHz for the $X$ direction and $4.05$ MHz for the $Y$ direction. (Unpublished data from the University of Maryland.) (b) Measured (black) and calculated (red) sideband spectrum for 2D crystal of $345 \pm 25$ $^9$Be$^+$ions in a Penning trap (see Fig. \ref{fig:traps}b) with rotation frequency $43.2$ kHz. As in the linear chain, the highest frequency sideband at $795$ kHz corresponds to COM motion. Features at the rotation frequency and its harmonics harmonics (green) are due to residual couplings to in-plane degrees of freedom from imperfect beam alignment.  Adapted from \citealp{sawyer2012spectroscopy}.}
\label{fig:modes}
\end{figure}

In general, the structure of transverse phonon modes of {\q a 1D or 2D} ion crystal (motion perpendicular to the axis or plane of the ions) has the center-of-mass (COM) mode as its highest frequency, with the lowest frequency corresponding to zig-zag motion where adjacent ions move in opposite directions, as shown in Fig. \ref{fig:modes} for a 1D linear chain of 32 ions and a 2D crystal of about 345 ions.  The bandwidth of the transverse modes can be controlled by tuning the spatial anisotropy of the trap.

\subsection{Programmable Magnetic Fields and Interactions between Trapped Ion Spins}
\label{sec:Eff_B_and_J}
Effective magnetic fields and spin-spin interactions can be realized by applying external microwave or optical fields to the ions. We consider the case of optical fields, since they can be used to not only provide effective site-dependent magnetic fields for the spins through tight focusing, but the strong dipole forces from laser beams can also drive effective Ising interactions between the spins \cite{cirac1995quantum,sorensen1999quantum,milburn2000ion,solano1999deterministic}.
Such forces can be applied to pairs of ions in order to execute entangling quantum gates applicable 
to quantum computing \cite{wineland2008entangled}.  When such forces are instead applied more globally, the resulting 
interaction network allows the quantum simulation of a wide variety of spin models such as the
Ising and Heisenberg spin chains \cite{porras2004effective,deng2005effective,taylor2008wigner}.  

Following Fig. \ref{fig:InitMeas}b and assuming the spins are encoded in stable (e.g., Zeeman or hyperfine) levels, the ion crystal is addressed with two laser beams detuned from the excited states by much more than the excited state radiative linewidth ($\Delta \gg \gamma$).  By adiabatically eliminating the excited states, this drives coherent Raman transitions between the spin states.  Alternatively, optically-defined spin levels can be coupled with a single laser beam \cite{InnsbruckFT2008}, but this is more difficult technically, as the spins acquire an optical phase that requires extreme positional stability of the optical setup.  With {\q rf- or microwave-defined spin states, the relevant phase is that of the microwave beat note between the Raman beams, which has much longer wavelength and is easier to control and stabilize.}

{\q The two Raman beams are oriented to have a projection $\delta k$ of their wavevector difference along the $X$-axis of motion.  The Raman beatnote is detuned by frequency $\omega_0+\mu_i$ from the resonance of spin $i$ with beat note phase $\delta\phi_i$.  The resonant Raman Rabi frequency on ion $i$ is $\Omega_i = g^i_1 g^i_2/2\Delta$, where $g^i_{1,2}$ are the direct (single field) Rabi frequencies of the associated transitions through the excited states (see Fig. \ref{fig:InitMeas}b), proportional to the respective applied optical electric fields. The atom-light interaction Hamiltonian  in a frame rotating at the spin resonance frequency $\omega_0 \gg \Omega_i$ (optical rotating wave approximation)} takes the form $(\hbar = 1)$
\begin{equation}
H =  \frac{1}{2}\sum_i \left[ \Omega_i \left( \sigma_+^ie^{i(\delta k \hat{X}_i-\mu_i t-\delta\phi_i)} + \sigma_-^ie^{-i(\delta k \hat{X}_i-\mu_i t-\delta\phi_i)} \right)
+ d_i \sigma^i_z \right].
\label{Hgen}
\end{equation}
The last term is an {\q AC} Stark shift of the $i$th spin by amount $d_i$ and arises {\q from differential AC Stark shifts between the two spin levels from the Raman beams. This shift includes the ``two-photon" differential Stark shift terms scaled by  $(g_1^2+g_2^2)\omega_0/4\Delta^2$ and summed over each excited state detuned by $\Delta$, where $\omega_0 \ll \Delta$ (see Fig. \ref{fig:InitMeas}).  There are also ``four-photon" Stark shifts scaled by $\Omega_i^2/4\mu_i$ and summed over each two-photon Raman resonance detuned by $\mu$, where $\Omega_i \ll \mu_i$. The magnitude of these shifts depends greatly on the atomic energy level structure and light polarization: see \citealp{lee2016engineering} for a discussion of Raman-coupled qubits (Fig \ref{fig:InitMeas}) and \citealp{InnsbruckStark} for direct optically-coupled qubits.  These Stark shifts can be either absorbed into the definition of the spin energy levels, or used as an effective axial magnetic field for simulations.}

Below, we will assume that the ions are confined to the ``Lamb-Dicke limit" \cite{wineland1998experimental,leibfried2003quantum} where the excursion of ion motion is much less than the associated wavelength of radiation driving transitions: $\delta k \langle \hat{x}_i^2 \rangle^{1/2} \ll 1$.  This is typically a good assumption for trapped ions laser-cooled to near their ground state{\q, with a zero-point spatial spread of all modes $\xi_m^{(0)}$ that is typically of order nanometers in experiments}.  

\subsubsection{Effective magnetic fields}
For a resonant ``carrier" interaction ($\mu=d_i=0$) and under the rotating wave approximation ($\omega_m \gg \Omega_i$), the time dependence of $\hat{X}$ averages to zero and the Hamiltonian of Eq. (\ref{Hgen}) is just
\begin{equation}
H_{B} =  \frac{1}{2}\sum_i \Omega_i \sigma^i_{\beta_i},
\label{field}
\end{equation}
where the transverse field spin operator is defined by
\begin{equation}
\sigma^i_{\theta_i} \equiv \sigma_+^i e^{-i\theta_i} + \sigma_-^i e^{i\theta_i} = \sigma_x^i \cos\theta_i + \sigma_y^i \sin\theta_i. \label{Isingphase}
\end{equation}
This Hamiltonian of Eq. (\ref{field}) describes the precession of spin $i$  about an effective transverse magnetic field in the $xy$ plane of the Bloch sphere at an angle
\begin{equation}
\beta_i=\delta\phi_i-\delta k \bar{X}_i \label{Bphase}
\end{equation}
that can be controlled through the phase $\delta\phi_i$. In cases where $\delta k \ne 0$, the Rabi frequency acquires a dependence on the motion of the ions through Debye-Waller factors \cite{wineland1998experimental}, but these are small in the Lamb-Dicke limit \cite{sorensen2000entanglement,leibfried2003quantum}. 

Tuning the Raman laser beat note away from the carrier {\q with $\mu \gg \Omega_i$ generally results in a four-photon AC Stark shift of the spin levels as discussed above}, given by the last term in Eq. (\ref{Hgen}) \cite{InnsbruckStark,lee2016engineering}. When each spin is exposed to a unique intensity of light and/or detuning {\q of a single beam}, parametrized by $d_i$, this gives rise to a site-dependent effective axial {\q ($z$)} magnetic field.

\subsubsection{Effective Ising interactions}
\label{subsec:Ising}
When the frequency $\mu_i$ is tuned to the neighborhood of the phonon modes $\omega_m$, the spin $i$ is coupled to the ion motion through the spatial variation of the phase factor in Eq. (\ref{Hgen}).  This will generate an effective spin-spin interaction between the {\q exposed} ions, mediated by the collective transverse vibrations of the chain. For most simulation experiments, the transverse modes of motion are used to mediate the spin-spin interaction because their frequencies are tightly packed and all contribute to the effective spin Hamiltonian, allowing control over the form and range of the interaction, described further below.  Transverse modes also oscillate at higher frequencies, leading to better cooling and less sensitivity to external heating and noise \cite{zhu2006trapped}. 

In general, when noncopropagating laser beams {\q form} bichromatic beat notes at frequencies $\omega_0 \pm \mu_i$ symmetrically detuned from the carrier with $\mu_i \approx \omega_m$, both upper and lower motion-induced sidebands of the normal modes of motion are driven in the ion crystal, giving rise to a spin-dependent force at frequency $\mu_i$ \cite{porras2004effective,sorensen1999quantum, milburn2000ion, solano1999deterministic}. {\q Owing to the symmetry of the detuned beatnotes, the four-photon Stark shift is generally negligible.  However,}
when the bichromatic beat notes are asymmetrically detuned from the carrier by $\omega_0 + \mu_{i+}$ and $\omega_0 - \mu_{i-}$, the effective spin-dependent force occurs at frequency $\mu_i = (\mu_{i+} + \mu_{i-})/2$ and the asymmetry provides a {\q Stark shift that gives rise to an} effective uniform axial magnetic field in Eq. (\ref{Hgen}) with $d_i = (\mu_{i+} - \mu_{i-})/2$.

Under the rotating wave approximation ($\omega_0 \gg \mu_i \gg \Omega_i$) with symmetric detuning $\mu_i=\mu_{i+}=\mu_{i-}$ and within the Lamb-Dicke limit, {\q the exponential function in Eq. (\ref{Hgen}) can be expanded to lowest order, resulting in} \cite{zhu2006trapped}
\begin{equation}
H(t) = \frac{1}{2}\sum_{i,m} \eta_{im}\Omega_i \sigma_{\theta_i}^i \left[a_m^\dag e^{-i(\delta_{im}t+\psi_i)} + a_m e^{i(\delta_{im}t+\psi_i)}\right].
\label{eq:sdf}
\end{equation}
{\q Here the beatnote detuning on ion $i$ from the $m$th motional sideband is defined as $\delta_{im}=\mu_i-\omega_m$ and the Lamb-Dicke parameter $\eta_{im} = b_{im} \delta k \xi_m^{(0)}$ describes the coupling between ion $i$ and mode $m$.

The above expression includes two phases: a ``spin phase" $\theta_i$ that determines the angle of the $i$th spin operator in the XY plane of the Bloch sphere that diagonalizes the spin-dependent force; and a ``motion phase" $\psi_i$ that determines the phase of the optical forces (but does not play a role in the spin-spin interactions as developed below).
These phases depend on the geometry of the bichromatic laser beams and there are two cases written in Eqs. (\ref{phasesens})-(\ref{phaseinsens}) \cite{lee2005phase, haljan2005spin}. When the upper and lower sideband running wave beat notes propagate in the same direction ($\delta k$ is the same sign for both), this is termed the ``phase-sensitive" geometry. 
On the other hand, when the upper and lower sideband running waves propagate in opposite directions (opposite sign of $\delta k$ for the two beat notes), this is called ``phase-insensitive."  The phases for each configuration are written, \begin{align}
&\, & &\text{SPIN PHASE} & &\text{MOTION PHASE} \nonumber \\
&\text{PHASE-SENSITIVE} &
&\theta_i =\left(\frac{\delta\phi_{i+}+\delta\phi_{i-}}{2}\right)-\delta k \bar{X}_i - \frac{\pi}{2} &
&\psi_i = \left(\frac{\delta\phi_{i+}-\delta\phi_{i-}}{2}\right)  \label{phasesens} \\
&\text{PHASE-INSENSITIVE} &
&\theta_i =\left(\frac{\delta\phi_{i+}+\delta\phi_{i-}}{2}\right) &
&\psi_i = \left(\frac{\delta\phi_{i+}-\delta\phi_{i-}}{2}\right)-\delta k \bar{X}_i - \frac{\pi}{2}.
\label{phaseinsens}
\end{align}
Here, $\delta\phi_{i+}$ and $\delta\phi_{i-}$ are the beat note phases associated with the upper and lower sideband fields, respectively. The additional $\pi/2$ phase factors compared with Eq. (\ref{Bphase}) originate from the imaginary linear term in the Lamb-Dicke expansion of $e^{ik\hat{x}_i}$.

The sensitivity of the spin phase to $\delta k \bar{X}_i$ has great importance on the practical implementation of spin-spin Hamiltonians. In the phase-sensitive geometry, this dependence may be desired when the phase of other Hamiltonian terms (such as the carrier spin flip operation of Eq. (\ref{field})) have the same form, as in Eq. (\ref{Bphase}). This allows a phase-sensitive diagnosis of individual spin operations.
However, sensitivity to $\delta k \bar{X}_i$ over sufficiently long times can lead to decoherence if there are drifts in the relative path length of non-copropagating beams or the ion chain position along the X-direction. 
The Phase Insensitive configuration is therefore useful for long simulation evolution times.

We note that the phase-insensitive geometry using Raman couplings (Fig. \ref{fig:InitMeas}b) can remove the spin-phase sensitivity to not only the absolute optical phase of the optical source, but also the relative optical path length difference between the counterpropagating beams, by setting $\delta\phi_{i+}=-\delta\phi_{i-}$ \cite{lee2005phase, haljan2005spin}.  This is not possible with direct optical upper and lower sideband transitions on spins with an optical energy splitting, as their optical phases add regardless of the geometry.}

For either phase configuration, the evolution operator under this Hamiltonian can be written from the Magnus expansion, which terminates after the first two terms 
\cite{zhu2006trapped},
\begin{eqnarray}
U(\tau) &=& \exp \left[ -i\int_0^{\tau} dt H(t) - \frac{1}{2}\int_0^{\tau} dt_1 \int_0^{t_1} dt_2 \; [H(t_1),H(t_2)]  \right] \label{beforeevolution}\\
&=& \exp \left[ \sum_{i}\hat{\zeta}_{i}(\tau) \sigma_{\theta_i}^i 
                   - i\sum_{i<j} \chi_{i,j}(\tau)      \sigma_{\theta_i}^i\sigma_{\theta_i}^j\right]. 
\label{evolution}
\end{eqnarray}
The first term of Eq. (\ref{evolution}) is a spin-phonon coupling with operator
\begin{equation}
\hat{\zeta}_i(\tau) = \sum_m
\left[\alpha_{i,m}(\tau)a_m^{\dagger}
-\alpha_{i,m}^*(\tau) a_m \right], \label{phonon}
\end{equation}
representing spin-dependent coherent displacements \cite{glauber1963coherent,leibfried2003quantum} of the $m$th motional mode through phase space by an amount  
\begin{eqnarray}
\alpha_{i,m}(\tau)= -\frac{i}{2} \eta_{im} \Omega_i\int_0^\tau dt  e^{-i(\delta_{im}t + \psi_i)} \label{alpha1} 
= -\frac{\eta_{im}\Omega_ie^{-i\psi_i}}{2\delta_{im}}\left(1-e^{-i\delta_{im}\tau}\right).
\label{alpha}
\end{eqnarray}
The second term of Eq. (\ref{evolution}) is the key result: a spin-spin interaction between ions $i$ and $j$ with coupling strength 
\begin{eqnarray}
\chi_{i,j}(\tau) &=& \frac{1}{2}\Omega_i\Omega_j \sum_{m} \eta_{im} \eta_{jm} \int_0^\tau dt_1  \int_0^{t_1} dt_2  
\sin(\delta_{im}t_1-\delta_{jm}t_2) \label{coupling1} \\
&=& \Omega_i\Omega_j \sum_m  \frac{\eta_{im}\eta_{j,m}}{2\delta_{im}\delta_{jm}} 
                  \left[\left(\frac{\delta_{im}+\delta_{jm}}{2}\right)\tau - \left(\frac{\sin\delta_{im}\tau+\sin\delta_{jm}\tau}{2}\right)\right]
             \label{coupling2} \\
&=& \Omega_i \Omega_j \sum_m  \frac{\eta_{im}\eta_{j,m}}{2\delta_m} 
                  \left(\tau - \frac{\sin\delta_{m}\tau}{\delta_{m}} \right),
     \label{coupling3}           \mbox{       for } \delta_{im}=\delta_m. 
\end{eqnarray}
{\q Fractional corrections to this expression arising from higher-order terms in the Lamb-Dicke expansion leading to Eq. (\ref{eq:sdf}) can be shown to be of order $\eta_{im}^2\eta_{j,m}^2\bar{n}_m^2$ for each mode \cite{sorensen2000entanglement}, where $\bar{n}_m$ is the average number of vibrational quanta in mode $m$.} 

{\q In this review, we generally consider global spin Hamltonians that are realized by exposing all the ions to the spin-dependent force. However, for the special case of applying a spin-dependent force to just two ions $i$ and $j$ in the chain, which is common for the execution of entangling two-qubit quantum logic gates \cite{sorensen1999quantum, sorensen2000entanglement}, the evolution operator in Eq. (\ref{evolution}) reduces to
\begin{eqnarray}
U_{ij}(\tau)= e^{-i\chi_{ij}(\tau)}&\ket{\uparrow_{\theta_i} \uparrow_{\theta_j}}\bra{\uparrow_{\theta_i} \uparrow_{\theta_j}}&\prod_m\hat{\mathcal{D}}_m\left[\alpha_{im}(\tau)+\alpha_{jm}(\tau)\right] \\
+e^{-i\chi_{ij}(\tau)}&\ket{\downarrow_{\theta_i} \downarrow_{\theta_j}}\bra{\downarrow_{\theta_i} \downarrow_{\theta_j}}&\prod_m\hat{\mathcal{D}}_m\left[-\alpha_{im}(\tau)-\alpha_{jm}(\tau)\right] \\
+e^{i\chi_{ij}(\tau)}&\ket{\uparrow_{\theta_i} \downarrow_{\theta_j}}\bra{\uparrow_{\theta_i} \downarrow_{\theta_j}}&\prod_m\hat{\mathcal{D}}_m\left[\alpha_{im}(\tau)-\alpha_{jm}(\tau)\right] \\
+e^{i\chi_{ij}(\tau)}&\ket{\downarrow_{\theta_i} \uparrow_{\theta_j}}\bra{\downarrow_{\theta_i} \uparrow_{\theta_j}}&\prod_m\hat{\mathcal{D}}_m\left[-\alpha_{im}(\tau)+\alpha_{jm}(\tau)\right].
\end{eqnarray}
In this expression, the spin projection operators are eigenvectors of $\sigma_{\theta_i}$:
$\ket{\uparrow_{\theta_i}}=(\ket{\uparrow_i}+e^{-i\theta_i}\ket{\downarrow_i})/\sqrt{2}$, $\ket{\downarrow_{\theta_i}}=(\ket{\uparrow_i}-e^{-i\theta_i}\ket{\downarrow_i})/\sqrt{2}$ and $\bra{\uparrow_{\theta_i}}\sigma_{\theta_i}\ket{\uparrow_{\theta_i}}=+1$, $\bra{\downarrow_{\theta_i}}\sigma_{\theta_i}\ket{\downarrow_{\theta_i}}=-1$. The coherent displacement operator on mode $m$ is $\hat{\mathcal{D}}_m(\alpha)=e^{\alpha a_m^{\dagger}
-\alpha^* a_m}$ \cite{glauber1963coherent}.}

There are two regimes where the collective modes of motion contribute to the spin-spin coupling,
taking evolution time $\tau$ to be much longer than the ion normal mode oscillation periods $(\omega_m \tau \gg 1)$.  
In the ``resonant" regime \cite{sorensen2000entanglement,milburn2000ion,solano1999deterministic}, the optical beatnote detuning $\mu$ is close to one or more normal modes and the spins become entangled with the motion through the spin-dependent displacements.  However, at certain times of the evolution $\alpha_{i,m}(\tau) \approx 0$ for all modes $m$ and the motion nearly decouples from the spin states, which is useful for applying synchronous entangling quantum logic gates between the spins.  {\q For closely-spaced modes such as the transverse modes as seen in Fig. \ref{fig:modes}, resolving individual modes becomes difficult and may require laser-pulse-shaping techniques \cite{zhu2006trapped}.}

{\q For generating pure spin Hamiltonians, we instead operate} in the ``dispersive" regime \cite{sorensen1999quantum,porras2004effective}, {\q where} the optical beatnote frequency is far from each normal mode compared to that mode's sideband Rabi frequency $(|\delta_{im}| \gg \eta_{im}\Omega_i)$.  In this case, the phonons are only virtually excited as the displacements become negligible $(|\alpha_{i,m}(t)| \ll 1)$. {\q The spin-phonon part of the evolution (Eq. (\ref{phonon})) is therefore negligible, and the spin-spin interaction evolution (Eq. (\ref{evolution})) is dominated by the secular terms of Eqs. (\ref{coupling2}) and (\ref{coupling3}) that are linear in time $\tau$.} The final result is an effective fully-connected Ising Hamiltonian,
\begin{equation}
H_{J_\theta} = \sum_{i<j} J_{ij}\sigma_{\theta_i}^i\sigma_{\theta_j}^j,
\label{Ham}
\end{equation}
where the Ising matrix is given by
\begin{eqnarray}
J_{ij} &=& \Omega_i\Omega_j \omega_{\rm rec}\sum_m \frac{b_{im}b_{jm}}{4\omega_m}\left(\frac{1}{\delta_{im}}+\frac{1}{\delta_{jm}}\right) \\
&=& \Omega_i\Omega_j \omega_{\rm rec}\sum_m\frac{b_{im}b_{jm}}{2\omega_m\delta_m} \mbox{       for } \delta_{im}=\delta_m. 
\label{Jij}
\end{eqnarray}
Here we have used $\omega_{\rm rec} = \hbar (\delta k)^2/2M$ as the recoil frequency associated with the transfer of momentum $\hbar (\delta k)$ to a single ion. 

\begin{figure}[h]
\includegraphics[width=\linewidth]{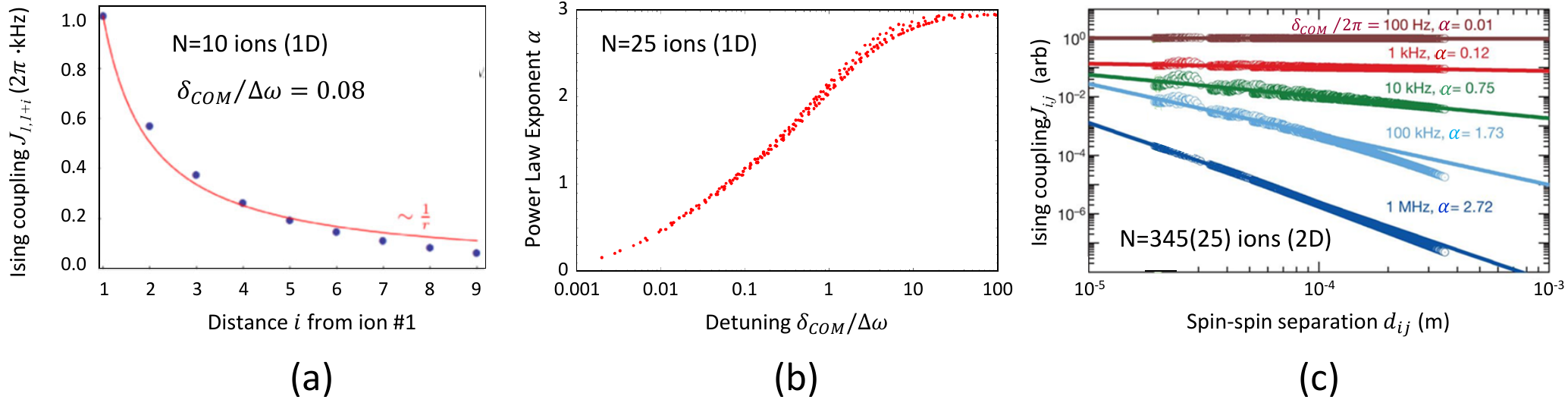}
\caption{{\q Theoretical comparison of exact Ising couplings to the inverse power law form of Eq. (ref{Jpowerlaw}).
(a) Calculated Ising couplings to edge ion in a 1D crystal of 10 ions confined with a harmonic external axial potential.  The Raman detuning from the center-of-mass (COM) mode, scaled to the bandwidth $\Delta\omega$ of the transverse modes, is $\delta_{\rm COM}/\Delta\omega$, with the best fit power law (red line) of $\alpha=1$. Adapted from \citealp{IslamThesis}.
(b) Best fit inverse power law exponent $\alpha$ for calculated Ising couplings within a 1D crystal of 25 ions with harmonic axial confinement, plotted as a function of the Raman detuning from the COM mode $\delta_{\rm COM}/\Delta\omega$. The spread of the points at a given detuning considers various bandwidths $\Delta\omega$ (axial COM confinement frequencies ranging from $100-350$ kHz with a transverse COM confinement frequency of $5$ MHz), and the COM sideband Rabi frequency $\eta_{\rm i,COM}\Omega$ is always less than $1/3$ of the detuning $\delta_{\rm COM}$ in order to limit direct phonon excitation.
Adapted from \citealp{IslamThesis}. (c) Calculated Ising couplings from Eq. (\ref{Jij}) in a 2D crystal of 217 ions versus a sampling of the distance $d_{ij}$ between ion pairs (circles). The lines are best-fit power law exponents $\alpha$ (lines) for various detunings from the center-of-mass (COM) mode of $795$ kHz. Adapted from \citealp{britton2012engineered}.}}
\label{fig:PowerLaw}
\end{figure}

Substituting the exact values for the normal mode matrix $b_{im}$ and assuming that the optical force is detuned at frequencies higher than all phonon modes ($\delta_m > 0$), we find that for a uniform Rabi frequency over the ions $\Omega_i=\Omega$, the Ising matrix is well approximated by a long-range antiferromagnetic (AFM) coupling that fall off as an inverse power law with distance
\begin{equation}
J_{ij} = \frac{J_0}{|i-j|^{\alpha}},
\label{Jpowerlaw}
\end{equation}
with nearest-neighbor Ising coupling $J_0$.  
The exponent $\alpha$ that determines the range of the Ising interaction can be set to $0<\alpha<3$ by simply adjusting the laser detuning $\mu > \omega_m$ \cite{porras2004effective,islam2011onset}.
The true asymptotic long-chain behavior of a trapped ion chain is more subtle \cite{nevado16}, but the power law approximation is very good, as shown in Fig. \ref{fig:PowerLaw}, comparing numerically exact couplings with best-fit power laws for various detunings in both a linear and 2D crystal.  

When the detuning $\mu$ is tuned between the modes of motion, many other patterns of the Ising graph can be realized \cite{korenblit2012quantum,lin2011sharp}. The spin-spin interaction profile $J_{ij}$ can in principle be programmed arbitrarily with sufficient control of the individual spin-phonon couplings on $N$ trapped ions. For example, $N$ unique bichromatic Raman beatnotes applied at frequencies $\omega_0\pm\mu_n$, with $\mu_n \approx \omega_n$ ($n=1,2,...,N)$ can be used with local intensity control over each ion to create an arbitrary interaction graph:
\begin{equation}
J_{ij} =\sum_n\Omega_{i,n}\Omega_{j,n} \omega_{\rm rec}\sum_m\frac{b_{im}b_{jm}}{2\omega_m(\mu_n - \omega_m)}.
\label{eqn:Jij_arbitrary}
\end{equation}
Here, $\Omega_{i,n}$ is the Rabi frequency corresponding to the $n^{\rm{th}}$ beatnote on the $i^{\rm{th}}$ ion. Note that $J_{ij}$ is nonlinear in the $\mathcal{O}(N^2)$ experimental control parameters $\Omega_{i,n}$ and $\mu_n$, and hence tuning the quantum simulator requires non-linear optimization methods \cite{korenblit2012quantum, Teoh_2020}. Alternative approaches to realize a target interaction graph without tuning the full Rabi frequency matrix, $\Omega_{i,n}$ include modifying a global  M\o lmer-S\o rensen coupling profile (such as Eq.(\ref{Jpowerlaw})) by local spatial control of spins in hybrid analog-digital ways \cite{Hayes2014programmable, Rajabi2019dynamical}.

This review will mainly consider Ising interactions in the slow dispersive regime in order to engineer pure spin Hamiltonians given by Eqs. (\ref{field}) and (\ref{Ham}) that do not directly involve the bosonic phonon operators.  An important class of models in quantum magnetism that will appear throughout this review is the transverse field Ising model, which is one of the simplest physical models that admits a quantum phase transition \cite{sachdev2011quantum}, owing to its noncommunting spin operators:
\begin{equation}
    H_{TI} = \sum_{i<j} J_{ij} \sigma_x^i \sigma_x^j + B_y \sum_i \sigma_y^i.
\label{eqn:TransversIsing}
\end{equation} 

In ion trap systems, this model can be generated with a combination of the effective magnetic field in Eq. (\ref{field}) and the Ising interactions in Eq. (\ref{Ham}) {\q with appropriate settings of the spin phases $\theta_i$ in Eq. (\ref{Bphase}) and Eqs. (\ref{phasesens})-(\ref{phaseinsens}).  When both (noncommuting) terms are simultaneously applied, additional spin-dependent ($\sigma_z^i$) phonon terms appear in higher orders of the Magnus expansion of Eq. (\ref{beforeevolution}) \cite{wang2010disentangling, wang2012intrinsic}, which is discussed below in Section \ref{sec:errors}.}  

{\q It is often desired to implement interacting spin models with multiple components of Ising interactions along various axes of the Bloch sphere, with the most general being the anisotropic Heisenberg model involving a sum of all three Ising terms 
\begin{equation}
H_{\rm Heis} = J^x_{ij}\sigma_x^i \sigma_x^j + J^y_{ij}\sigma_y^i \sigma_y^j + J^z_{ij}\sigma_z^i \sigma_z^j.
\end{equation}
Subclasses of the Heisenberg model arise naturally in physics and can possess useful symmetries \cite{sachdev2011quantum}. 
For example, the isotropic Heisenberg model $(J^x_{ij}=J^y_{ij}=J^z_{ij})$ is relevant to natural 3D magnetic interactions.  The XXZ model $(J^x_{ij}=J^y_{ij})$ and the XY model $(J^x_{ij}=J^y_{ij}; J^z_{ij}=0)$ result from standard transformations of Fermionic Hubbard models to spin models \cite{JW,BK}, which is discussed in Sec. \ref{sec:Hsequencing}. 
The XXZ and XY models also conserve the z-component of the total spin in the system, allowing simplifications to the spin dynamics and the interpretation of measurements.

Ion trap quantum simulators can generate generic Heisenberg models with the primitive of the Ising interaction discussed above, using a variety of extensions.  First, the three separate Ising terms in the Hamiltonian can exploit three independent modes (or spatial directions) of motion \cite{porras2004effective}, although this may not easily produce the same form or range of Ising interactions for all three axes. Second, the desired interactions can be applied sequentially in a Trotter expansion of the desired Hamiltonian \cite{trotter1959, suzuki1985, lloyd1996universal, lanyon2011universal, Johri2017}, as will be discussed in section \ref{sec:Hsequencing}. Third, a transverse field Ising model (Eq. \ref{eqn:TransversIsing}) can be applied with $B_y \gg J_{ij}$ so the field overwhelms the Ising interactions.  This can be seen by expanding the Ising term as 
$\sigma_x^i \sigma_x^j = (\sigma_+^i + \sigma_-^i)(\sigma_+^j + \sigma_-^j) \approx \sigma_+^i\sigma_-^j + \sigma_-^i\sigma_+^j = \sigma_x^i \sigma_x^j + \sigma_y^i \sigma_y^j$.  Here the strong field energetically forbids double spin-flips, or equivalently bestows fast oscillations to the $\sigma_{\pm}^i \sigma_{\pm}^j$ terms, which average to zero in an effective rotating-wave approximation \cite{cohen2015simulating}.} 

\section{Spin Hamiltonian Benchmarking and Many-Body Spectroscopy \label{sec:MB_hamil}}

A compelling Hamiltonian quantum simulation usually results in some type of nontrivial ground state or dynamics that may elude classical computation. It is therefore important to verify that the desired Hamiltonian is indeed being faithfully implemented by the quantum simulator \cite{cirac2012goals,hangleiter2017direct}.  For systems that are small enough and tractable for a direct comparison between the simulator's results and a theoretical calculation, this will provide some confidence that the proper simulation has been run. But scaling up the system can introduce additional imperfections that may call into question the accuracy of the applied Hamiltonian.

Two approaches for verifying quantum simulators beyond classically computability are the use of fault-tolerant techniques in the expression of the simulation in terms of discrete error-corrected gates \cite{preskill1998reliable,gottesman1998theory}, and the comparison of the results of multiple quantum simulators built upon different platforms \cite{leibfried2010could}.  Though this list is not comprehensive (see, e.g. \cite{hangleiter2017direct}), we examine these two approaches briefly below.

Universal Hamiltonian digital quantum simulators (DQS) {\q as will be discussed in section \ref{sec:seq}} break up the simulation evolution into a series of time steps, and the error $\epsilon$ introduced by this ``Trotter" expansion is bounded and inversely proportional to the number of steps $M$ \cite{lloyd1996universal}.  
This approach was notably employed for the implementation of various Hamiltonians, including many-body magnetic couplings, in a system of trapped ions \cite{lanyon2011universal}.
Because DQS relies on discrete quantum gate sets, one possibility for its verification is that this can in principle be accomplished through fault-tolerant error correction on the gates \cite{terhal2015quantum}. This increases the number of operations required, and {\q it has been shown that the run time can scale exponentially with the desired precision of the parameter being computed, in which case} the resource cost for {\q fault tolerant} DQS {\q for parameter estimation becomes} similar to that of universal quantum computing \cite{brown2006limitations,clark2009resource}.  Further, simulations with systems that lack universal gate sets or the digital simulation of open systems may render fault-tolerance unavailable in a DQS \cite{hauke2012can}. Since the cost of introducing currently known methods for fault-tolerance is too high for precision DQS \cite{kendon2010quantum}, other methods for verifying non-fault-tolerant-DQS are needed.

Another way one might test quantum simulators is to compare the results of two simulations.  This could involve comparing the results of simulations performed on different platforms \cite{leibfried2010could}, or even comparing the results obtained by using different simulation methods on the same machine.  A variant on this second theme is to run a {\q Hamiltonian} simulation forward and then backward in time \cite{cirac2012goals}, which may reveal flaws that are not undone by the time reversal, such as dissipation. An initial experiment demonstrated this time-reversal technique for a trapped ion quantum simulation by adiabatically ramping from an initial state of high magnetization along $y$, through a phase transition, and then back again \cite{islam2013emergence}.  Measurements of the magnetization at all three extrema in this time sequence revealed a revival in the magnetization, achieving an average of $\langle S_y \rangle = 68(4)\%$ of the initial value, in agreement with closed-system numerical integration.

Recently, a variational eigensolver approach has been combined with an ion trap quantum simulator to perform variational quantum simulation (VQS) of the lattice Schwinger model that combines some ways to verify some features of the result \cite{InnsbruckLGT2}.  The VQS uses feedback with a classical computer that translates measurement results from the AQS into expectation values of a software Hamiltonian to find energy eigenstates of the underlying Hamiltonian.  Since the conversion between the measurement and its interpretation happens in classical software, it provides a way to perform some verification of the resulting states because both the eigenvalues and their variances are accessible.  For instance, the VQS demonstrated in \cite{InnsbruckLGT2} measured the expectation value of the simulated Hamiltonian $E = \langle H \rangle$, as well as the expectation value of $\left( H - E \right)^2$, which should be zero if the state is an energy eigenstate of $H$ with eigenvalue $E$.  While this does not guarantee that the state found by the VQS is the ground state, this verification can be used to assess the confidence in the state being an eigenstate.

\subsection{Sources of Error} \label{sec:errors}
Quantum simulations with trapped ions can be susceptible to unwanted interactions that lead to inaccuracies in the simulation.  Many such error sources are common to both simulations and trapped ion quantum computing gates, such as spontaneous emission from the lasers driving spin transitions, and have been examined in detail elsewhere \cite{wineland1998experimental,Ozeri2007}.  Further, the simulation protocol itself may have known approximations (such as the Trotterization errors and non-adiabatic evolution) that may be rigorously bounded, though their effects may not be fully understood. 


The inclusion of a transverse effective magnetic field to the spin-dependent force of Eq. (\ref{eq:sdf}) includes higher order terms beyond the simple transverse Ising model of Eq. (\ref{eqn:TransversIsing}) \cite{wang2012intrinsic}. These additional terms can create substantial spin-motion entanglement that can affect measurements in bases other than the Ising direction.  For transverse field strengths that exceed the Ising coupling, the system begins to attain the character of an XY model with both $\sigma_x^i\sigma_x^j$ and $\sigma_y^i\sigma_y^j$ couplings, and the strong-field Ising model can break down.  The spins in this case do not strictly decouple from the phonons at any point in time, but it has been shown that it can typically be made small for experimentally accessible timescales  \cite{Wall2017}.

Protocols that rely on adiabatic ramping through a small gap can be susceptible to the breakdown of the adiabatic approximation in the region where the gap is small.  As we discuss below, for a linear ramp of $B(t)$ through a system energy gap of size $\Delta$, the adiabaticity criterion is approximately $|\dot{B}_y/\Delta^2| \ll 1$.  For cases where the gaps are known, the ramp rate can be adaptively matched to the gap to maximize adibaticity, a technique known as local adiabatic evolution \cite{roland2002quantum}.  However, since repeated experiments can be used to gather statistics about the final state, it has been shown that significant non-adiabaticity can be present and still allow the ground state spin configuration to be found due to its statistical prevalence \cite{richerme2013experimental}.

State preparation and measurement (SPAM) errors, which are to some degree common to simulators and gate-based quantum computers, are another source of error in lattice spin simulators.  Given an uncorrelated, single-shot, single ion SPAM fidelity $\mathcal{F}$, the probability of a single SPAM error is $1-\mathcal{F}^N$.  {\q The current state of the art SPAM fidelity for single qubits is $\mathcal{F}=0.99971(3)$, which will produce one error on average for single-shot projective measurements of a register with $N \ge \ln{2}/(1-\mathcal{F}) \approx 2400$ qubits \cite{Christensen2020}.}  In some special cases, repeated experiments can be used to mitigate this error through statistical methods \cite{shen2012correcting}.  Cross-talk between neighboring ions can also lead to measurement errors, and the ion positions must be calibrated to define a region of interest on the camera for each ion's fluorescence detection.  {\q State detection infidelity from neighboring ions is} typically no more than a few percent {\q per qubit \cite{zhang2017observation}, although this error can be suppressed to well below $1\%$ \cite{cetina2020quantum} and can be made even smaller with ion-shuttling \cite{Kielpinski:2002}.}

There are various sources of decoherence quantum simulations in the trapped ion platform, such as stray magnetic and electric fields, mode frequency drifts, off-resonant motional excitation and spontaneous emission.  While these errors have been analyzed in the context of quantum computing \cite{wineland1998experimental}, decoherence can set a practical time limit for quantum simulations that may lead to other errors, such as diabatic errors.  Since many of these error sources increase with system size, it may be necessary to employ methods of mitigation such as magnetic field shielding \cite{Ruster2016}.

{\q Off-resonant excitation to motional modes is nominally already considered in the spin-motional coupling described by Eq. (\ref{alpha}). The probability of motion-induced spin-flip errors in a chain of $N$ ions can be estimated by $\varepsilon=\sum_{i,m}^N|\alpha_{i,m}|^2$,
summing over all modes. By neglecting the time dependence and assuming equally-contributing modes, this error is expected to scale as $\varepsilon \sim N(\eta \Omega/\delta)^2$, where $\delta$ is the smallest detuning from any mode. This same approximation was made in the derivation of Eq. (\ref{Jij}), leading to the scaling $J_{ij}\sim (\eta\Omega)^2/\delta$. Motion induced spin-flip errors can therefore be mitigated by increasing both $\delta$ (Eq. (\ref{eq:sdf})) and the Raman Rabi frequency $\Omega=g_1g_2/2\Delta$ to keep the same spin-spin interaction. However, increasing the Rabi frequency may increase spontaneous emission errors, which grow linearly with Rabi frequency for large detuning as $\Gamma=\gamma\Omega/4\Delta$, where the atomic linewidth $\gamma$ is defined in section \ref{sec:atomic_ion_spin} and $\Delta\gg\gamma$ is assumed. Given a level of motion-induced spin flip error $\varepsilon$, it can be shown that the spontaneous emission error during the interaction time $t_{J_0}=2\pi/J_0$ scales as $\sqrt{N/\varepsilon}$ \cite{kim2010quantum}}.

\subsection{Benchmarking Ising couplings}
For the Ising spin models considered in this review, it is crucial to validate the strength of the Ising coupling matrix $J_{ij}$ of Eqs. (\ref{Ham})-(\ref{Jij}) and effective magnetic fields in Eq. (\ref{field}). For small numbers of spins, it is possible to directly extract the Ising couplings and fields by preparing the spins in a $\sigma_z$ eigenstate and subjecting them to the $\sigma_\theta$ Ising interactions or field terms. The resulting oscillations in population are given by the energies of the occupied states, so performing a Fourier transform on these oscillations directly provides the energy differences.

\begin{figure}[ht]
\includegraphics[width=0.8\linewidth]{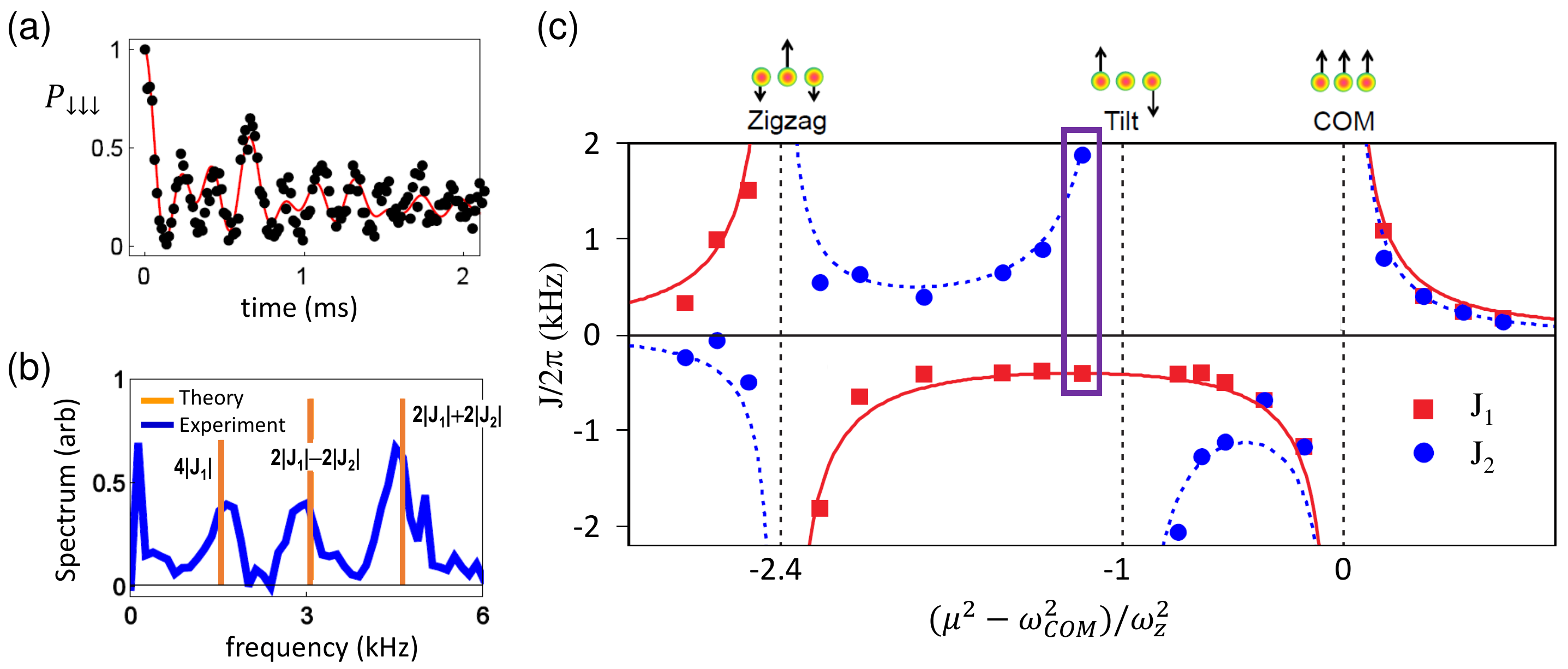}
\caption{Direct measurement of Ising nearest-neighbor ($J_1$) and next-nearest-neighbor ($J_2$) interactions for $N=3$ trapped ion spins. (a) Measured time evolution of the probability of state $\ket{\downarrow\downarrow\downarrow}$ subject to global Ising interactions, with the spins initialized in the state $\dn\dn\dn$.  The solid (red) line is a fit to theory with an empirical exponential decay. (b) The Fourier spectrum of the oscillations in (a) exposes the frequency splittings of the Ising Hamiltonian. (c) Extracted Ising couplings $J_1$ and $J_2$ from the Fourier spectra as a function of the applied beat-note detuning $\mu$, scaled so that the center-of-mass (COM), Tilt, and Zigzag modes of transverse motion occur at $({\mu}^2-\omega_{\rm COM}^2)/\omega_z^2 = 0$, $-1$, and $-2.4$, respectively. The red squares ($J_1$) and blue circles ($J_2$) are experimentally measured couplings, and the lines are calculated from Eq. (\ref{Jij}) with no free fit parameters. The particular measurements in (a) and (b) correspond to the scaled laser beat note detuning $({\mu}^2-\omega_{\rm COM}^2)/\omega_z^2 = -1.2$ indicated by the points highlighted in the violet rectangle in (c). Adapted from \citealp{kim2009entanglement}}
\label{fig:SpinInteraction}
\end{figure}

An example of directly-measured Ising oscillations and their resultant Fourier transform is shown in Figs. \ref{fig:SpinInteraction}a-b for $N=3$ spins. The extracted interaction strengths are shown in Fig. \ref{fig:SpinInteraction}c. This technique is effective in contexts of both continuous \cite{kim2011quantum} and digital \cite{lanyon2011universal} simulations of Ising models. A similar method has been used to extract the strengths of a magnetically induced spin-spin couplings \cite{Khromova2012,piltz2016versatile}. Neither technique scales well to more than a few spins, owing to the difficulty in extracting many closely-spaced frequencies in the spin oscillations.

Individual Ising couplings within a given spin chain can be measured using an auxiliary state of the ions, even for large numbers of spins. Because the Ising couplings depend on the vibrational mode spectrum, all ions must be physically present in the trap to obtain a meaningful result, but the spectrum of the population oscillations illustrated in Fig. \ref{fig:SpinInteraction} would be difficult to obtain if all spins participate in the many-body dynamics. An alternative approach is to perform a separate measurement for each individual Ising coupling, by ``hiding" all ions except the pair of interest into an auxiliary internal state that does not experience the spin-dependent force giving rise to Ising couplings. In this manner, the frequency with which the ions $i$ and $j$ of interest oscillate between correlated states $\ket{00}$ and $\ket{11}$ for example, allow the determination of the Ising matrix $J_{ij}$ \cite{jurcevic2014quasiparticle}.

\begin{figure}[ht]
\includegraphics[width=0.6\linewidth]{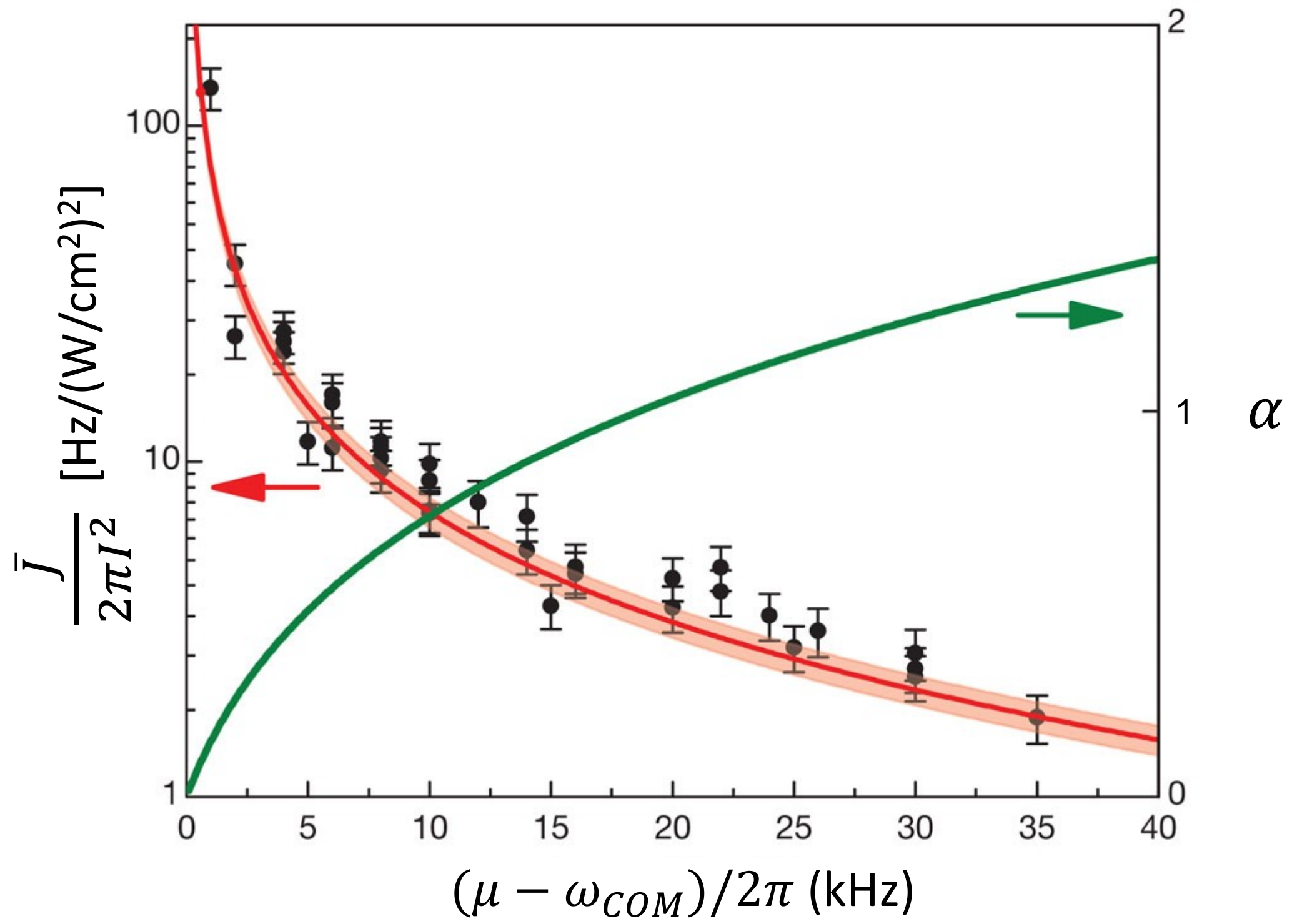}
\caption{Mean-field benchmarking of Ising couplings in a 2D crystal of 206(10) ions confined in a Penning trap. The spins are all initialized in a separable product state slightly tipped from the axis of a subsequently applied (weak) long-range Ising interaction.  The resulting global spin-precession of the ions is measured as a function of the detuning of the optical dipole force laser beams from the axial center-of-mass (COM) mode (see Sec. \ref{sec:intro} and Fig. \ref{fig:modes}b). Each point is generated by measuring the mean Ising coupling at a particular the laser beam intensity $I$. The solid line (red) is the prediction of mean-field theory that accounts for couplings to all N transverse modes, with no free parameters. The line’s breadth reflects experimental uncertainty in the initial tipping angle of the spins. The mean-field prediction for the average value of the power-law exponent $\alpha$ from Eq. (\ref{Jpowerlaw}), is drawn in green (right axis, linear scale). For stronger Ising interactions, the mean-field approach breaks down.  Adapted from \citealp{britton2012engineered}.}
\label{fig:globalspinprec}
\end{figure}

{\q For large collections of spins with long-range interactions, benchmarking of weak Ising interactions can be accomplished by measuring the global precession of the spins. Here, the spins are each prepared in an identical state that is tipped away from the Bloch sphere axis of the Ising interaction and the resulting dynamics of the global Ising interaction can be recorded. Figure \ref{fig:globalspinprec} shows measurements of this type of benchmarking in a collection of over 200 trapped ion spins \cite{britton2012engineered}, agreeing well with mean-field theory. For sufficiently strong Ising interactions, as shown in this experiment, the mean-field approximation breaks down, indicating the entanglement between the spins.}

\begin{figure}[ht]
\includegraphics[width=0.8\linewidth]{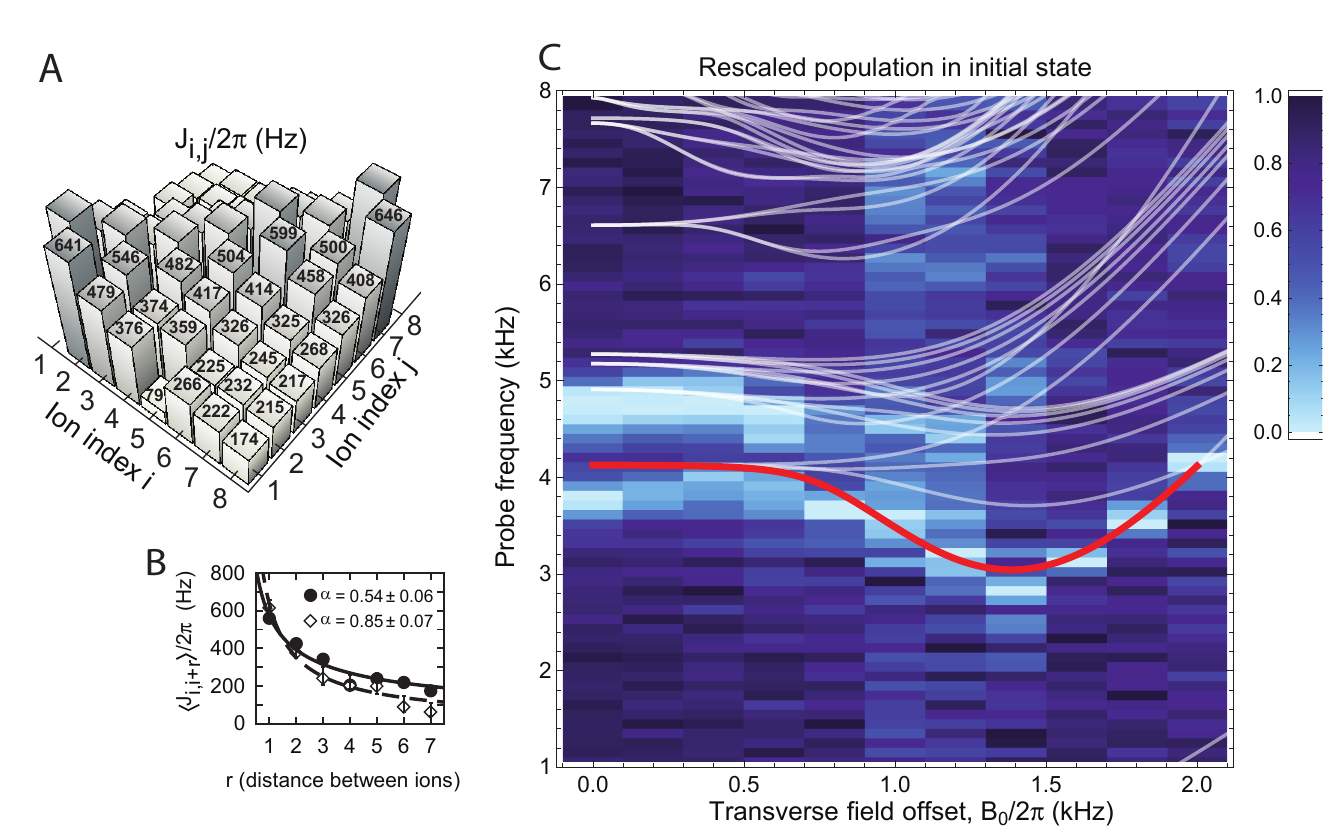}
\caption{(a) All elements of a Ising coupling
matrix measured with a spectroscopic probe in the form of a modulated transverse field. (b) Measurements of two sets of Ising coupling matrices, demonstrating two different effective interaction ranges across the chain, with the solid lines best fits to inverse power law form expected in Eq. (\ref{Jpowerlaw}). (c) Rescaled populations in the approximate ground state versus a static transverse field offset $B_0$ and the modulation frequency of an small additional transverse field. Calculated energy levels, based on measurements of trap and laser parameters, are overlaid as thin white lines, and
the lowest coupled excited state as a thick red line, showing the critical gap near $B_0/2\pi=1.4$  kHz. The energy of the ground state is always taken to be zero.  From \citealp{senko2014coherent}.}
\label{fig:MBS}
\end{figure}

Other spectroscopic techniques for probing the energy spectrum of the bare Ising Hamiltonian are also possible \cite{senko2014coherent}. For instance, in the tranverse Ising model of Eq. (\ref{eqn:TransversIsing}), modulating the effective field $B_y(t)$ at a frequency commensurate with an energy difference in the full spin Hamiltonian will drive transitions between the two differing states. Specifically, taking $ B_y(t) = B_0 + B_p \sin (\omega_{mod} t)$ with $B_p<<J$, the frequency $\omega_{mod}$ at which such transitions occur are directly related to the Ising couplings $J_{ij}$ \cite{senko2014coherent}. Figure \ref{fig:MBS}a-b illustrates examples where the Ising matrix is directly measured using this technique to confirm the validity of power-law approximation described in Eq. (\ref{Jpowerlaw}) for a handful of spins. The technique of applying a small oscillating term can be generalized to other cases, for example, measuring the critical (minimum) gap between ground and first excited state of the transverse field Ising model, as shown in Fig. \ref{fig:MBS}c.

\section{Equilibrium studies \label{sec:equil}}
Finding ground states of a non-trivial Hamiltonian has tremendous importance in various disciplines across condensed matter physics, quantum chemistry and computer science. In condensed matter physics, the rich phenomena of complex quantum systems can be understood by finding the ground states of the corresponding many-body Hamiltonian \cite{kohn1999nobel,foulkes2001quantum,schollwock2005the}. In quantum chemistry and molecular physics, the central problems is to determine the electronic structure and the ground-state energy of atoms and molecules \cite{jensen1989an}. In computer science, the ground state of the complex quantum Hamiltonian can encode other computational problems such as satisfiability or optimization \cite{kolsgaard2001a,lloyd2008quantum,albash2018adiabatic}. 

The computational tasks of finding {\q the ground state of non-trivial Hamiltonians} are classically demanding because of the exponentially increasing Hilbert space of the Hamiltonian. A quantum simulator is expected to provide a solution beyond the limitations of classical computation. Recently various theoretical schemes for the ground-state problem have been proposed and proof-of principle experimental demonstrations have been performed, including adiabatic preparation \cite{friedenauer2008simulating,kim2010quantum,edwards2010quantum,islam2011onset,kim2011quantum,schneider2012experimental,islam2013emergence,richerme2013experimental,richerme2013quantum}, direct cooling by bath-engineering \cite{barreiro2011open,lin2013dissipative}, and algorithmic cooling schemes \cite{baugh2005experimental,xu2014demon,zhang20}. For the case of adiabatic method, it has been shown to be closely related to adiabatic quantum computation, which is proved to be equivalent to a universal quantum computer \cite{albash2018adiabatic}.

We focus on the adiabatic preparation of the ground state of quantum spin models with trapped atomic ion spins, including a description of the general scheme of the experimental procedure and various adiabatic ramping protocols. Following a discussion of the adiabatic protocol applied to varying numbers of trapped ion spins, we consider how this protocol can be optimized and applied to broader classes of spin models, and briefly discuss the case of a spin-1 system. These quantum spin models clearly show the essences of the adiabatic quantum simulation with wide applications. Moreover, these quantum spin models can describe a large class of many-body quantum physics in condensed matter such as quantum magnetism \cite{moessner2006geometrical}, spin glasses \cite{binder1986spin}, and spin liquids \cite{balents2010spin}. The solutions of certain spin Hamiltonians are also connected to many other computational problems including optimization problems when the system is extended to 2 dimensions \cite{albash2018adiabatic}.    

\subsection{Adiabatic Ground State Preparation \label{sec:adiabatic_prep}}
Adiabatic ground state preparation is analogous to that of adiabatic quantum computation \cite{farhi2000quantum,albash2018adiabatic}: a quantum system is initialized to the ground state of a trivial Hamiltonian $H_{\rm triv}$.  Next, the Hamiltonian is adiabatically deformed into the Hamiltonian of interest $H_{\rm prob}$, whose ground state encodes the solution of a problem that has been mapped to this final Hamiltonian. The adiabatic evolution is generated by 
\begin{equation}
H (s)= (1-s) H_{\rm triv} + s H_{\rm prob},
\label{eqn:adiabaticevolution}
\end{equation}
where $s=s(t)$ is a time dependent parameter changing from 0 to 1 during the time interval from $t=0$ to $t=t_{f}$. 
In the context of trapped ion spin models, the trivial spin Hamiltonian can be an effective magnetic field as described by Eq. (\ref{field}) and the Hamiltonian of interest can be a fully-connected transverse Ising model of Eq. (\ref{eqn:TransversIsing}).
The determination of ground states of the long range transverse field Ising model cannot always be predicted, even with just a few dozen spins \cite{sandvik2010ground}. 

Although the fidelity of remaining in the ground state of Eq. (\ref{eqn:adiabaticevolution}) can always be improved by evolving more slowly, a practical upper limit on the transition time is enforced by the finite coherence time of the chosen experimental platform. Given a fixed transition time, it is possible to further optimize the preparation fidelity by adjusting the transition rate based on the local energy gap to the nearest excited state \cite{richerme2013experimental}.
Such ``local adiabatic evolution" can be used for improved preparation and determination of many-body ground states in a trapped-ion quantum simulator. Compared with other adiabatic methods, local adiabatic evolution \cite{roland2002quantum} yields the highest probability of maintaining the ground state in a system that is made to evolve from an initial Hamiltonian to the Hamiltonian of interest. Compared with optimal control methods \cite{kjaneja2005optimal,krotov1996global}, local adiabatic evolution may require knowledge of only the lowest $\sim N$ eigenstates of the Hamiltonian rather than all $2^N$ values. These methods have been used in both linear Paul traps \cite{richerme2013experimental} and Penning traps \cite{safavi-naini2018verification} to demonstrate optimized ground state preparation as well as a method to find the ground state spin ordering, even when the evolution is non-adiabatic.

\begin{figure}[ht]
\includegraphics[width=0.75\linewidth]{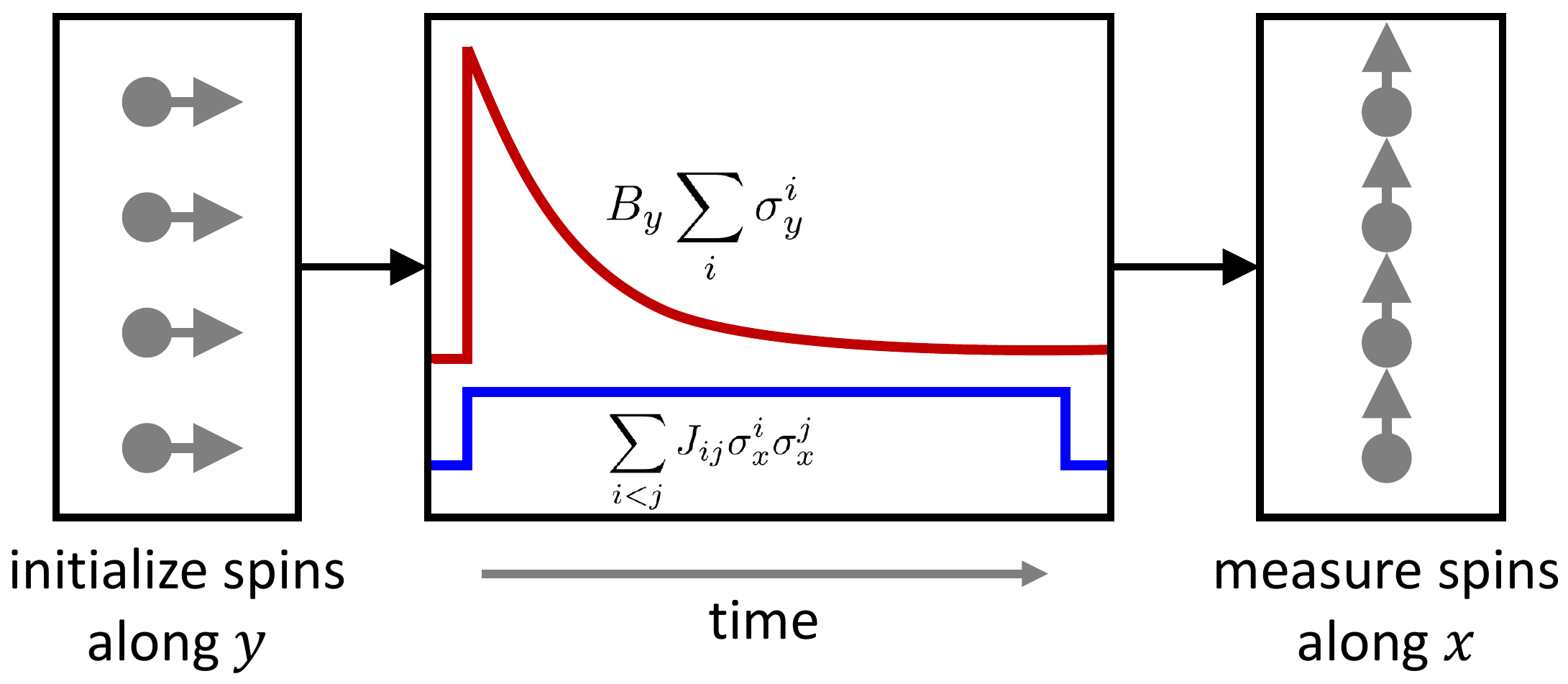}
\caption{After preparing the spins in the ground state of a $y-$polarized field $B_y$, the spins evolve subject to a long-range transverse Ising model described by Eq. (\ref{eqn:TransversIsing}), with the condition $B_y(t=0) \gg J_{ij}$. During the evolution, the strength of the field (red curve) is reduced to zero or an intermediate value compared to the Ising couplings (blue curve).  Finally, the spins are measured along any axis of the Bloch sphere. (In much of the work reported below, the measurements are taken to be along the Ising-coupling (x) direction of the Bloch sphere for each ion.) If the evolution is adiabatic, the resulting state should have remained in the ground state of the Hamiltonian throughout. Adapted from \citealp{islam2011onset}.}
\label{fig:AdibaticProtocol}
\end{figure}

For example, to find the ground state of a fully-connected Ising Hamiltonian in Eq. (\ref{eqn:TransversIsing}) via an adiabatic protocol, the spins can be initialized to point along the transverse magnetic field direction with $B_y \gg \text{Max}(J_{ij})$. This initial state is, to good approximation, the instantaneous ground state of the full Eq. (\ref{eqn:TransversIsing}). After initialization, the (time-dependent) transverse field $B_y(t)$ can then be ramped adiabatically from $B_y(t=0)=B_0$ to $B_y(t=t_f)=0$, ensuring that the system remains in its instantaneous ground state during its evolution, as depicted in Fig. \ref{fig:AdibaticProtocol}. At the conclusion of the ramp, the ground state spin ordering of the Ising Hamiltonian [first term in Eq. (\ref{eqn:TransversIsing})] may be either directly read out or used as a starting point for further experiments.

\begin{figure}[htbp!]
\includegraphics*[width=.5\linewidth]{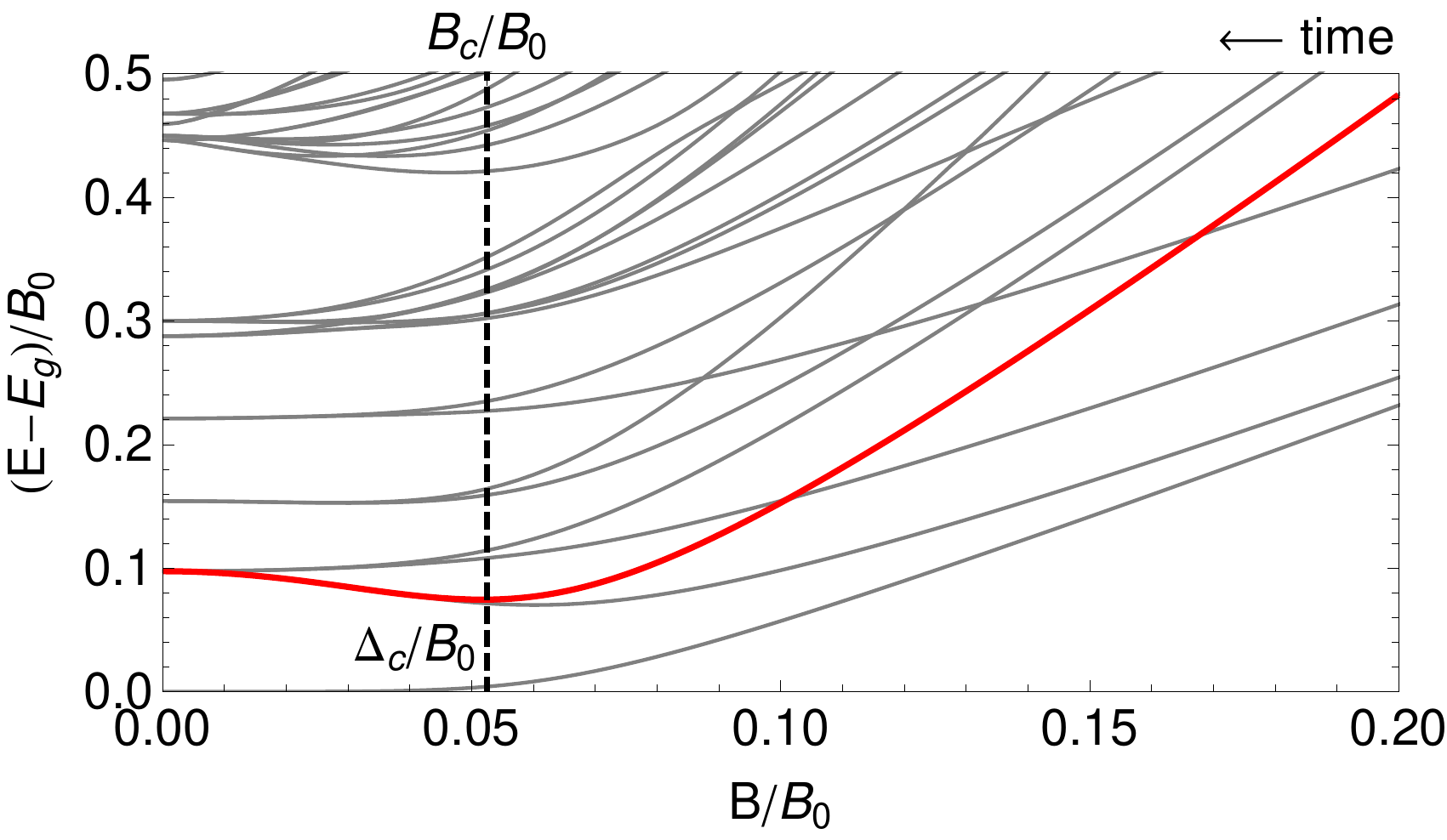}
\caption{Low-lying energy eigenvalues of Eq. (\ref{eqn:TransversIsing}) for $N=6$, with the ground state energy $E_g$ set to 0, $B_0=5J_{\text{max}}$, and the long-range $J_{ij}$ couplings determined from experimental conditions (see text). Indicated in bold red is the first coupled excited state, the minimum of which determines the critical field $B_c$ and the critical gap $\Delta_c$. From \citealp{richerme2013experimental}.}
\label{fig:OptRampEnergyLevels}
\end{figure}

\subsubsection{Adiabatic Ramp Profiles}
Fig. \ref{fig:OptRampEnergyLevels} shows the energy level spectrum for the Hamiltonian in Eq. (\ref{eqn:TransversIsing}) for $N=6$ spins. Since the Hamiltonian obeys $Z_2$ symmetry \cite{sachdev2011quantum} (as well as parity symmetry in the experiments), the ground state $\ket{g}$ is coupled to only a subset of the excited energy eigenstates. The first coupled excited state, shown in red in Fig. \ref{fig:OptRampEnergyLevels}, is the lowest energy excited state $\ket{e}$ for which $\bra{e}\sigma_y\ket{g}\neq0$. This state displays a general property seen in most adiabatic quantum simulations -- namely, the existence of a critical gap $\Delta_c$ that is central to parameterizing the adiabaticity of a given ramp. Many different ramp profiles allow one to transform from the initial Hamiltonian to the Ising Hamiltonian, each with different implications for adiabaticity and ground state preparation. Three possibilities are discussed below.

\begin{figure}[t!]
\includegraphics[width=.426\linewidth]{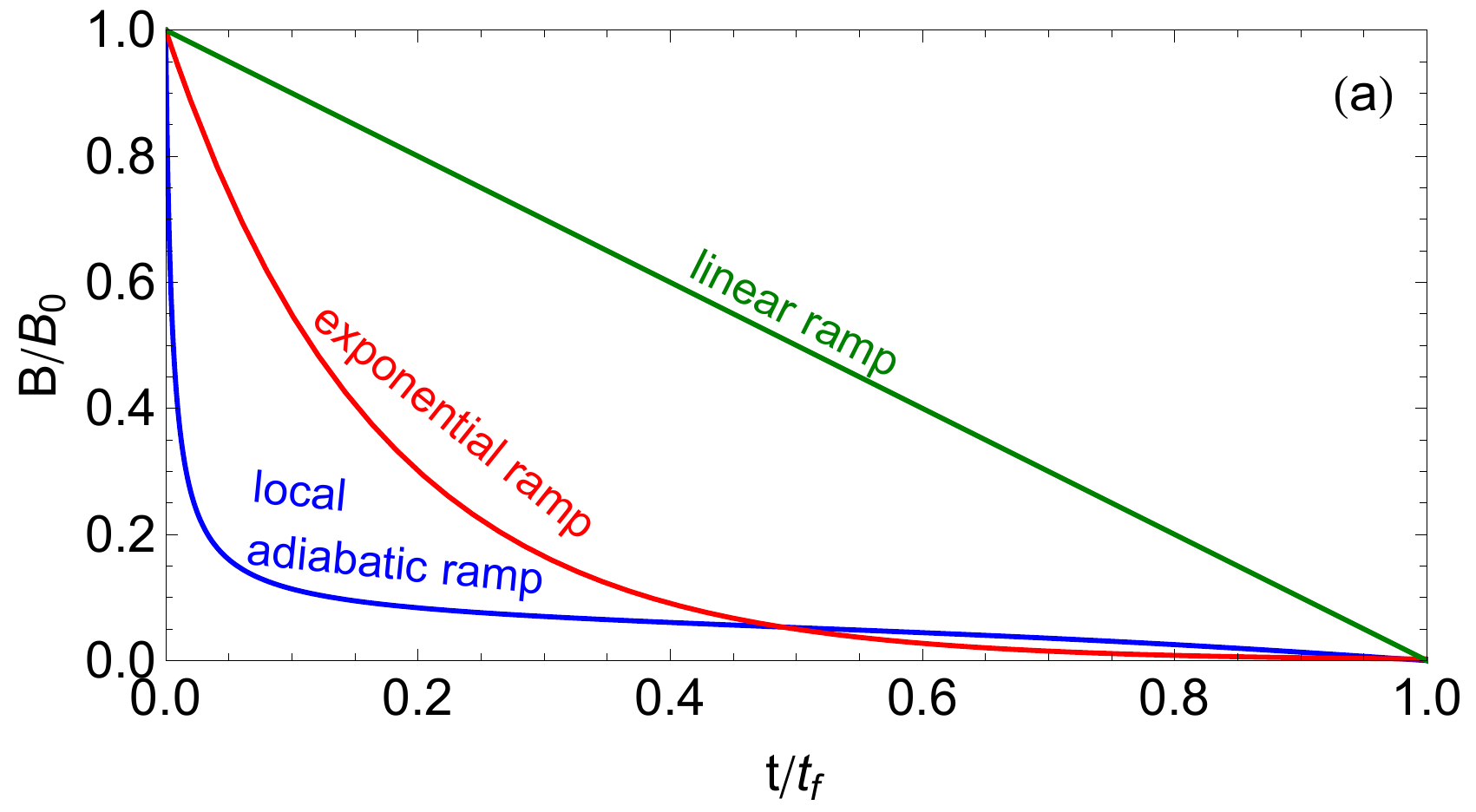}
\includegraphics[width=.278\linewidth]{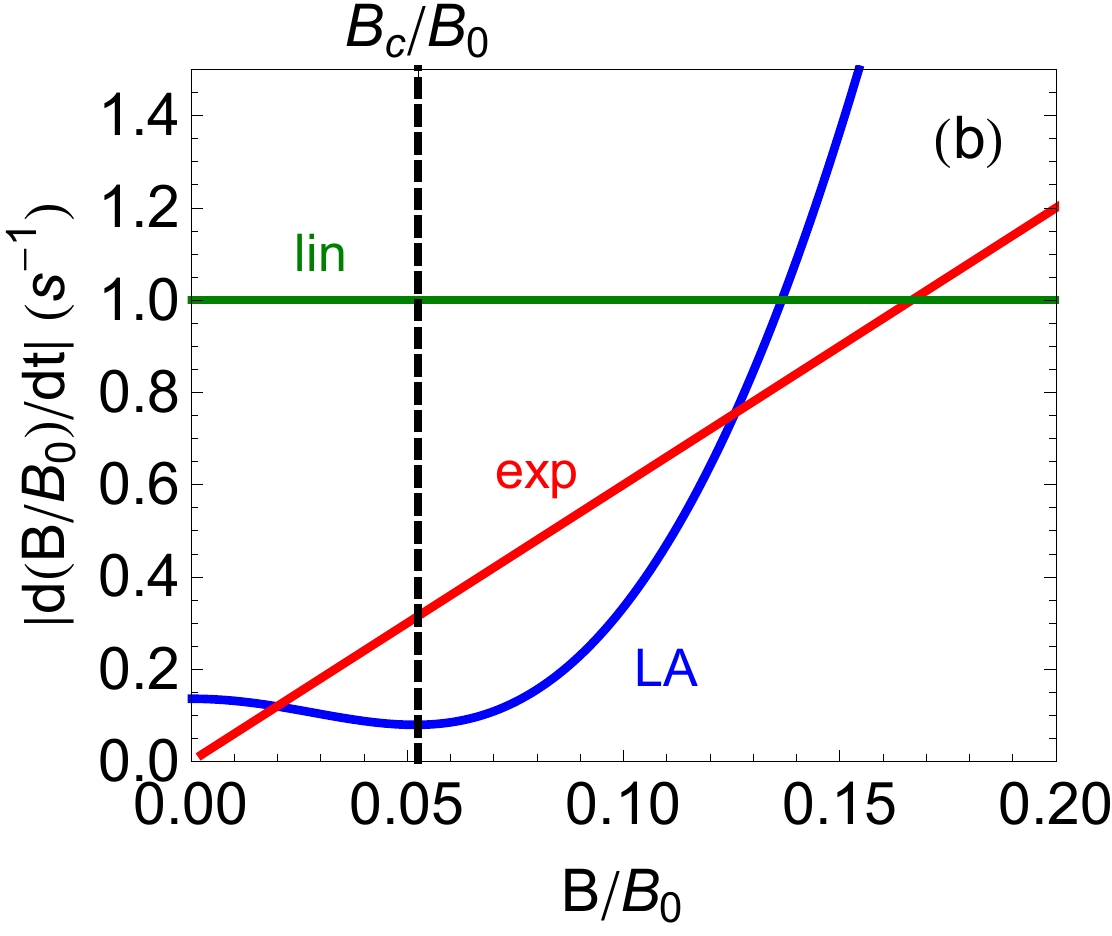}
\includegraphics[width=.278\linewidth]{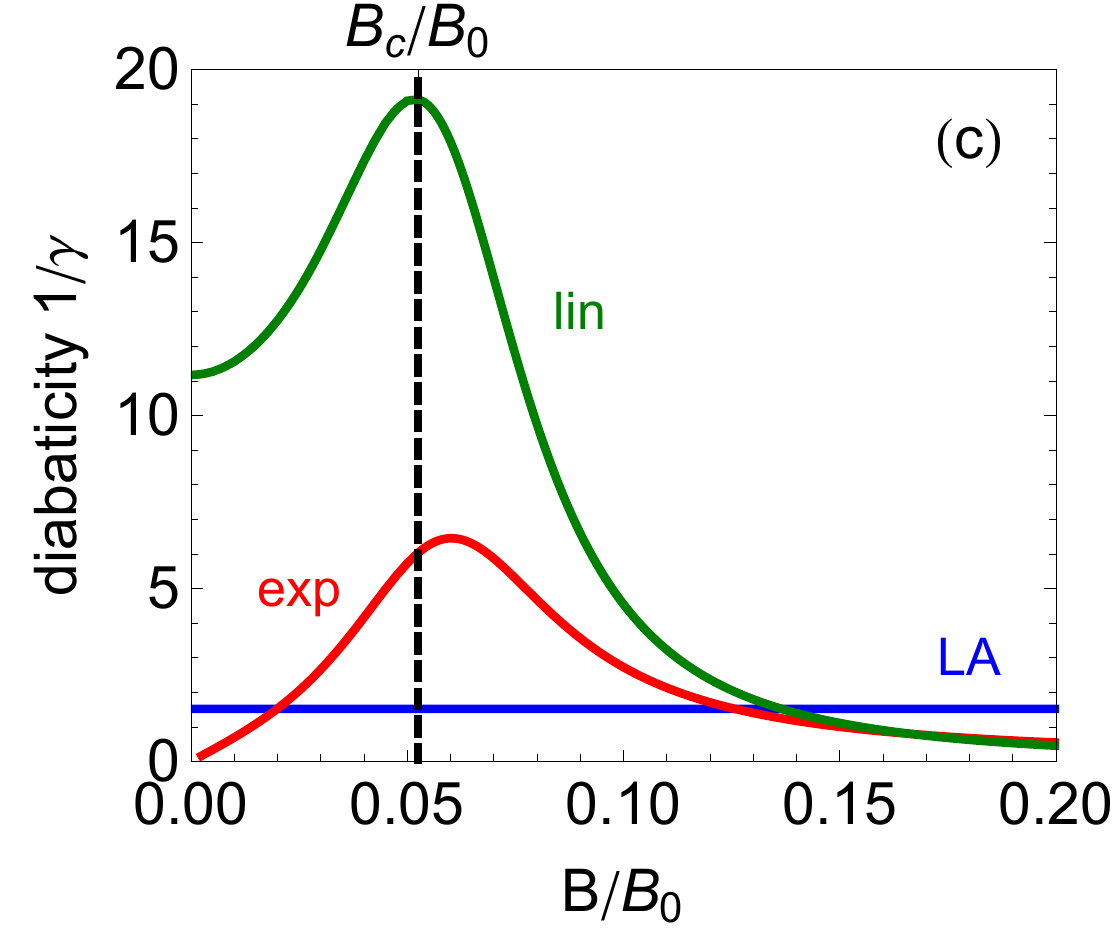}

\caption{(a) Local adiabatic ramp profile calculated for the energy levels in Fig. \ref{fig:OptRampEnergyLevels}, along with a linear ramp and an exponential ramp with decay constant $\tau=t_f/6$. (b) The slope of the local adiabatic (LA) ramp is minimized at the critical field value $B_c$, and is smaller than the slopes of the exponential and linear ramps at the critical point. (c) The inverse of the adiabaticity parameter $\gamma$ (see text) is peaked near the critical point for exponential and linear ramps but constant for the local adiabatic profile. Adapted from \citealp{richerme2013experimental}.}
\label{fig:OptRampProfiles}
\end{figure}

\textbf{Linear Ramps.}  For a linear ramp, the time-dependent transverse field $B_y$ in Eq. (\ref{eqn:TransversIsing}) takes the form $B_y^{\text{lin}}(t)=B_0(1-t/t_f)$, with a ramp profile shown in Fig. \ref{fig:OptRampProfiles}a. To determine whether such a ramp is adiabatic or not, it can be compared to the adiabatic criterion \cite{messiah1962quantum}
\begin{equation}
\label{eqn:adiabatic}
\left|\frac{\dot{B}_y(t) \epsilon}{\Delta_c^2}\right| \ll 1
\end{equation}
where $\dot{B}_y(t)$ is the rate at which the transverse field is changed and 
$\epsilon=\text{Max}[\bra{e}dH/dB_y\ket{g}]$ is a number of order unity that parametrizes the coupling strength between the ground state $\ket{g}$ and the first coupled excited state $\ket{e}$. Eq. (\ref{eqn:adiabatic}) highlights that fast ramps and small critical gaps can greatly decrease adiabaticity.

To satisfy the adiabatic criterion, a linear ramp must proceed slowly enough so that the total time \mbox{$t_f \gg B_0/\Delta_c^2$}. For the $N=6$ Ising Hamiltonian shown in Fig. \ref{fig:OptRampEnergyLevels}, $B_0=3.9$ kHz and $\Delta_c=0.29$ kHz, giving the adiabaticity requirement \mbox{$t_f\gg46$ ms}. This time is long compared with the typical coherence time of ion trap quantum simulation experiments. It is therefore desirable to seek alternative ways to decrease $B(t)$ more quickly while maintaining adiabaticity.

\textbf{Exponential Ramps.} Decreasing the transverse field exponentially according to $B_y^{\text{exp}}(t)=B_0\exp(-t/\tau)$, with $t_f=6\tau$, can yield a significantly more adiabatic evolution than linear ramps for the same $t_f$. Figure \ref{fig:OptRampEnergyLevels} shows that the instantaneous gap $\Delta$ between the ground and first coupled excited state is large at the beginning of the ramp and small only when $B$ approaches $0$. Exponential ramps exploit this gap structure by quickly changing the field at first, then gradually slowing the rate of change as $t\rightarrow t_f$.Such ramps have been used to produce ground states in several of the previously discussed experiments, such as \cite{kim2010quantum},\cite{islam2011onset} and \cite{islam2013emergence}.

At the critical point of the Hamiltonian shown in \mbox{Fig. \ref{fig:OptRampEnergyLevels}}, $|\dot{B}_{exp}(t)|=0.3B_0/t_f$. The adiabaticity criterion of Eq. (\ref{eqn:adiabatic}) then requires $t_f \gg 14.5$ ms, a factor of 3 less time than the requirement found for linear evolution. Note that the adiabaticity gains of exponential ramps can be realized whenever the critical gap occurs towards the end of the ramp ($B_c/B_0 < \tau/t_f$), which is generally the case for the transverse Ising Hamiltonian of Eq. (\ref{eqn:TransversIsing}).

\textbf{Local Adiabatic Ramps.} Local adiabatic ramps seek to keep the adiabaticity fixed at all points along the evolution by adjusting $\dot{B}_y(t)$ based on the instantaneous gap $\Delta (B_y(t))$ that varies with the field profile $B_y(t)$ \cite{roland2002quantum, quan2010testing}. If the adiabaticity parameter is defined as
\begin{equation}
\label{eqn:gamma}
\gamma=\left|\frac{\Delta (B_y(t))^2}{\dot{B}_y(t)}\right|
\end{equation}
then a local adiabatic ramp would follow the profile $B_y(t)$ that solves the differential equation \ref{eqn:gamma} with $\gamma$ fixed. Adiabaticity then requires $\gamma\gg 1$.

To solve Eq. (\ref{eqn:gamma}), it is necessary to know $\Delta(t)$ everywhere along the evolution. This requires knowledge of the first \textit{coupled} excited state of the $N$-spin Hamiltonian of Eq. (\ref{eqn:TransversIsing}), {\q which is always the first excited state at $B_y=0$ and the $(N+1)^{\text{st}}$ excited state at large $B_y$}. Determining the local adiabatic evolution profile therefore relies on calculation of only the lowest $\sim N$ eigenvalues, which is much more computationally approachable than direct diagonalization of a $2^N\times 2^N$ matrix \cite{lanczos1950iteration}.

For a local adiabatic ramp, {\q the total evolution time $t_f$} may be calculated by integrating Eq. (\ref{eqn:gamma}). Since $\dot{B}_y(t)$ is negative throughout the evolution,
\begin{equation}
\label{eqn:tf}
t_f=\gamma\int_0^{B_0}\frac{dB}{\Delta^2 (B)}
\end{equation}
which shows a linear relationship between the total time $t_f$ and the adiabaticity parameter $\gamma$. Satisfying the adiabaticity condition $\gamma \gg 1$ for the Hamiltonian in \mbox{Fig. \ref{fig:OptRampEnergyLevels}} implies $t_f \gg 3.6$ ms, a factor of 4 and 12 less time than exponential and linear ramps, respectively. The fact that local adiabatic evolution can lead to faster ramps while satisfying adiabaticity has been well-explored in \cite{roland2002quantum}, where it was shown that local adiabatic ramps could recover the quadratic speedup of Grover's quantum search algorithm \cite{nielsen2000quantum}. In contrast, it was found that linear ramps offer no improvement over classical search algorithms \cite{farhi2000quantum}.

Fig. \ref{fig:OptRampProfiles}a compares a linear, exponential, and local adiabatic ramp profile for the Hamiltonian shown in \mbox{Fig. \ref{fig:OptRampEnergyLevels}}. The local adiabatic ramp spends much of its time evolution in the vicinity of the critical point, since the transverse field changes slowly on account of the small instantaneous gap. This is further illustrated in Fig. \ref{fig:OptRampProfiles}b, which shows that at the critical point, the slope of the local adiabatic ramp is minimized and smaller than slopes of the exponential or linear ramps. As a result, the inverse adiabaticity $1/\gamma$ is peaked near the critical point for exponential and linear ramps, greatly increasing the probability of non-adiabatic transitions away from the ground state (see Fig. \ref{fig:OptRampProfiles}c). By design, the local adiabatic ramp maintains constant adiabaticity for all values of $B$ and does not suffer from large non-adiabaticities near $B_c$.

\subsubsection{General Adiabatic Simulation Issues}
To determine the effects of a chosen adiabatic ramp protocol, the probability of creating the ground state can be measured following any chosen ramp profile of identical times. \cite{richerme2013experimental} used $N=6$ ions and chose the trap voltages and the laser detuning $\mu$ to give AFM spin-spin interactions of the form $J_{ij}\approx(0.77~\text{kHz})/|i-j|$. These long-range AFM interactions lead to a fully-connected, frustrated system as all couplings cannot be simultaneously satisfied. Nevertheless, the ground state of this system reduces to an equal superposition of the two N\'eel-ordered AFM states, $(\ket{\downarrow\uparrow\downarrow\uparrow\downarrow\uparrow}+\ket{\uparrow\downarrow\uparrow\downarrow\uparrow\downarrow})/\sqrt{2}$.

\begin{figure}[h]
\includegraphics*[width=.5\linewidth]{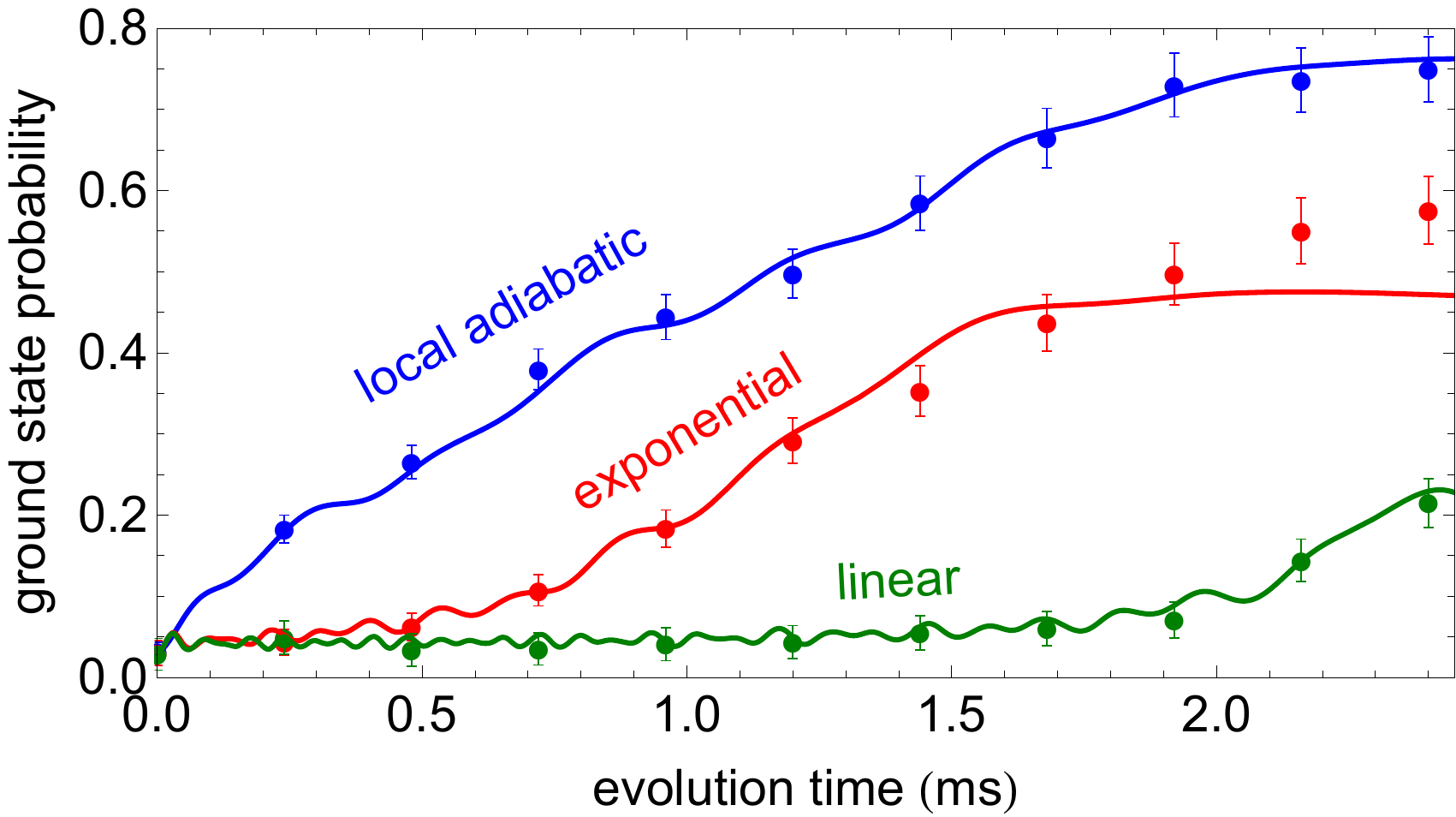}
\caption{Probability of preparing the AFM ground state for various times during $t_f=2.4$ ms simulations with three different ramp profiles. The linear ramp takes $\sim2.3$ ms to reach the critical point, while the local adiabatic and exponential ramps need only 1.2 ms. Locally adiabatic ramps yield the highest preparation fidelity at all times. Adapted from \cite{richerme2013experimental}.}
\label{fig:OptRamprampdata}
\end{figure}

The data in Fig. \ref{fig:OptRamprampdata} show how the AFM ground state probability grows during a single 2.4 ms linear, exponential, or local adiabatic ramp. Each data point is the result of 4000 repetitions of the same experiment, with error bars that account for statistical uncertainty as well as estimated drifts in the Ising coupling strengths. In agreement with the arguments above, the data show that local adiabatic ramps prepare the ground state with higher fidelity than exponential or linear ramps. The ground state population grows quickly under local adiabatic evolution since the transverse field $B(t)$ is reduced quickly at first. In contrast, the linear ramp does not approach the paramagnetic to AFM phase transition until $\sim 2$ ms, and the AFM probability is suppressed until this time.

The solid lines in Fig. \ref{fig:OptRamprampdata} plot the theoretical prediction of the ground state probability with no free parameters. In each case, the Schr\"odinger equation is numerically integrated using Hamiltonian (\ref{eqn:TransversIsing}), the desired $B(t)$, and the initial state $\ket{\psi(0)}=\ket{\downarrow\downarrow\downarrow\ldots}_y$. At the end of the ramp, the overlap between the final state $\ket{\psi(t_f)}$ and the AFM ground state $(\ket{\downarrow\uparrow\downarrow\ldots}+\ket{\uparrow\downarrow\uparrow\ldots})/\sqrt{2}$ is calculated to extract the probability of the ground state spin configuration. Effects of decoherence-induced decay in the ground state probability are included by multiplying the calculated probability at time $t$ by $\exp[-t/t_d]$, where $t_d$ is the measured $1/e$ coherence time of the spin-spin interactions. 

{\q The key to all adiabatic protocols is that the ramp rate must remain slow when compared to the critical energy gap, as shown above in Eq. (\ref{eqn:adiabatic}). However, determining the scaling of the critical gap with system size is itself a notoriously difficult problem in the general case \cite{albash2018adiabatic}. For simple problems, such as the ground state of a Lipkin-Meshkov-Glick model, the gap is known to shrink only polynomially for large-$N$ \cite{caneva2008adiabatic}. For more complex problems, such as those in the $NP$-complexity class, the gap may close exponentially quickly with increasing system size \cite{das2008quantum,morita2008mathematical}. In these cases, or in cases for which the gap scaling is unknown, experimental implementations will require exponentially longer ramp times to ensure that adiabaticity is maintained as the problem size grows.

One obvious question is whether quantum adiabatic protocols are useful for solving complex computation or quantum many-body type problems. There are some reasons to be hopeful. For certain problems, such as the Grover search algorithm, the use of local adiabatic ramp profiles have been shown to provide the same quadratic quantum advantage as found in the circuit model \cite{roland2002quantum}. For other problems, such as finding the ground state of a transverse-field Ising model, adiabatic quantum protocols can provide a polynomial speedup over simulated annealing \cite{kadowaki1998quantum}. Although exponential speedups are more elusive, finding such examples is nevertheless an active area of research.
}

While adiabatic simulation protocol allows for the preparation of the ground state of non-trivial spin models, maintaining the adiabatic condition (Eq.(\ref{eqn:adiabatic})) for a large system within the constraint of an experimentally realistic coherence time will be challenging. Alternate protocols have been explored to bypass the strict requirements of adiabaticity, while achieving high ground state probability. For example, a `Bang-bang' control of the Hamiltonian has been suggested \cite{Viola1998dynamical, Balasubramanian2018bang-bang}, where the initial trivial Hamiltonian can be quenched to an intermediate Hamiltonian, followed by a final quench to the problem Hamiltonian. In another approach, a classical-quantum hybrid protocol (the Quantum Approximate Optimization Algorithm, or QAOA \cite{Farhi2014quantum}) theoretically enables ultra-fast creation of ground states \cite{Ho2019ultrafast}. Implementations of this method are discussed in detail in section \ref{subsec:qaoa}. Here, we restrict our discussions to adiabatic simulation protocols.

\subsection{Experimental Progress in Adiabatic Quantum Simulation}

\subsubsection{Transverse Ising model with a small number of spins}
\label{sec:smallspins}
The adiabatic preparation of the ground state for the transverse Ising model was first demonstrated with two trapped ion spins \cite{friedenauer2008simulating,schneider2012experimental}, followed by experiments with three spins \cite{kim2009entanglement,kim2010quantum,edwards2010quantum,kim2011quantum, Khromova2012}. These entry experiments demonstrated the adiabatic evolution from para-magnetic initial state to magnetically ordered ground states and allowed tests of adiabaticity \cite{edwards2010quantum} and direct measures of entanglement in the ground state \cite{kim2010quantum,kim2011quantum}. The three-spin system moreover supports spin frustration, or a competition between the nearest and next-nearest couplings in the case of antiferromagnetic (AFM) ground states.  By tuning the system to have either ferromagnetic (FM) and anti-ferromagnetic (AFM) ground states, two different types of magnetic order were indeed measured, paralleling the two different classes of entanglement known to exist with exactly three spins \cite{Acin2001}.

\begin{figure}[ht]
\includegraphics[width=1.0\linewidth]{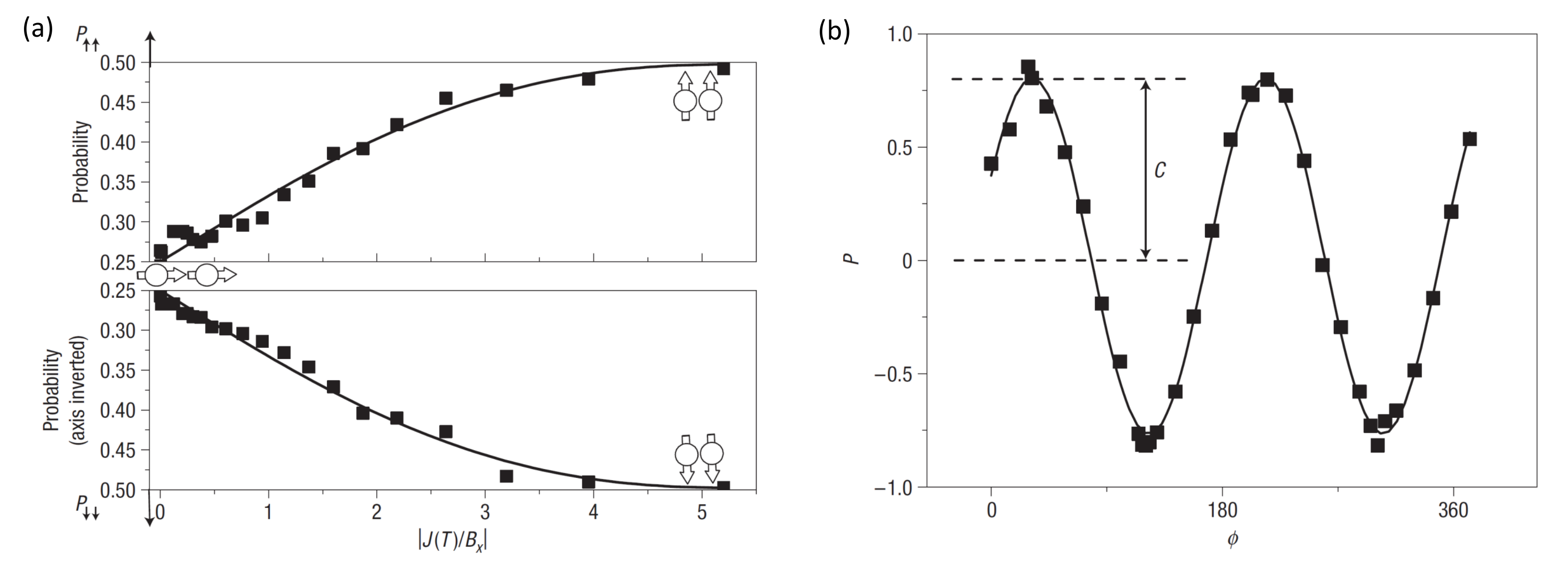}
\caption{(a) Observed adiabatic evolution of two trapped ion spins from a paramagnetic state to ferromagnetic (FM) state. (b) The observed parity oscillation signal of the resulting FM state upon a subsequent rotation of both spins reveals the amount of entanglement in the ground state. From \citealp{friedenauer2008simulating}.}
\label{fig:2qubit}
\end{figure}

For the case of three Ising spins, the transverse Ising Hamiltonian (\ref{eqn:TransversIsing}) is reduced to
\begin{equation}
H_3 = J_1 ( \sigma_x^{(1)}\sigma_x^{(2)}+\sigma_x^{(2)}\sigma_x^{(3)}) +J_2 \sigma_x^{(3)}\sigma_x^{(1)}  + B_y(t) (\sigma_y^{(1)}+\sigma_y^{(2)}+\sigma_y^{(3)}),
\label{Ham3}
\end{equation}
where the transverse field and the Ising interaction are chosen to act along the $y$-axis and $x$-axis, respectively. This is the simplest Hamiltonian that can exhibit frustration in the ground state due to a compromise between the various Ising couplings. 

As seen in Eq. (\ref{Jij}), the sign and the strength of the Ising couplings $J_{ij}$ can be controlled by the proper choice of the driving field detuning $\delta_m$ from the motional modes. For three spins, the expected and measured nearest-neighbor (NN) interactions, $J_1 \equiv J_{1,2}=J_{2,3}$ and the next-nearest-neighbor (NNN) interaction $J_2 \equiv J_{1,3}$ are shown in Fig. \ref{fig:SpinInteraction}. For certain ranges of the drive field detuning, both NN and NNN couplings have anti-ferromagnetic (AFM) interactions ($J_1, J_2 >0$), and for other domains both show ferromagnetic (FM) interactions ($J_1, J_2 <0$).

\begin{figure}[ht]
\includegraphics[width=1.0\linewidth]{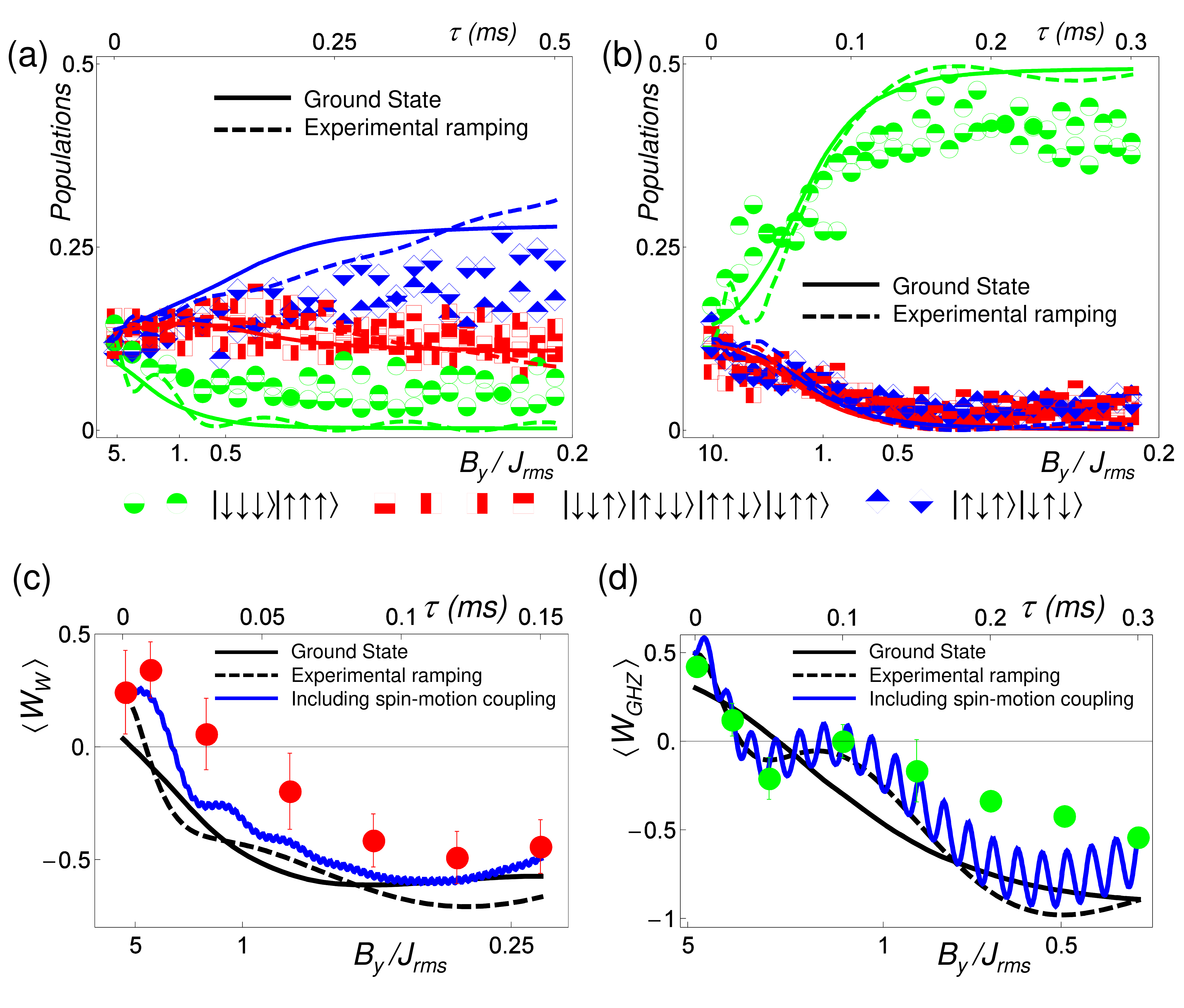}
\caption{Evolution of each of the eight spin states, measured with a CCD camera, plotted as $B_y/J_{rms}$ is ramped down in time. The dotted lines correspond to the populations in the exact ground state and the solid lines represent the theoretical evolution expected from the actual ramp. (a) All interactions are AFM. The FM-ordered states vanish and the six AFM states are all populated as $B_y \rightarrow 0$. Because $J_2 \approx 0.8 J_1$, a population imbalance also develops between symmetric and asymmetric AFM. (b) All interactions are FM, with evolution to the two ferromagnetic states as $B_y \rightarrow 0$.(c) Entanglement generation for the case of all AFM interaction, where symmetric W-state witness $W_{\rm W}$ is used. The entanglement emerges for $B_y/J_{rms} <$1.1.(d) Entanglement generation for the all FM interactions, where the GHZ witness $W_{\rm GHZ}$ is used. The entanglement occurs when $|B_y|/J_{rms} < 1$.  In both (c) and (d) the error bars represent the spread of the measured expectation values for the witness, likely originating from the fluctuations of experimental conditions. The black solid lines are theoretical witness values for the exact expected ground states, and the black dashed lines describe theoretically expected values at the actual ramps of the transverse field $B_y$. The blue lines reveal the oscillation and suppression of the entanglement due to the remaining spin-motion couplings, showing better agreement to the experimental results. Adapted from \citealp{kim2011quantum}.}
\label{FrustSim}
\end{figure}

Figure \ref{FrustSim}a shows the time evolution for the Hamiltonian frustrated with nearly uniform AFM couplings and gives almost equal probabilities for the six AFM states $\ket{\downarrow \downarrow \uparrow}, \ket{\uparrow \downarrow \downarrow},  \ket{\downarrow \uparrow \uparrow},  \ket{\uparrow \uparrow \downarrow},  \ket{\downarrow \uparrow \downarrow}$, and  $\ket{\uparrow \downarrow \uparrow}$ (labeled in the x-basis of the Bloch sphere and accounting for $3/4$ of all possible spin states) at $B_y \approx 0$. Because $J_2 <0.8 J_1$ for this data, a population imbalance also develops between symmetric ($\ket{\downarrow \uparrow \downarrow}$ and $\ket{\uparrow \downarrow \uparrow}$) and asymmetric ($\ket{\downarrow \downarrow \uparrow}, \ket{\uparrow \downarrow \downarrow},  \ket{\downarrow \uparrow \uparrow}$, and $\ket{\uparrow \uparrow \downarrow}$) AFM states. Figure \ref{FrustSim}b shows the evolution to the two ferromagnetic states ($\ket{\downarrow \downarrow \downarrow}$ and $\ket{\uparrow \uparrow \uparrow}$) as $B_{y} \rightarrow 0$, where all interactions are FM.

The adiabatic evolution of the ground state of Hamiltonian (\ref{Ham3}) from $B_y \gg J_{{rms}}$ to $B_y \ll J_{{rms}}$ should result in an equal superposition of all ground states and therefore be entangled. For instance, for the isotropic AFM case, the ground state is expected to be $\ket{\downarrow \downarrow \uparrow}+ \ket{\uparrow \downarrow \downarrow}+ \ket{\downarrow \uparrow \uparrow} - \ket{\uparrow \uparrow \downarrow} - \ket{\downarrow \uparrow \downarrow} - \ket{\uparrow \downarrow \uparrow}$. For the FM case, the ground state is a GHZ as $\ket{\downarrow \downarrow \downarrow} - \ket{\uparrow \uparrow \uparrow}$. 
The entanglement in the system at each point in the adiabatic evolution can be characterized by measuring particular entanglement witness operators \cite{guhne2009entanglement}. When the expectation value of such an operator is negative, this indicates entanglement of a particular type defined by the witness operator. For the AFM (frustrated) case as shown in Fig. \ref{FrustSim}c, the expectation of the symmetric W state witness, $W_{\rm W} = (4+ \sqrt{5})\hat{I} - 2(\hat{\mathcal{J}}_x^2 + \hat{\mathcal{J}}_y^2)$ is measured \cite{guhne2009entanglement}. For the FM case as shown in Fig. \ref{FrustSim}d, the expectation of the symmetric GHZ witness operator
$W_{\rm GHZ}= 9\hat{I}/4 - \hat{\mathcal{J}}_x^2 - \sigma_y^{(1)} \sigma_y^{(2)} \sigma_y^{(3)}$ \cite{sackett2000experimental,guhne2009entanglement} is measured, where $\hat{I}$ is the identity operator and $\hat{\mathcal{J}}_i \equiv \frac{1}{2}(\sigma_l^{(1)} + \sigma_l^{(2)} + \sigma_l^{(3)})$ is proportional to the $l$th projection of the total effective angular momentum of the three spins. In both cases, as shown in Figs. \ref{FrustSim}c and (d), entanglement of the corresponding form is clearly observed during the adiabatic evolution.

\subsubsection{Onset of quantum many-body effects with increasing system size}

The ground state in the transverse field Ising model, Eq.~(\ref{eqn:TransversIsing}) undergoes a crossover between polarized/paramagnetic and magnetically ordered spin states, as the relative strengths of the transverse field $B_y$ and the Ising interactions $J_{ij}$ are varied. For $|B_y/J_{ij}|\gg 1$, the ground state has the spins independently polarized (paramagnetic phase). For $|B_y/J_{ij}|\ll 1$, the ground state is magnetically ordered for $|B_y/J_{ij}|\ll 1$, with ferromagnetic order for $J_{ij}<0$ in Eq.~(\ref{eqn:TransversIsing}). A second order quantum phase transition is predicted for this model in the thermodynamic limit \cite{sachdev2011quantum} when the magnitude of the transverse field is comparable to the interaction strength. 

\begin{figure}[h]
\includegraphics*[width=.9\linewidth]{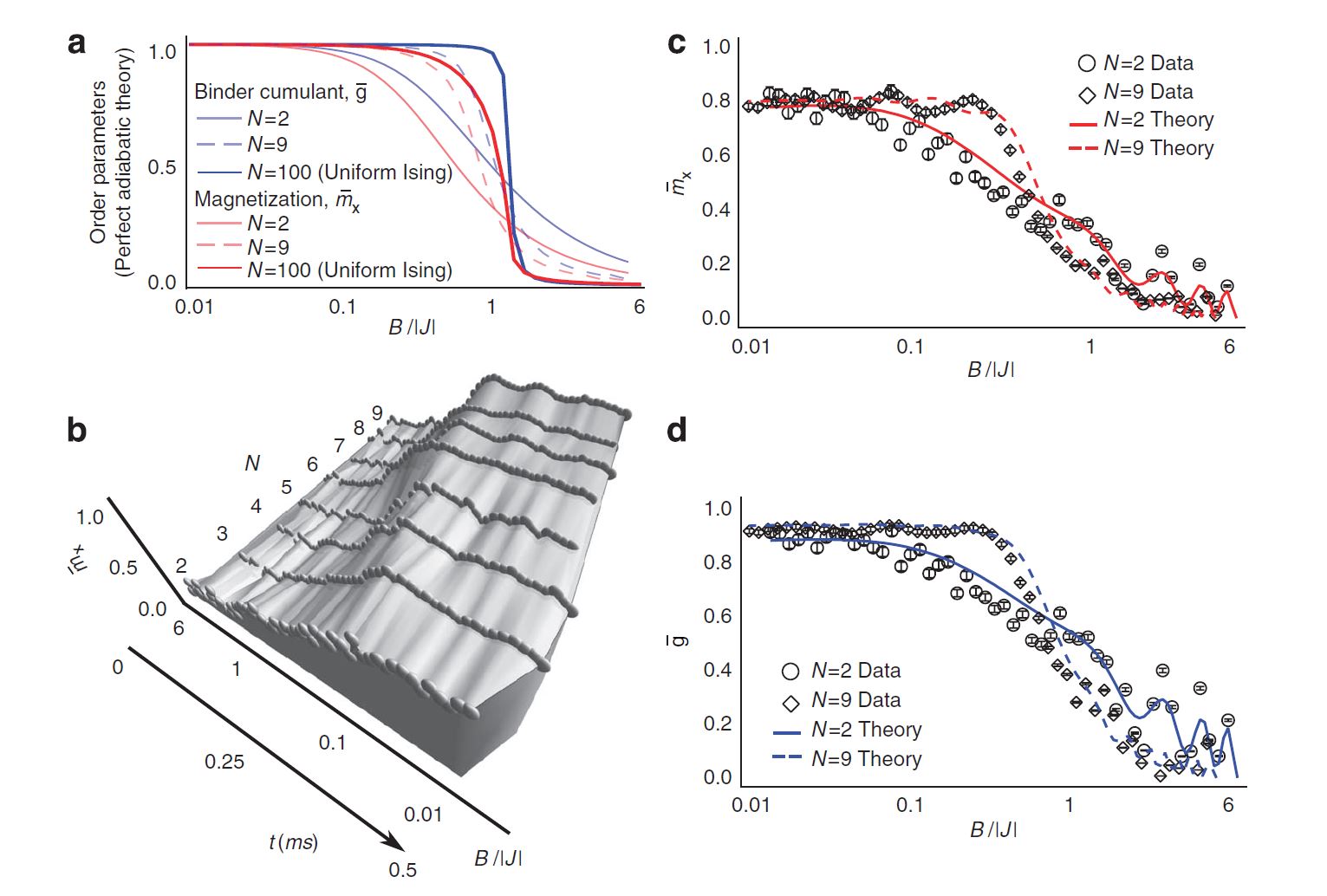}
\caption{Paramagnetic-to-Ferromagnetic crossover in a small collection of trapped ion spins. (a) Theoretical values of order parameters versus vs ratio of transverse field to average Ising coupling $B/|J|$ for $N=2$ and $N=9$ spins with non-uniform Ising couplings following the experiment of \cite{islam2011onset} and assuming perfect adiabatic time evolution. The order parameters, the Binder Cumulant and the magnetization are calculated by directly diagonalizing the relevant Hamiltonian (Eq. \ref{eqn:TransversIsing}). Order parameters are also calculated for a moderately large system (N=100) with uniform Ising couplings, to show the difference between these order parameters. (b) Measured magnetization vs $B/|J|$ (and simulation time) plotted for N=2 to N=9 spins, and scaled to the number of spins. As $B/|J|$ is lowered, the spins undergo a crossover from a paramagnetic to ferromagnetic phase. The crossover curves sharpen as the system size is increased from $N=2$ to $N=9$, prefacing a phase transition in the limit of infinite system size. The oscillations in the data arise from the imperfect initial state preparation and non-adiabaticity due to finite ramping time. Measured magnetization (c) and Binder cumulant (d) vs. $B/|J|$ for for $N=2$ (circles) and $N=9$ spins (diamonds) with representative detection error bars. The data deviate from unity at $B/|J|=0$ owing to decoherence driven by the Raman transitions creating the Ising couplings. The theoretical curves (solid line $N=2$ and dashed line for $N=9$ spins) are calculated by averaging over 10,000 quantum trajectories. [from \cite{islam2011onset}].}
\label{fig:islam2011onset}
\end{figure}

{\q The presence or absence of a spin-order can be quantified by adopting a suitable order parameter. For example, the average absolute magnetization per site along the Ising direction, 
 \begin{equation}
 m_x=\frac{1}{N}\sum_{s=0}^N |N-2s|P(s)
 \label{eq:magnetization}
 \end{equation}
differentiates between a ferromagnetic state and paramagnetic state. Here, $P(s)$ is the probability of finding $s$ spins in the $\up$ state ($s=0,1,2,\cdots,N)$ along x. To remove a finite size effect due to the difference between Binomial
and Gaussian distributions, a scaled order parameter, $\bar{m}_x=(m_{x,N}^0-m_x)/(m_{x,N}^0-1)$ can be adopted. Here, $m_{x,N}^0=\frac{1}{N 2^N}\sum_{s=0}^N {{N}\choose{s}}|N-2s|$ is the average absolute magnetization of the paramagnetic state.
This scaled order parameter assumes a value of $\bar{m}_x=1$ for the ideal ferromagnetic state while $\bar{m}_x\approx 0$ in the paramagnetic state.  
A finite system does not support a phase transition, but shows a smooth crossover from the paramagnetic to the spin-ordered phases that becomes sharper as the system size is increased.  
Higher order moments of the distribution of measured spins may be more suitable to extract the phase transition point from experiments performed on finite system sizes. For instance, the fourth magnetization moment is known as the Binder cumulant,
 \begin{equation}
g=\frac{\sum_{s=0}^N (N-2s)^4 P(s)}{\left[\sum_{s=0}^N(N-2s)^2
 P(s)^2\right]^2}. \label{eq:binder}
 \end{equation}
The Binder cumulant can also be scaled to remove the finite size effect, as before, by defining $\bar{g}=(g_N^0-g)/(g_N^0-1)$, where $g_N^0=3-2/N$ is the Binder cumulant for the paramagnetic phase. 

Fig. \ref{fig:islam2011onset} shows measurements of the mean magnetization and Binder cumulanant in a transverse Ising system ranging from $N=2$ to $N=9$ atomic ion spins \cite{islam2011onset}.
The observed sharpening of the crossover from paramagnetic to ferromagnetic spin-order with system size (Fig. \ref{fig:islam2011onset}) is consistent with an onset of the quantum phase transition. 
In Fig. \ref{fig:islam2011onset}a, theoretical values of both order parameters are shown for up to $N=100$ spins in an all-to-all coupled ferromagnetic transverse Ising model, with measurements and comparison to theory in Fig. \ref{fig:islam2011onset}b-d. Both metrics have been scaled to take into account finite size effects \cite{islam2011onset}.} 
{\q Ferromagnetic spin order was also observed in adiabatic quantum simulation experiments with up to $N=16$ ions by directly measuring a bimodal distribution of magnetization \cite{islam2013emergence}.}

\begin{figure}[h]
\includegraphics*[width=.9\linewidth]{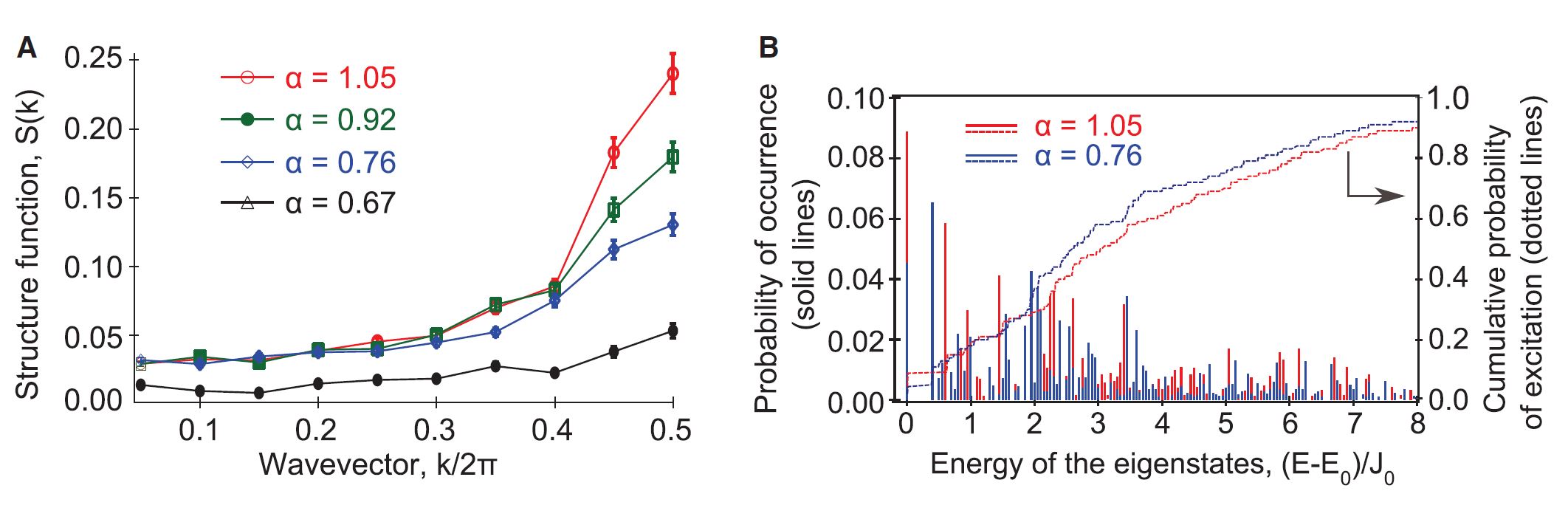}
\caption{(a) Structure function $S(k)$ for various ranges of AFM interactions, for $B/J_0 = 0.01$ in a system of $N = 10$ spins. The increased level of frustration for the longer-range interactions reduces the observed antiferromagnetic spin order. The detection errors may be larger than shown here for the longest range of interactions, owing to spatial crosstalk from their closer spacing. (b) Distribution of observed states in the spin system, sorted according to their energy $E_i$ (with $E_0$ denoting the ground state energy) calculated exactly from Eq. (\ref{eqn:TransversIsing}) with B = 0. Data are presented for two ranges (red for $\alpha = 1.05$ and blue for $\alpha = 0.76$). The dashed lines indicate the cumulative energy distribution functions for these two ranges. Adapted from \cite{islam2013emergence}.}\label{fig:islam2013emergence}
\end{figure}

AFM ground states of the transverse field Ising model (Eq. (\ref{eqn:TransversIsing}) with $J_{ij}>0$) are more difficult to prepare because the long-ranged AFM interactions lead to competing pairwise spin order, or frustration, as detailed with $N=3$ spins \cite{kim2010quantum} in section \ref{sec:smallspins}. Intuitively, the longer the range of AFM interactions, the less energy it takes to create spin-flip excitations. Thus the critical field required to destroy the AFM spin order is less than that with a relatively shorter range interaction. The `critical gap' in the many-body energy spectra also decreases with increasing range of the AFM couplings (Fig. \ref{fig:islam2013emergence}). The reduction of the critical gap with increased range of interaction was experimentally probed in a quantum simulation of the transverse field Ising model, Eq.\ref{eqn:TransversIsing} with the interaction profile following an approximate AFM power law, Eq. (\ref{Jpowerlaw}) for $N=10$ spins \cite{islam2013emergence}. The ratio of the transverse field to the Ising couplings was varied quasi-adiabatically from a high transverse field to a final value of $B/J_0=0.01$. As the interaction range was increased and the critical gap closed, more excitations were created, resulting in a reduction in the ground state order. This was observed through a decrease in the measured structure function,
\begin{equation}
S(k)=\frac{1}{N-1}\left|\sum_{r=1}^{N-1}C(r)e^{ikr}\right|,
\end{equation}
where the average correlation of spins along the Ising direction $x$ and separated by $r$ sites is 
\begin{equation}
C(r)=\frac{1}{N-r}\sum_{m=1}^{N-r}\left(\langle \sigma_x^{(m)}\sigma_x^{(m+r)}\rangle-\langle\sigma_x^{(m)}\rangle\langle\sigma_x^{(m+r)}\rangle\right).
\end{equation}
The measured structure function for $N=10$ spins (Fig. \ref{fig:islam2013emergence}a) shows a clear decline at $k=\pi$ as the interaction range is made longer, thus quantifying the degradations of nearest-neighbor antiferromagnetic spin order as the ground state gap shrinks.

\subsubsection{Ground state identification} 
Interestingly, the ground state spin ordering may still be determined experimentally even when the ramp is non-adiabatic. The key to ground state identification is to examine the probability distribution of all spin configurations at the conclusion of the ramp and select the most prevalent state in the final eigenbasis. Consider an experiment where the spins are initialized into $\ket{\downarrow\downarrow\downarrow\ldots}_y$ (as usual) and the transverse field $B(t)$ is instantly switched from $B=B_0$ to $B=0$. Measurement along the $x$-direction would yield an equal superposition of all spin states; in this instance, the ground state is just as probable as any other state. If the transverse field $B(t)$ is instead ramped at a fast but finite rate, the quantum simulation is slightly more adiabatic than the instantaneous case, and the ground state becomes slightly more prevalent than any other state. When $B(t)$ is ramped slowly enough, the ground state population is nearly $100\%$ and dominates over that of any other state.

A close analogy may be drawn with a Landau-Zener process \cite{Zener1932non} in a two-level system comprised of the ground and first coupled excited states. Adiabatic ramps correspond to half of a Landau-Zener process, in which $B(t)$ starts with $B \gg J$ and ends at $B=0$. One can write an analytic expression to calculate the transition probability for this half-Landau-Zener evolution \cite{damski2006adiabatic}, which has a maximum value of 0.5 for an instantaneous ramp. Any fast but finite ramp will give a transition probability $< 0.5$, so the ground state will always be more prevalent than the excited state.

The technique of identifying the most prevalent state as the ground state is subject to some limitations. First, the initial state (before the ramp) should be a uniform superposition of all spin states in the measurement basis -- a condition satisfied by preparing the state $\ket{\downarrow\downarrow\downarrow\ldots}_y$ and measuring along $\hat{x}$. If some spin states are more prevalent than the ground state initially, then some non-zero ramp time will be necessary before the ground state probabilities ``catch up'' and surpass these initially prevalent states. Second, the ramp must not cross any first-order transitions between ordered phases, as non-adiabatic ramps may not allow sufficient evolution time towards the new ground state order. In addition, the initial and final states must share the same symmetry properties.

Finally, a good determination of the ground state requires that the difference between the measured ground state probability $P_g$ and next excited state probability $P_e$ be large compared with the experimental uncertainty, which is fundamentally limited by quantum projection noise $\sim1/\sqrt{n}$ after $n$ repetitions of the experiment \cite{itano1993quantum}. This implies that the most prevalent ground state can be determined reliably after repeating the measurement $n > (P_g^2+P_e^2)/(P_g-P_e)^2$ times.  Assuming an exponential distribution of populated states during the ramp (as may be expected from Landau-Zener-like transitions), the number of required runs should then scale as $n \sim (\bar{E}/\Delta)^2$ in the limit $\bar{E}\gg\Delta$, where $\bar{E}$ is the mean energy imparted to the spins during the ramp, and $\Delta$ is the energy splitting between the ground and first coupled excited state. 

If the gap shrinks exponentially with the number of spins $N$ (i.e. $\Delta\sim e^{-N}$), ground state identification thus requires an exponential number of measurements $n$ in the simulation. However, in cases where the gap shrinks like a power law ($\Delta\sim N^{-\alpha}$), the most prevalent state can be ascertained in a time that scales polynomially with the number of spins.  Regardless of the scaling, techniques that improve the ground state probability (such as local adiabatic evolution) can greatly increase the contrast of the most prevalent state and reduce the number of necessary repetitions.

\begin{figure}[h!]
\includegraphics[width=.5 \linewidth]{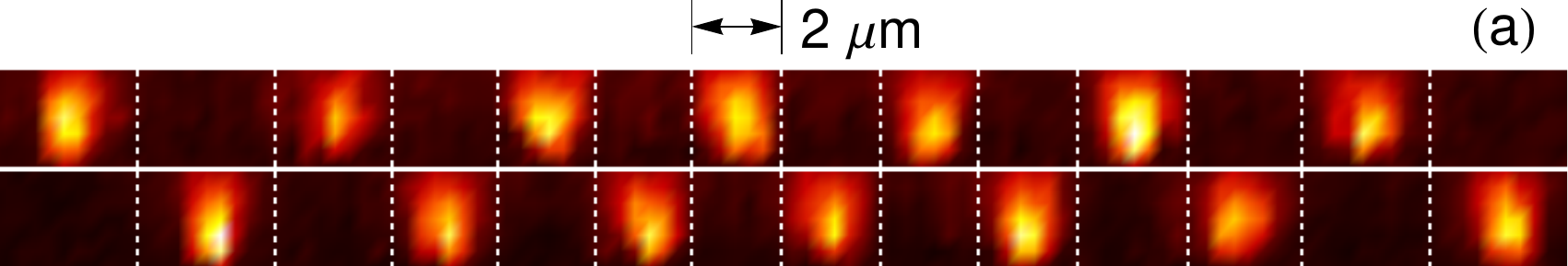} 
\includegraphics[width=.5 \linewidth]{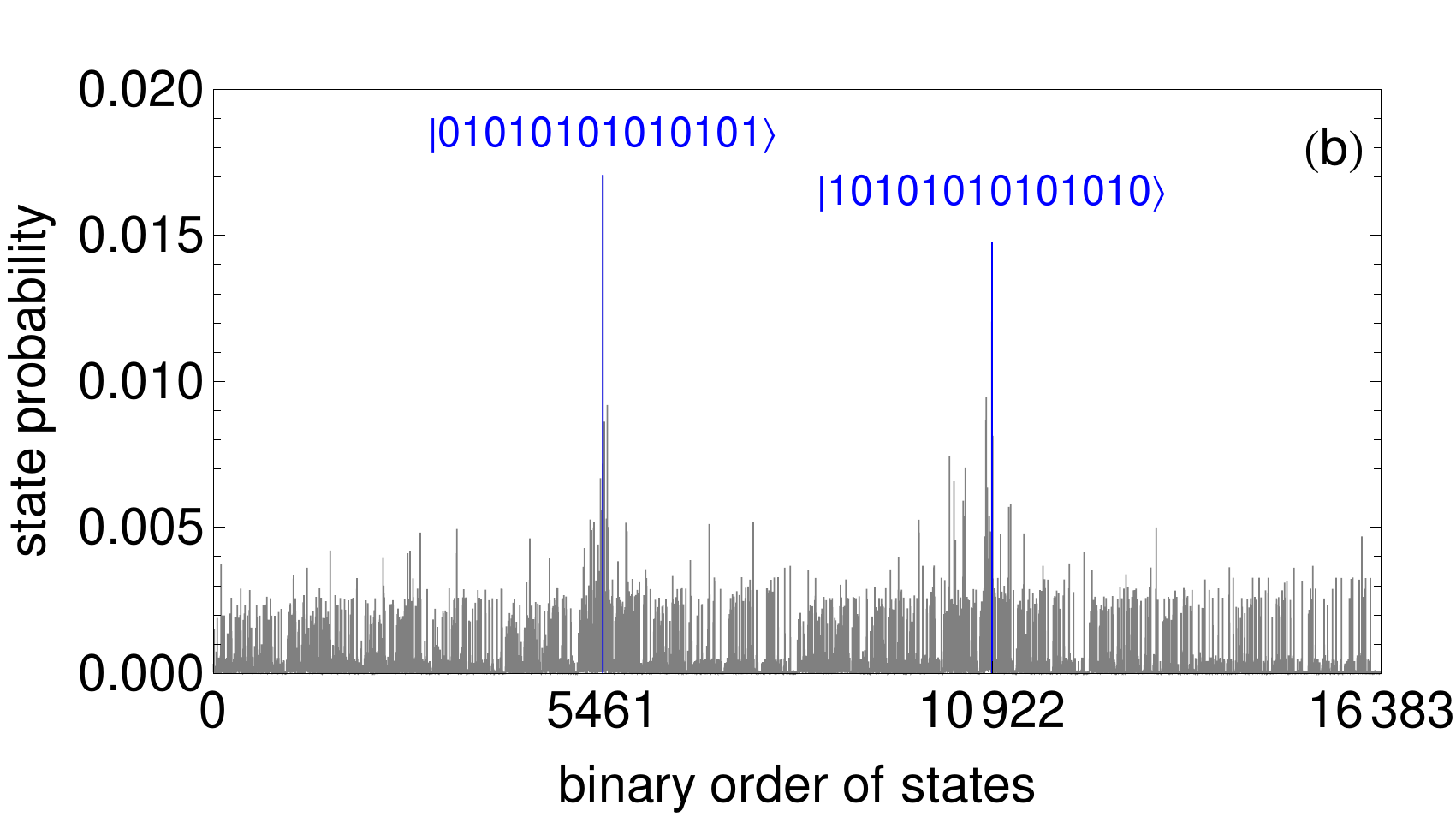}
\caption{(a) Fluorescence images of both AFM ground states prepared for $N=14$ trapped ion spins, with bright = $\up$ and dark = $\dn$ along the x-direction of the Bloch sphere of each spin. (b) State probabilities of all $2^{14}=16384$ spin configurations for a 14-ion system following a local adiabatic ramp, ordered by binary index of the states (i.e., $\ket{\downarrow\downarrow...\downarrow}$ = 0 and $\ket{\uparrow\uparrow...\uparrow}$ = 16,383). The N\'eel-ordered ground states $5,461$ and $10,922$ are unambiguously the most prevalent, despite a total probability of only 3\%. From \cite{richerme2013experimental}.}
\label{fig:ProbsFourteen}
\end{figure}

Fig. \ref{fig:ProbsFourteen} shows a direct identification of the ground state AFM order of $N=14$ trapped ion spins by imaging the most prevalent state created after a ramp. Each box in Fig. \ref{fig:ProbsFourteen}a contains an ion that scatters many photons when in the state $\ket{\uparrow}$ and essentially no photons when in the state $\ket{\downarrow}$.

demonstrates the resiliency of most-prevalent state selection to ramps that are far from adiabatic. Identification of the ground state is clear, even though the total ground state probability is only $\sim 3\%$. The requirement of satisfying the adiabatic criterion is replaced only by the requirement that the most prevalent state probabilities are accurately resolvable compared with those of any other states. While the method should remain robust for even larger $N$, ramps that are more adiabatic (by using longer ramp times or stronger spin-spin couplings) will decrease the number of experimental repetitions needed to clearly resolve the state probabilities.

\subsubsection{Classical Ising model}

Adiabatic protocols can also be used to create the ground states of a classical spin model, catalyzed by quantum fluctuations. Consider a system described by the transverse Ising model of Eq. \ref{eqn:TransversIsing} accompanied by a longitudinal field $B_x$:
\begin{equation}
H=\sum_{i<j} J_{ij} \sigma_x^{i} \sigma_x^{j} + B_x \sum_{i} \sigma_x^{i} + B_y(t) \sum_{i} \sigma_y^{i}
\label{eqn:axialfield}
\end{equation}
When the transverse field $B_y$ is set equal to 0, and the longitudinal field $B_x$ is varied, this Hamiltonian exhibits many distinct ground state phases separated by first-order classical phase transitions. Yet even for just a few spins, the various ground states at different $B_x$ are classically inaccessible in a physical system at or near zero temperature due to the absence of thermal fluctuations to drive the phase transitions \cite{sachdev2011quantum}. {\q Quantum fluctuations are therefore required to reach these various ground states in a physical system.} 

Such quantum fluctuations can be introduced to the system by applying a transverse magnetic field, which does not commute with the longitudinal-field Ising Hamiltonian. Using $N=6$ or $N=10$ spins, this technique has been used to experimentally identify the locations of the multiple classical phase transitions and to preferentially populate each of the classical ground states that arise for varying strengths of the longitudinal field \cite{richerme2013quantum}. The ground state spin ordering reveals a Wigner-crystal spin structure \cite{wigner1934on} that maps to particular energy minimization problems \cite{katayama2001performance} and shows the first steps of the complete ``devil's staircase" \cite{bak1982one} expected to emerge in the $N\rightarrow\infty$ limit.

\begin{figure}[h]
\includegraphics*[width=.5\linewidth]{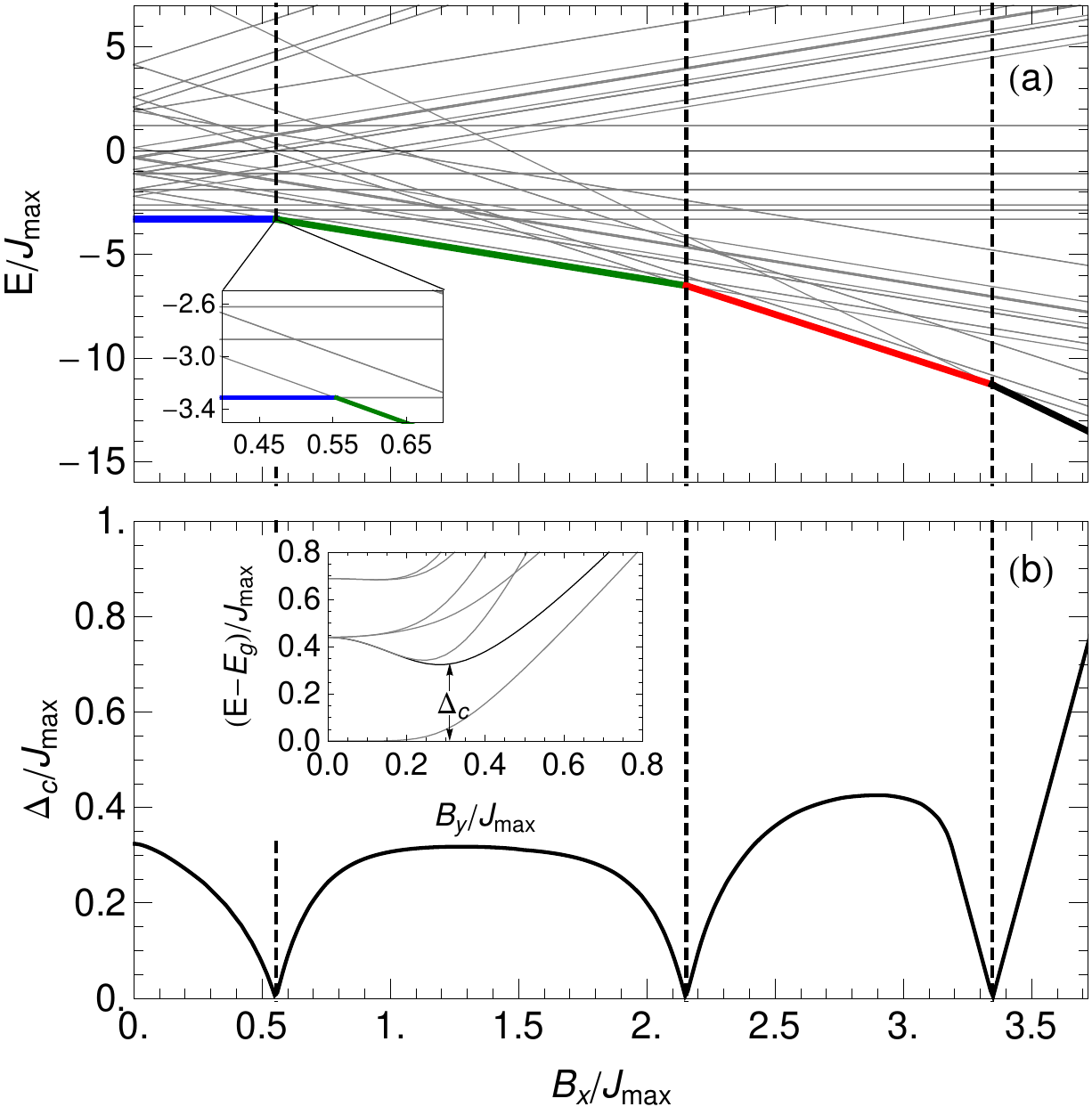}
\caption{(a) Low-lying energy eigenvalues of Eq. (\ref{eqn:axialfield}) for $B_y=0$ for $N=6$ spins, with the long-range $J_{ij}$ couplings determined from experimental conditions (see text). Level crossings (inset) indicate the presence of first-order phase transitions in the ground state. (b) The critical gap $\Delta_c$ shrinks to zero at the three phase transitions (vertical dashed lines). Inset: low-lying energy levels of Eq. (\ref{eqn:axialfield}) with $B_x=0$. From \cite{richerme2013quantum}.}
\label{fig:AFEnergyLevels}
\end{figure}

Fig. \ref{fig:AFEnergyLevels}a shows the energy eigenvalues of the Hamiltonian given in Eq. (\ref{eqn:axialfield}) with $B_y=0$ for a system of 6 spins. The ground state passes through three level crossings as $B_x$ is increased from 0, indicating three classical first-order phase transitions separating four distinct spin phases. For each $B_x$, there is a critical point at some finite $B_y$ characterized by a critical gap $\Delta_c$ (inset of Fig. \ref{fig:AFEnergyLevels}b). When $B_x$ is near a classical phase transition, the near energy-degeneracy of spin orderings shrinks the critical gap, as shown in Fig. \ref{fig:AFEnergyLevels}b.

Long-range interactions give rise to many more ground state spin phases than does a local Ising model. Consider an $N$-spin nearest-neighbor AFM model (Ising coupling $J$) and a ground-state ordering $\ket{..\downarrow\uparrow\downarrow\uparrow\downarrow\uparrow..}$. An excited state at longitudinal field $B_x=0$ may have an additional spin polarized along $\ket{\downarrow}$, either by making a kink of type $\ket{..\downarrow\uparrow\downarrow\downarrow\uparrow\downarrow..}$ or a spin defect of type $\ket{..\downarrow\uparrow\downarrow\downarrow\downarrow\uparrow..}$. The interaction energy gain of making $n$ kinks is $2nJ$, while the field energy loss is $2nB_x$. At $B_x/J=1$, multiple energy levels intersect to give a first-order phase transition. Similarly, the energy gain of making $n$ spin defects is $4nJ$ and the loss is $2nB_x$, so a second phase transition occurs at $B_x/J=2$. Only three different ground state spin phases are observable as $B_x$ is varied from $0 \rightarrow \infty$, independent of $N$, and there is a large degeneracy of spin eigenstates at the phase transitions. The presence of long-range interactions lifts this degeneracy and admits $[N/2]+1$ distinct spin phases with $\{0,1,\ldots,[N/2]\}$ spins in state $\ket{\uparrow}$, where $[N/2]$ is the integer part of $N/2$.

To create these various spin phases, the experiment from \cite{richerme2013quantum} begins by optically pumping the effective spins to the state $\ket{\downarrow\downarrow\downarrow..}_z$. The spins are then coherently rotated into the equatorial plane of the Bloch sphere so that they point along $\vec{B}=B_x \hat{x}+B_y(0) \hat{y}$, with $B_x$ varied between different simulations. The Hamiltonian of Eq. (\ref{eqn:axialfield}) is then switched on at $t=0$ with the chosen value of $B_x$ and $B_y(0)=5 J_{\text{max}}$. The transverse field (which provides the quantum fluctuations) is ramped down to  $B_y\approx0$ exponentially with a time constant of $600~\mu$s and a total time of 3 ms, which sacrifices adiabaticity in order to avoid decoherence effects. At $t=3$ ms, the Hamiltonian is switched off and the $x-$component of each spin is measured by applying a global $\pi/2$ rotation about the $\hat{y}$ axis, illuminating the ions with resonant light, and imaging the spin-dependent fluorescence using an intensified CCD camera. Experiments are repeated 4000 times to determine the probability of each possible spin configuration. 

\begin{figure}[t!]
\includegraphics[width=.46\linewidth]{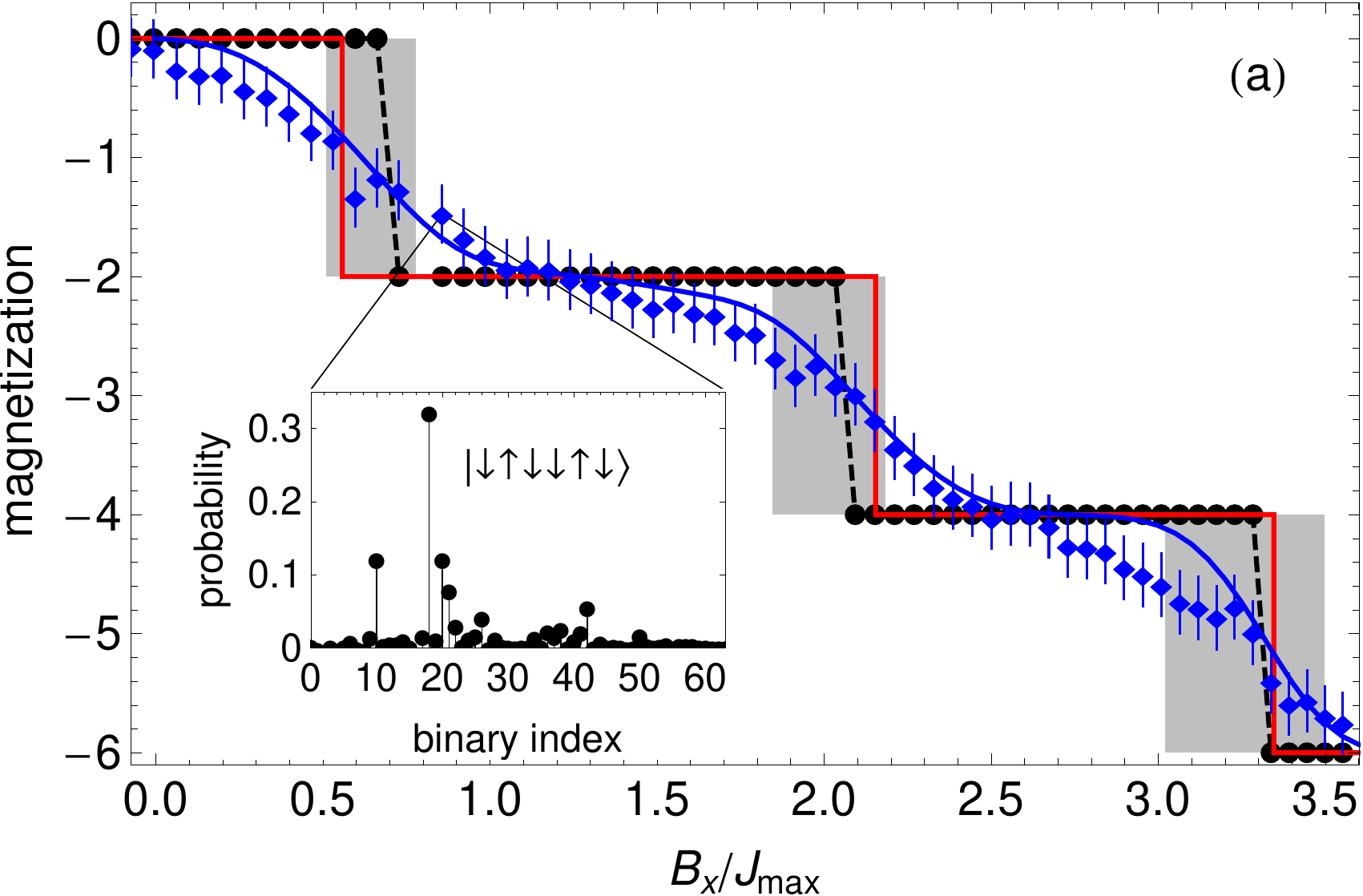}\hspace{.04\linewidth}\includegraphics[width=.5\linewidth]{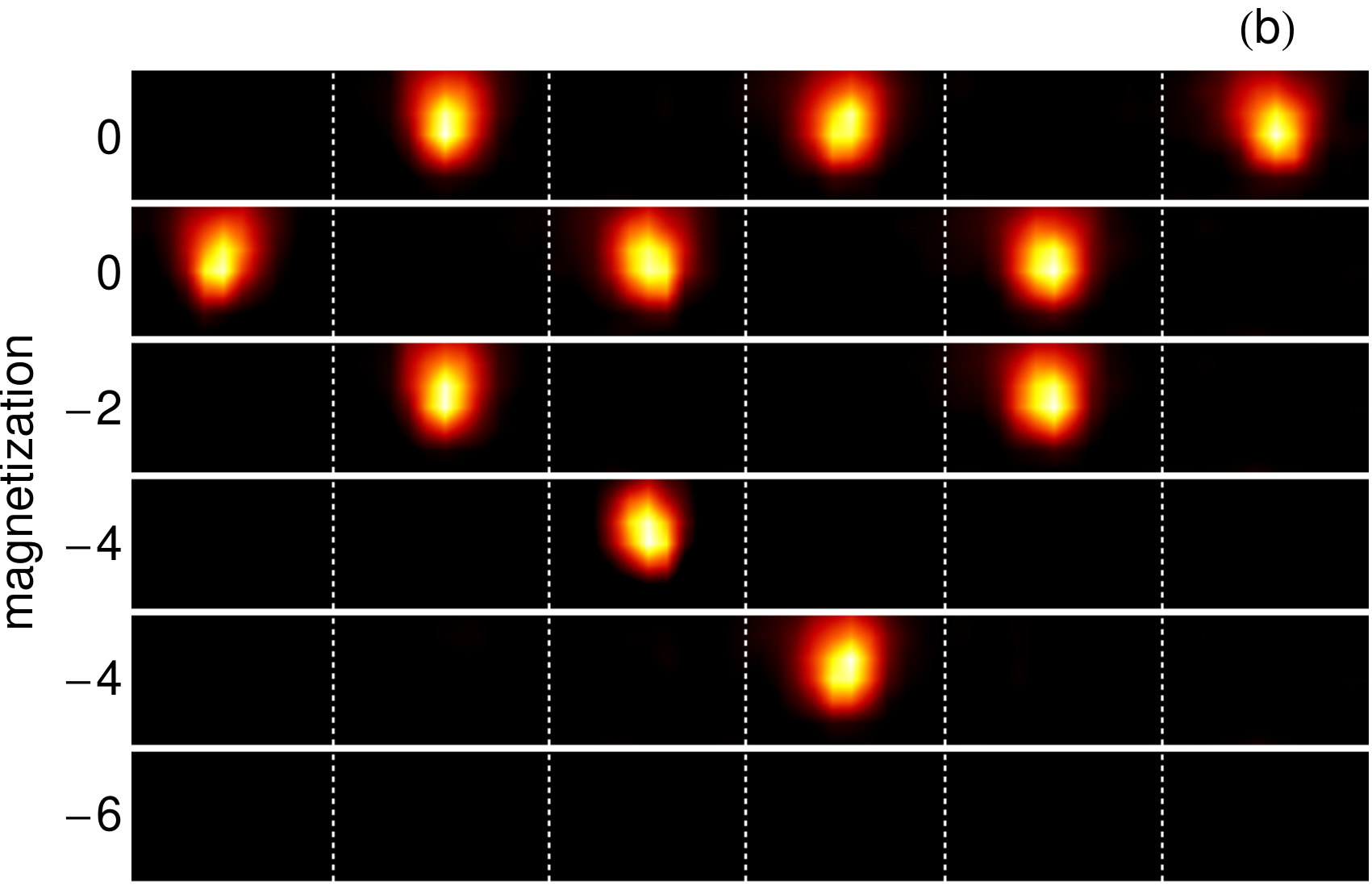}
\caption{(a) Magnetization ($M_x=N_{\uparrow}-N_{\downarrow}$) of $6$ ions for increasing axial field strength. Red, solid: magnetization of the calculated ground state, with the step locations indicating the first-order phase transitions. Blue diamonds: average magnetization of 4000 experiments for various $B_x$. Blue, solid: magnetization calculated by numerical simulation using experimental parameters. Black, dashed: magnetization of the most probable state (see inset) found at each $B_x$ value. Gray bands indicate the experimental uncertainty in $B_x/J_{\text{max}}$ at each observed phase transition. (b) Fluorescence images of the $6$ ions with bright = $\up$ and dark = $\dn$ along the x-direction of the Bloch sphere of each spin, showing the ground states found at each step in (a): $\ket{\downarrow\uparrow\downarrow\uparrow\downarrow\uparrow}$ and $\ket{\uparrow\downarrow\uparrow\downarrow\uparrow\downarrow}$ ($M_x=0$), $\ket{\downarrow\uparrow\downarrow\downarrow\uparrow\downarrow}$ ($M_x=-2$), $\ket{\downarrow\downarrow\uparrow\downarrow\downarrow\downarrow}$ and $\ket{\downarrow\downarrow\downarrow\uparrow\downarrow\downarrow}$ ($M_x=-4$), and $\ket{\downarrow\downarrow\downarrow\downarrow\downarrow\downarrow}$ ($M_x=-6$). Adapted from \cite{richerme2013quantum}.}
\label{fig:StepsSixIon}
\end{figure}

The order parameter of net magnetization along $x$, $M_x=N_{\uparrow_x}-N_{\downarrow_x}$, can then be investigated as a function of longitudinal field strength. The magnetization of the ground state spin ordering of Eq. (\ref{eqn:axialfield}) is expected to yield a staircase with sharp steps at the phase transitions (red line in Fig. \ref{fig:StepsSixIon}a) when $B_y=0$ \cite{bak1982one}. The experimental data (blue points in Fig. \ref{fig:StepsSixIon}a) show an averaged magnetization with heavily broadened steps due largely to the non-adiabatic exponential ramp of the transverse field. The deviation from sharp staircase-like behavior is predicted by numerical simulations (solid blue line in Fig. \ref{fig:StepsSixIon}a) which account for the implemented experimental parameters and ramp profiles. Differences between theory and experiment are largest near the phase transitions, where excitations are easier to make due to the shrinking critical gap (Fig. \ref{fig:AFEnergyLevels}b).

The ground state spin configuration at each value of $B_x$ can be extracted by looking at the probability distribution of all spin states and selecting the most prevalent state (inset of Fig. \ref{fig:StepsSixIon}a) \cite{richerme2013experimental}. The magnetization of the spin states found by this method (black points in Fig. \ref{fig:StepsSixIon}a) recover the predicted staircase structure. The steps in the experimental curve agree with the calculated phase transition locations to within experimental error (gray bands in Fig. \ref{fig:StepsSixIon}a), which accounts for statistical uncertainty due to quantum projection noise and estimated drifts in the strengths of $J_{ij}$, $B_x$, and $B_y$.

Fig. \ref{fig:StepsSixIon}b shows approximately 1000 averaged camera images of the most probable spin configuration observed at each plateau in Fig. \ref{fig:StepsSixIon}a. Each box contains an ion that scatters many photons when in the state $\ket{\uparrow}$ and essentially no photons when in the state $\ket{\downarrow}$ along the x-axis of the Bloch sphere. The observed spin orderings in Fig. \ref{fig:StepsSixIon}b match the calculated ground states at each magnetization, validating the technique of using quantum fluctuations to preferentially create these classically inaccessible ground states. (For magnetizations $M_x=0$ and $M_x=-4$, two ground state orderings are observed due to the left-right symmetry of the spin-spin interactions.)

\begin{figure}[t]
\raisebox{-.5\height}{\includegraphics[width=.5\linewidth]{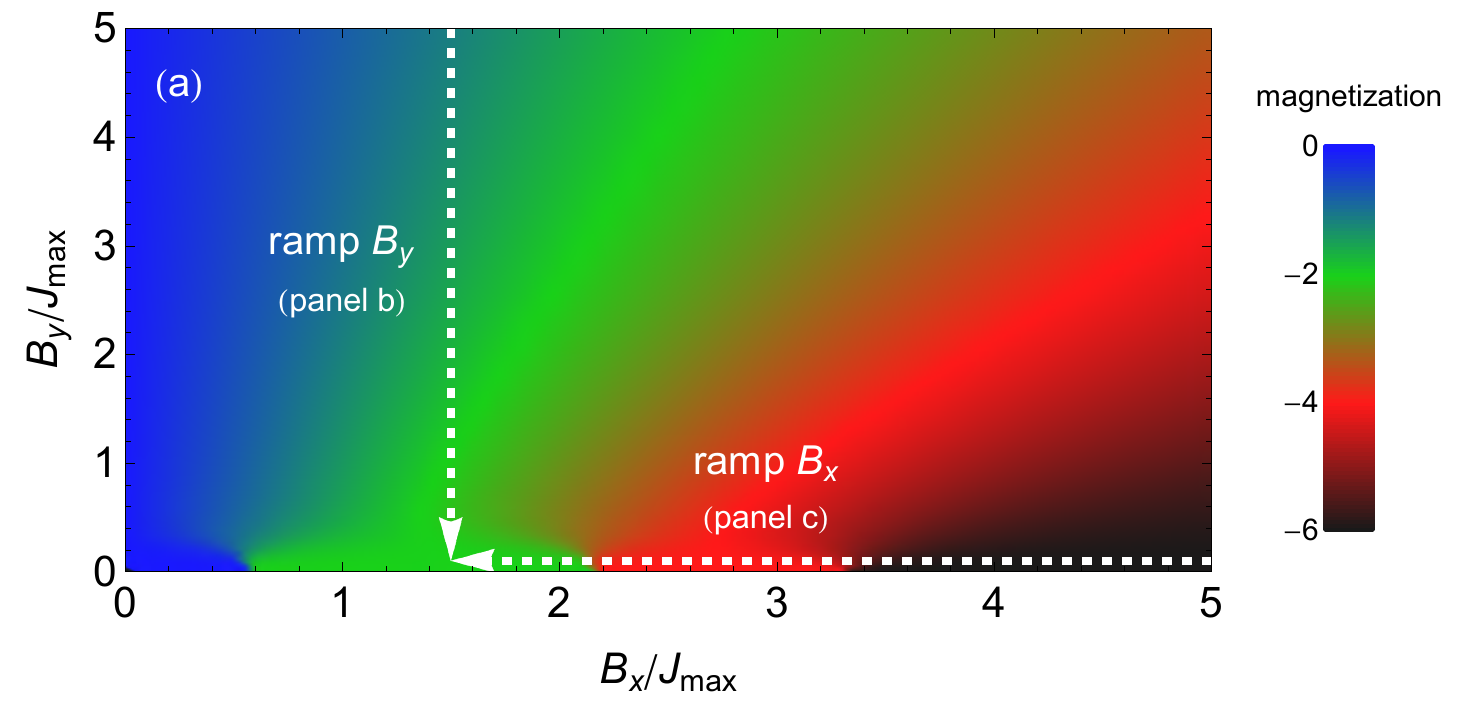}}\raisebox{-.5\height}{\includegraphics[width=.5\linewidth]{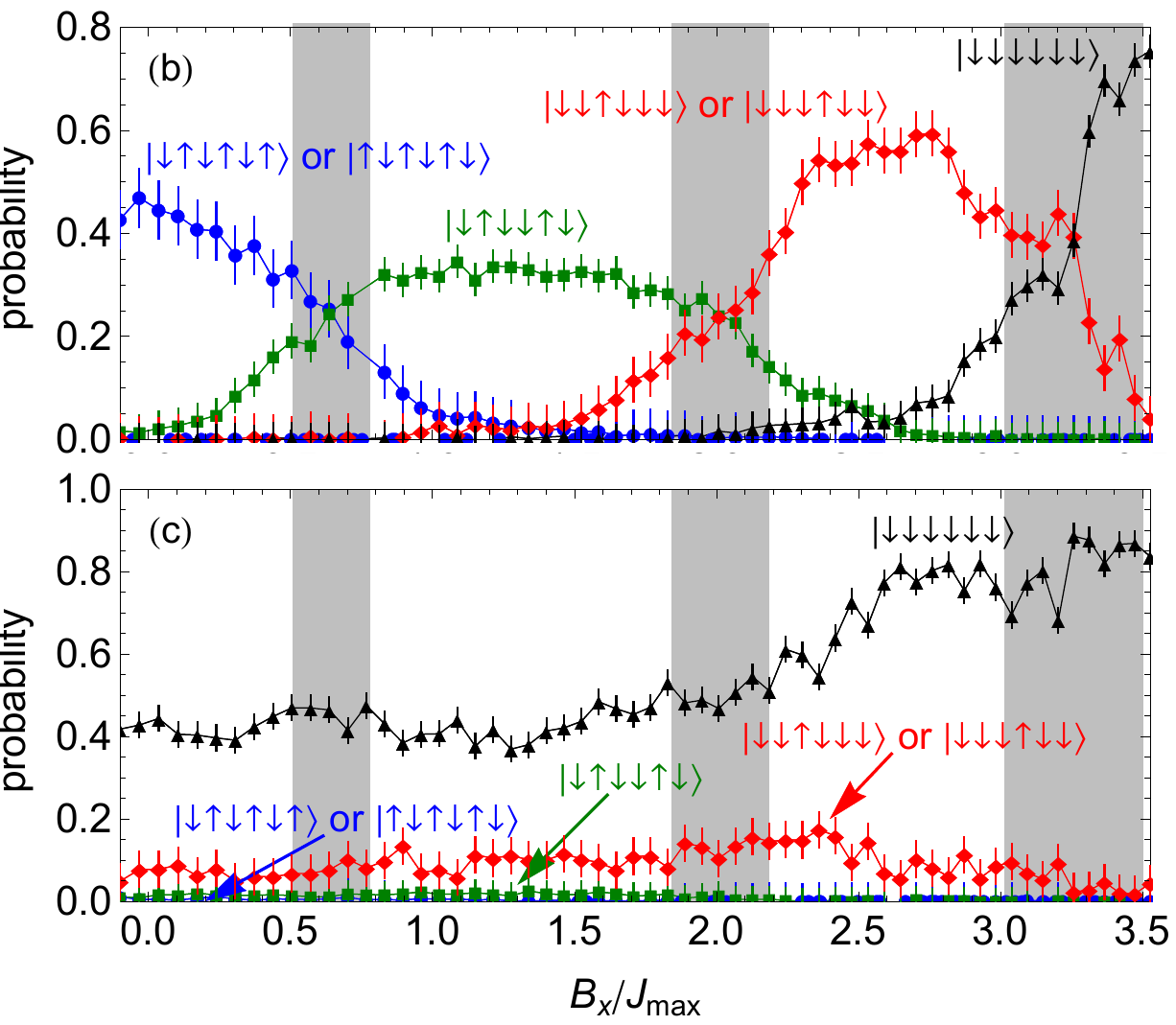}}
\caption{(a) Ground state phase diagram of the system, along with two different trajectories that end at the same value of $B_x$. (b) Probabilities of the 4 different ground state spin phases when $B_y$ is ramped in a 6-ion system. Blue dots: $\ket{\downarrow\uparrow\downarrow\uparrow\downarrow\uparrow}$ or $\ket{\uparrow\downarrow\uparrow\downarrow\uparrow\downarrow}$. Green squares: $\ket{\downarrow\uparrow\downarrow\downarrow\uparrow\downarrow}$. Red diamonds: $\ket{\downarrow\downarrow\uparrow\downarrow\downarrow\downarrow}$ or $\ket{\downarrow\downarrow\downarrow\uparrow\downarrow\downarrow}$. Black triangles: $\ket{\downarrow\downarrow\downarrow\downarrow\downarrow\downarrow}$. Gray bands are the experimental uncertainties of the phase transition locations. (c) Probabilities of creating the 4 different ground states when $B_x$ is ramped. Most of the ground states are classically inaccessible in our zero temperature system. Adapted from \cite{richerme2013quantum}.}
\label{fig:AFstateprobs}
\end{figure}

To further illustrate the necessity of using quantum fluctuations to catalyze the magnetic phase transitions, alternate ramp trajectories can be used to reach a final chosen value of $B_x$. Figure \ref{fig:AFstateprobs}a shows the ground state phase diagram of the Hamiltonian of Eq. (\ref{eqn:axialfield}), with the sharp classical phase transitions visible along the bottom axis ($B_y/J_{\text{max}}=0$). In addition, it shows two possible trajectories through the phase diagram that start in a paramagnetic ground state (which is easy to prepare experimentally) and end at the same value of $B_x$ with $B_y=0$.

The first trajectory, in which $B_x$ is fixed and $B_y$ is ramped from $5J_{\text{max}}$ to 0, was the one used in Fig. \ref{fig:StepsSixIon} to experimentally verify the locations of the 3 classical phase transitions and to experimentally create the 4 different ground state phases. Along this trajectory, Fig. \ref{fig:AFstateprobs}b plots the probability of creating each ground state as a function of $B_x$ and find populations of $\sim40-80\%$. A smooth crossover between the four ground state phases was observed, with the classical phase transitions occurring at the crossing points. This arises since distinct spin eigenstates have degenerate energies at the phase transition, causing the critical gap between them to close and allowing quantum fluctuations to populate both states equally (see Fig. \ref{fig:AFEnergyLevels}).

The second trajectory in Fig. \ref{fig:AFstateprobs}a is purely classical, with $B_y$ set to 0. The spins are initialized into the state $\ket{\downarrow\downarrow\downarrow\downarrow\downarrow\downarrow}$ along x, and $B_x$ is ramped from $5J_{\text{max}}$ to its final value at a rate of $5J_{\text{max}}/3$ ms. Figure \ref{fig:AFstateprobs}c shows that in a classical system without thermal or quantum fluctuations, the phase transitions remain undriven and the initial state $\ket{\downarrow\downarrow\downarrow\downarrow\downarrow\downarrow}$ remains dominant for all values of $B_x$. The ground state phases with magnetization 0 and $-2$ (blue and green in Fig. \ref{fig:AFstateprobs}c) are separated from the initial state by several classical phase transitions and have essentially zero probability of being created.

\subsubsection{Spin-1 simulations}

As with the spin-1/2 systems described above, spin-1 systems (spanned by the three basis kets $\ket{+}, \ket{0}$, and $\ket{-}$) can likewise exhibit a variety of interesting new physics and ground-state phases. As a notable example, Haldane conjectured \cite{haldane1983} that integer-spin Heisenberg chains with nearest-neighbor AFM interactions are gapped, in contrast to gapless half-integer spin chains. This energy gap in integer-spin systems corresponds to short-range exponentially decaying correlation functions, as opposed to long-range power-law decaying correlations in half-integer systems. It was later suggested \cite{den1989preroughening} that this Haldane phase of the spin-one chain is governed by a hidden order, which can be characterized by a non-local string order parameter and is consistent with a full breaking of a hidden $Z_2 \times Z_2$ symmetry \cite{kennedy1992hidden}. The Haldane phase can also be described by a doubly-degenerate entanglement spectrum \cite{pollmann2010entanglement}, hinting at a topologically protected phase in one-dimension.

\begin{figure}
\includegraphics[width=.6\linewidth]{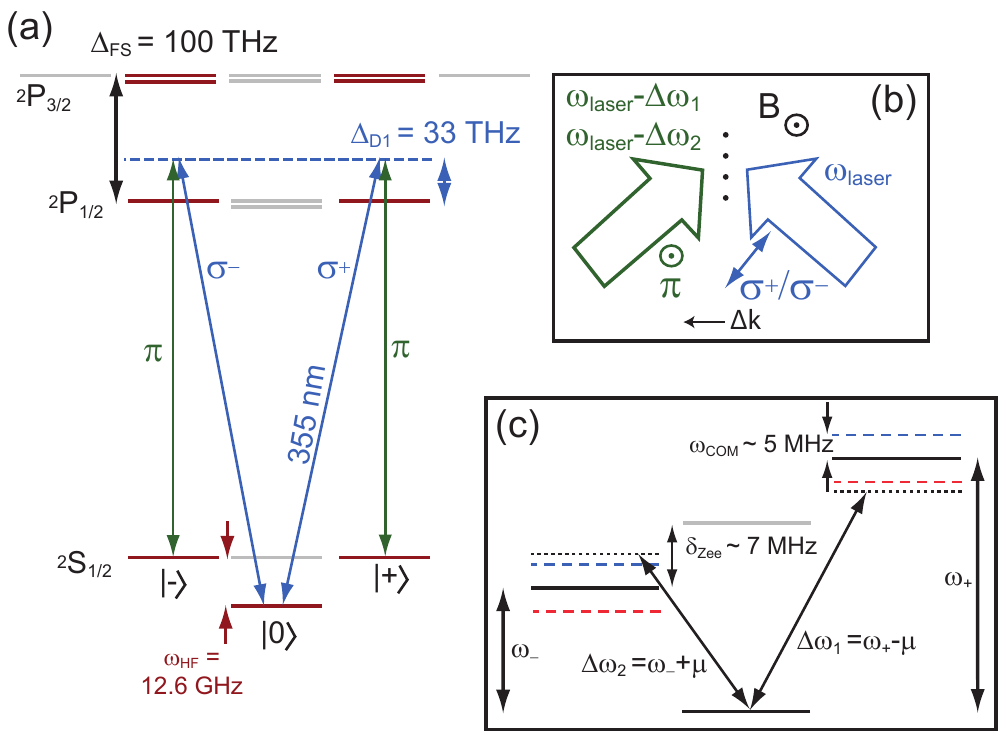}
\caption{(a): Level diagram for $^{171}$Yb$^+$, highlighting relevant states for spin-1 physics. (b): Sketch of experimental geometry, showing the directions of the laser wavevectors and the real magnetic field relative to the ion chain. Both beams are linearly polarized, one along the $\vec{B}$ field (providing $\pi$ light) and one orthogonal to the $\vec{B}$ field (providing an equal superposition of $\sigma^+$ and $\sigma^-$ light). Multiple beatnotes are applied by imprinting multiple frequencies onto one beam (in this case, the $\pi$-polarized beam). (c): Detailed level diagram of the $^2$S$_{1/2}$ ground state, showing Raman beatnotes in relation to Zeeman splittings and motional sidebands for the center-of-mass mode. Level splittings are not to scale. From \cite{senko2015realization}}
\label{fig:spin1levels}
\end{figure}

{\q Effective spin-1 particles can be represented by three hyperfine levels in the $^2S_{1/2}$ ground manifold of $^{171}$Yb$^+$ ions: $\ket{+} \equiv \ket{F=1,m_F=1}$, $\ket{-} \equiv \ket{F=1,m_F=-1}$, and $\ket{0} \equiv \ket{F=0,m_F=0}$. The $\ket{0}$ and $\ket{\pm}$ states are separated by frequencies $\omega_{\pm}$, as shown in Fig. \ref{fig:spin1levels}. Here, $\ket{+}$, $\ket{-}$, and $\ket{0}$ are the eigenstates of $S_z$ with eigenvalues +1, -1, and 0 respectively; $F$ and $m_F$ are quantum numbers associated with the total angular momentum of the atom and its projection along the quantization axis, defined by a magnetic field of $\sim$5 G. }

{\q Spin-1 Ising couplings can be generated analogously to the spin-1/2 case detailed in Section \ref{sec:intro}. Laser beams are applied to the ion chain with a wavevector difference along a principal axis of transverse motion, but here driving stimulated Raman transitions between both the $\ket{0}$ and $\ket{-}$ states and the $\ket{0}$ and $\ket{+}$ states with balanced Rabi frequencies $\Omega_i$ on ion $i$ \cite{kim2009entanglement}. To generate spin-1 XY interactions, two beat frequencies are applied at $\omega_- + \mu$ and $\omega_+ - \mu$ to these respective transitions, where $\mu-\omega_m = \delta_m$ is the detuning from the transverse motional mode $m$ sideband, as shown in Fig. \ref{fig:spin1levels}c. Under the approximations that the beatnotes are far detuned ($|\delta_m| \gg \eta_{i,m}\Omega_i$) and that $\omega_\pm \gg \mu \gg \Omega_i$ (the rotating-wave approximation), the resulting interaction Hamiltonian in the Lamb-Dicke regime is \cite{senko2015realization}

\begin{eqnarray}
H_{\mathrm{eff}} = \sum_{i<j} \frac{J_{ij}}{4} \left(S_+^i S_-^j + S_-^i S_+^j \right) 
+ \sum_{i,m} V_{i,m} \left[ \left(2 a_m^\dagger a_m +1 \right) S_z^i - \left( S_z^i \right)^2 \right],
\label{eq:spin1XY}
\end{eqnarray}
where $S^i_\pm$ are the spin-1 raising and lowering operators.
The pure ``XY" or ``flip-flop" spin-spin interaction in the first term of Eq. (\ref{eq:spin1XY}) follows the same formula as for generating spin-1/2 Ising interactions in Eq. (\ref{Jij}) \cite{kim2009entanglement},
\begin{equation}
J_{ij} = \Omega_i \Omega_j \sum_m \frac{\eta_{i,m} \eta_{j,m}}{2\delta_m}.
\label{eq:Jij}
\end{equation}
As in the case for spin-1/2, when $\delta_m > 0$ is larger than transverse mode frequencies, $J_{ij}$ falls off with distance approximately with a power law form from Eq. (\ref{Jpowerlaw}), where nearest-neighbor Ising coupling $J_0$ is typically of order $\approx 1$ kHz and $\alpha$ can be tuned between 0 and 3 as discussed above \cite{porras2004effective,islam2013emergence}. 

The second term in Eq. (\ref{eq:spin1XY}) represents additional spin-phonon terms and is parameterized by the factor $V_{i,m} =  (\eta_{i,m} \Omega_i)^2/(8\delta_m)$.
For long-range spin-spin interactions with $\alpha \lesssim 0.5$, or for small numbers of ions, the $V_{i,m}$ terms are approximately uniform across the spin chain. In these instances, the $V_{i,m}$ coefficient can be factored out of the sum over ions in Eq. (\ref{eq:spin1XY}), leaving only global $S_z^i$ and $(S_z^i)^2$ terms. For shorter-range interactions or for longer chain lengths, the $V_{i,m}$ terms can be eliminated by adding an additional set of beat frequencies at $\omega_- - \mu$ and $\omega_+ + \mu$, which would generate Ising-type interactions between effective spin-1 particles using the M$\o$lmer-S$\o$rensen gate \cite{molmer1999multiparticle}.}

As theoretically proposed in both \cite{cohen2015simulating} and \cite{gong2016kaleidoscope}, a range of spin-1 phenomena discussed above can be accessed in trapped ion quantum spin simulators. In \cite{cohen2015simulating}, for instance, it is shown how to generate the full spin-1 XXZ Hamiltonian
\begin{equation}
    H=\sum_{i<j} J_{ij}(S_x^i S_x^j+S_y^i S_y^j + \lambda S_z^i S_z^j) + D\sum_i (S_z^i)^2
    \label{spin1Hamiltonian}
\end{equation}
where the $S_\gamma^i$ terms are the spin-1 Pauli operators on site $i$ along the $\gamma$ direction, $\lambda$ is the $ZZ$ Ising anisotropy, and $D$ is analogous to a magnetic field $B$ term of Eq. (\ref{field}) for spin-1/2 systems. The interacting terms in Eq. (\ref{spin1Hamiltonian}) arise from a generalization of the M\o lmer-S\o rensen gate \cite{molmer1999multiparticle} to spin-1 systems, followed by a transformation to the interaction picture; the on-site $D$-term can be generated by imposing frequency detunings $D$ on all the previous driving fields. 

Generating the ground state of the Haldane phase \cite{haldane1983} can be realized by an adiabatic ramp procedure \cite{cohen2015simulating}. To begin, the spin-1 system can be initialized into a product state of $\ket{0}$ on each site, which is the trivial ground state when $D \gg J$. Adiabatically reducing $D$ will then drive the system towards the Haldane phase. As the system size increases and the critical gap between the $D$ and Haldane phase closes, a symmetry-breaking perturbation can be implemented to circumvent the phase transition. For example, adding a site-specific term $H_\text{pert}=-h\sum_i (-1)^{i}S_z^i$ will break all symmetries of the Haldane phase, allowing for a finite energy gap along the entire ramp path. The ground state can then be characterized using site-specific measurements to determine the spin correlation functions $\langle S_i^z S_j^z\rangle$ and the string-order correlation $\mathcal{S}_{ij}^z\equiv \langle S_i^z S_j^z \prod_{i<k<j} (-1)^{S_k^z}\rangle$.

Since the interactions $J_{ij}$ in Eq. (\ref{spin1Hamiltonian}) are long-ranged, this can lead to both quantitative and qualitative differences in the phase diagram as compared to the nearest-neighbor XXZ model \cite{gong2016kaleidoscope}. For instance, the positions of the phase boundaries shift for long-ranged AFM interactions, whereas long-ranged FM interactions can destroy the Haldane phase and support a new continuous symmetry-broken phase. Each of these possible phases can be distinguished by comparing measured values of the spin and string-order correlation functions described above. 

\begin{figure}
\raisebox{-.5\height}{\includegraphics[width=.5\linewidth]{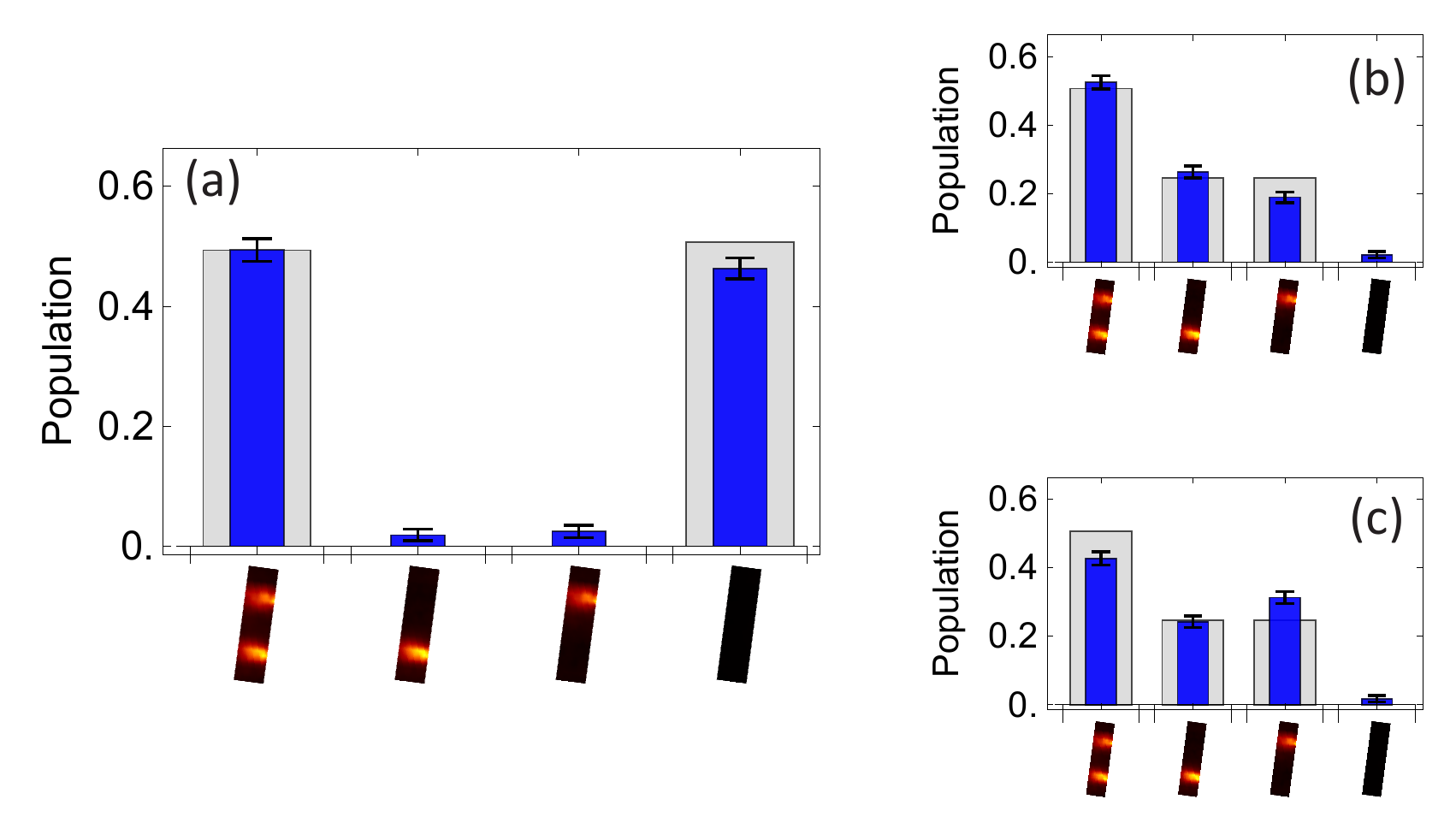}}\raisebox{-.5\height}{\includegraphics[width=.5\linewidth]{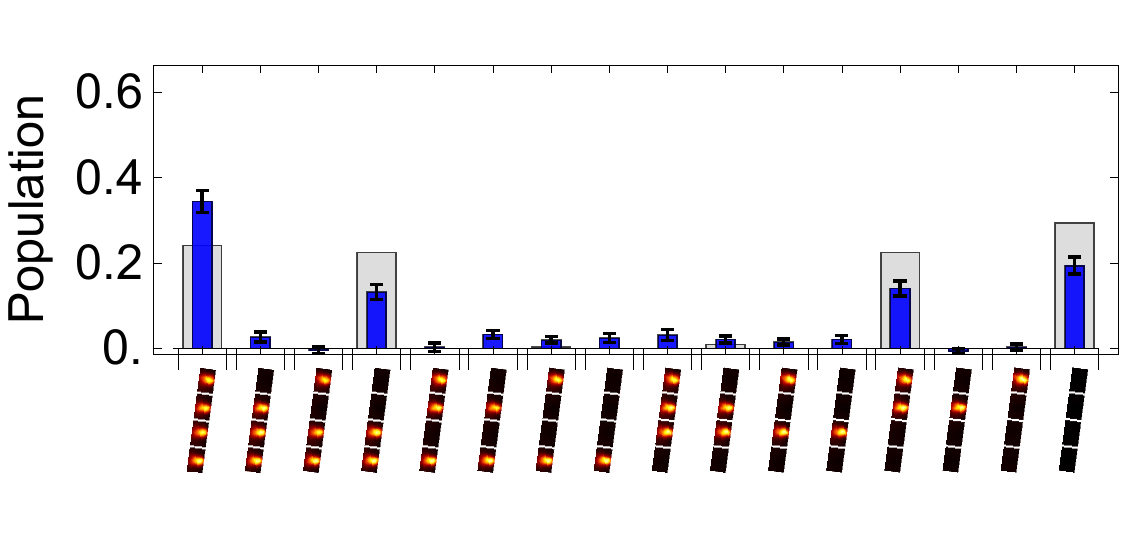}}
\caption{Measurements of the prepared 2-spin (a-c) and 4-spin (d) states after ramping an $S_z^2$ field (narrow blue bars) compared to the values expected for the calculated ground state (gray bars). Panels (a), (b), and (c) show the measured populations when the `dark' state is set to be $\ket{0}$, $\ket{-}$, or $\ket{+}$, respectively. The dark state is set to $\ket{0}$ in part (d). Adapted from \cite{senko2015realization}.}
\label{fig:EvenIonGroundState}
\end{figure}

The first experimental steps towards Haldane physics in an ion-trap quantum simulator implemented the model in Eq. (\ref{spin1Hamiltonian}) with $\lambda=0$ \cite{senko2015realization}. To generate the ground states of this effective spin-1 XY model, for 2- and 4-ion spin chains, the spins were initially prepared in the state $\ket{00\cdots}$. This is the approximate ground state of Eq. (\ref{spin1Hamiltonian}) in the presence of a large $D$-field. This field was then ramped down slowly until $D\approx 0$; the resulting state populations, shown in Fig. \ref{fig:EvenIonGroundState}, match reasonably well with the exactly-calculated ground state.

\begin{figure}
\includegraphics[width=.6\linewidth]{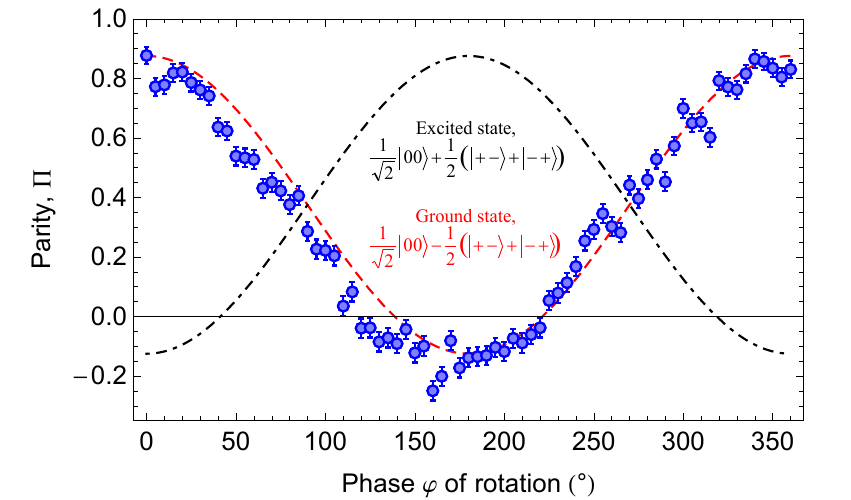}
\caption{Following an adiabatic ramp, the parity of the final state is measured as a function of the final rotation phase $\varphi$ as a witness for spin-1 entanglement (see text for rotation protocol). Dashed and dot-dashed lines represent the theoretically expected values for the ground state, $\ket{00}/\sqrt{2} - \left( \ket{-+} + \ket{+-} \right)/2$, and highest excited state, $\ket{00}/\sqrt{2} + \left( \ket{-+} + \ket{+-} \right)/2$, respectively. The amplitude and phase of the measures oscillation reveals that the prepared state are consistent with the expected ground state. From \citealp{senko2015realization}.}
\label{fig:GroundStatePhase}
\end{figure}

{\q Detection of the spin-1 states was accomplished by imaging the spin-dependent fluorescence \cite{olmschenk2007manipulation} onto an intensified CCD camera and observing which ions were `dark,' indicating the presence of the $\ket{0}$ state. Because both of the $\ket{\pm}$ states appear `bright' during the detection process and are scattered into an incoherent mixture of the $\ket{F=1}$ states, such a setup does not allow discrimination among all three possible spin states in a single experiment. However, the population of either $\ket{+}$ or $\ket{-}$ can be measured by repeating the experiment and applying a $\pi$ rotation to the appropriate $\ket{0}\leftrightarrow\ket{\pm}$ transition before the fluorescence imaging. For instance, measuring an ion in the `dark' state after a $\pi$ pulse between $\ket{0}\leftrightarrow\ket{+}$ indicates that the spin was in the $\ket{+}$ state before detection. This binary discrimination is not a fundamental limit to future experiments, since populations could be ``shelved" into atomic states that do not participate in the detection cycle \cite{Christensen2020}.

Measurements of populations in the $S_z$ basis necessarily discard phase information about components of the final state. This can be important in many spin models, including the XY model, where such measurements alone cannot discriminate between different eigenstates. For example, the ground state of an XY model with two spin-1 particles is $\ket{00}/\sqrt{2} - \left( \ket{-+} + \ket{+-} \right)/2$, while the highest excited state is $\ket{00}/\sqrt{2} + \left( \ket{-+} + \ket{+-} \right)/2$, differing only by a relative phase. In \cite{senko2015realization}, verification of ground state production was accomplished by employing a modified parity entanglement witness procedure \cite{sackett2000experimental}.  First, global $\pi/2$ rotations were applied on both the $\ket{0}\leftrightarrow\ket{+}$ and $\ket{0}\leftrightarrow\ket{-}$ transitions, with a relative phase $\varphi$.  Then parity $\Pi(\varphi) = \sum_{n=0}^2 (-1)^n P_n$ of the number of spins in state $\ket{0}$ was measured, where $P_n$ is the probability of $n$ spins appearing in state $\ket{0}$. This is expected to result in $\Pi(\varphi) = \frac{3}{8} \pm \frac{1}{2}\cos\varphi$, where the $+$ and $-$ signs correspond to the ground and highest excited states, respectively.  The data shown in Fig. \ref{fig:GroundStatePhase} clearly show the phase of the parity oscillations to be consistent with having prepared the 2-spin ground state of the spin-1 XY model.}

\section{Nonequilibrium Phases of Matter and Dynamics \label{sec:nonequil}}

In contrast to section \ref{sec:equil} dealing with equilibrium properties of quantum spin models, trapped ion simulators are well-suited for studying non-equilibrium phenomena. 
Non-equilibrium dynamics might even be considered more natural, since the study of equilibrium-like properties requires a specific protocol for preparing the corresponding ground or thermal state, unlike conventional condensed matter materials that are directly cooled through phonon interactions. 
The simplest non-equilibrium studies, on the other hand, can start with an initial product state and then simply evolve the system under a (time-dependent) Hamiltonian of interest. 

Trapped ion quantum simulators allow the study of non-equilibrium dynamics over a broad range of both spatial and temporal resolution. The effective long-range spin-spin interactions described in section~\ref{subsec:Ising} can be modulated in time by turning on or off the laser-ion interactions. This allows non-equilibrium states to be prepared via quenches or stroboscopic application of the Ising Hamiltonian while their subsequent dynamics are observed over timescales both shorter and longer than the natural timescale of interactions $1/J_{ij}$. Single spin resolution can be achieved even as the system size is scaled to many particles, allowing access to non-trivial observables such as spin-spin correlations and magnetic domain sizes.
Since strongly interacting and highly-frustrated Ising spin models are often employed in analytical and numerical studies of non-equilibrium quantum dynamics, the results of trapped ion spin simulations serve as an important benchmark for these theoretical predictions.

Perhaps the most natural non-equilibrium experiments are global and local quenches. In a global quench experiment, a simple initial state evolves under a time-independent Hamiltonian. A global quench originates from situations where the simple initial state can be naturally thought of as the ground state of some simple Hamiltonian, in which case the dynamics ensues when the Hamiltonian is changed (quenched). A local quench allows the comparison of the non-equilibrium dynamics between two initial states that differ by the application of a locally applied unitary operation. A particular example of a local quench is a situation where one of the two initial states is an eigenstate of the Hamiltonian.

Short-time spin dynamics (Sec.\ \ref{sec:prop}) allows the investigation of time scales over which quantum information can propagate across the system. Long-time dynamics, on the other hand, can indicate whether the system eventually approaches some effective steady state and if so, how this steady-state is approached. Effective thermalization is often expected, even in a closed system of spins where a subset of the spin system uses its complement as the bath~\cite{dalessio2016from}).  However, as discussed in Sec.\ \ref{sec:mbl}, disorder can often prevent such thermalization leading to many-body localization. Similarly, as discussed in Sec.\ \ref{sec:pretherm}, even in cases where the system eventually thermalizes, it is possible that dynamics dramtically slows and actual thermalization takes a very long-time to occur, in a phenomenon known as prethermalization.

There are many forms of inducing and probing spin dynamics in trapped ion systems. One example is the periodic modulation of a Hamiltonian, which gives rise to stroboscopic Floquet dynamics. In Sec.\ \ref{sec:dtc}, we will discuss such dynamics in trapped ion systems in two complementary contexts. The first will focus on the application of a Hamiltonian and its negative counterpart in order to measure so-called out-of-time-ordered correlation (OTOC) functions. The second will focus on the spontaneous breaking of discrete time translation symmetry, leading to the emergence of time crystalline order.

\subsection{Information Propagation \label{sec:prop}}

Many properties of a many-body quantum system depend on how quickly quantum information can propagate in that system. Indeed, the speed of a quantum computation through an array of qubits will be boosted by sending quantum information faster across the array. Similarly, fast information propagation through qubits may allow for faster preparation of highly entangled states of the array. With its tunable approximately-power-law-decaying interactions, a trapped ion chain is an ideal testbed for studying how much such long-range interactions can speed up quantum information propagation relative to nearest-neighbor interactions and for elucidating the implications of such speedups.

To make connection with the native interactions in trapped ion spin crystals, we assume that the interactions fall off with distance as a power-law $J_0/|i-j|^\alpha$ between ions $i$ and $j$, with $0 \leq \alpha \leq 3$, as derived in Eq. (\ref{Jpowerlaw}) and discussed above. For the purposes of studying the bounds on the speed of information propagation, it is the convenient to express the spin Hamiltonian in the following general form:
\begin{eqnarray} \label{eq:hlr}
H = \sum_{i < j} h_{ij} + \sum_i h_i,
\end{eqnarray}
where $h_i$ is a Hamiltonian acting on spin $i$ and where the two-spin Hamiltonian $h_{ij}$ acting on spins $i$ and $j$ is subject to the bound
\begin{eqnarray} \label{eq:power}
||h_{ij}|| \leq \frac{J_0}{|i - j|^\alpha}.
\end{eqnarray}
Here $|| O ||$ indicates the operator (or spectral) norm of operator $O$, or the magnitude of its largest absolute eigenvalue.
We are interested in studying how quickly quantum information can propagate when the system evolves unitarily under Eq.\ (\ref{eq:hlr}). Various notions of information propagation can be defined, depending on the particular problem \cite{tran20}. 

We first focus on the dynamics following a local-quench \cite{jurcevic2014quasiparticle}. Let $B$ be a unitary operator acting on a single site, while $A$ is a single-site observable acting on another site a distance $r$ away. Let $\ket{\psi}$ be an arbitrary initial state, and let $A(t)$ be the Heisenberg evolution of $A$ under the Hamiltonian $H$ in Eq.\ (\ref{eq:hlr}). Then the effect on observable $A$ due to the disturbance $B$ can be defined as the difference between the expectation values of $A(t)$ in the original state $\ket{\psi}$ and in the quenched state $B \ket{\psi}$,
\begin{eqnarray} \label{eq:local}
|\bra{\psi} B^\dagger A(t) B \ket{\psi} - \bra{\psi} A(t) \ket{\psi}| = |\bra{\psi} B^\dagger [A(t),B] \ket{\psi}| \leq ||[A(t),B]||.
\end{eqnarray}
The signal after time $t$ a distance $r$ away is thus bounded by the unequal-time commutator $||[A(t),B]||$. 
Intuitively, the operator $A(t)$ can be thought to originate on a given site at $t=0$ (and commuting with operator $B$ on another site at distance $r$ away) and then spreading in time until its support significantly overlaps with the support of $B$ and thus allows for a substantial commutator $||[A(t),B]||$.

Upper bounds on $||[A(t),B]||$ subject to the Hamiltonian in Eq. (\ref{eq:hlr}) are referred to as Lieb-Robinson-type bounds, named after the original work considering nearest-neighbor interactions ($\alpha = \infty$) \cite{lieb72}. The region in the $r$-$t$ plane outside of which $||[A(t),B]||$ must be small is called the causal region, while its boundary is called the (effective) light cone. While the Hamiltonian in Eq.\ (\ref{eq:hlr}) is time-independent,  Lieb-Robinson bounds also hold for time-dependent $h_{ij}$ subject to Eq.\ (\ref{eq:power}) and for arbitrary time-dependent $h_i$. A growing body of theoretical literature places upper bounds on $||[A(t),B]||$ and therefore derives tighter and tighter light cones for different values of $\alpha$ \cite{hastings06,lashkari13,gong14,foss-feig15,storch15,matsuta16,tran18,else18,guo19,sweke19,tran19c,chen19b,kuwahara19,tran20}. At the same time, a complementary growing body of theoretical literature considers specific Hamiltonians and protocols demonstrating larger and larger causal regions \cite{hazzard13,hauke13b,schachenmayer13,knap2013probing,junemann13,eisert13,hazzard14,storch15,nezhadhaghighi14,rajabpour15,eldredge17,guo19,maghrebi16b,luitz19,gong14,buyskikh16,cevolani15,cevolani16,van-regemortel16,lepori17,cevolani18,frerot18b,chen18d,chen18b,kastner11,bachelard13,worm13,kloss19}. While these upper and lower bounds on information propagation are appearing to converge, there is still no provably tight light cone shape other than at large $\alpha$, where the light cone is strictly linear \cite{chen19b,kuwahara19,tran20}.

\begin{figure}[t!]
\includegraphics*[width=\columnwidth]{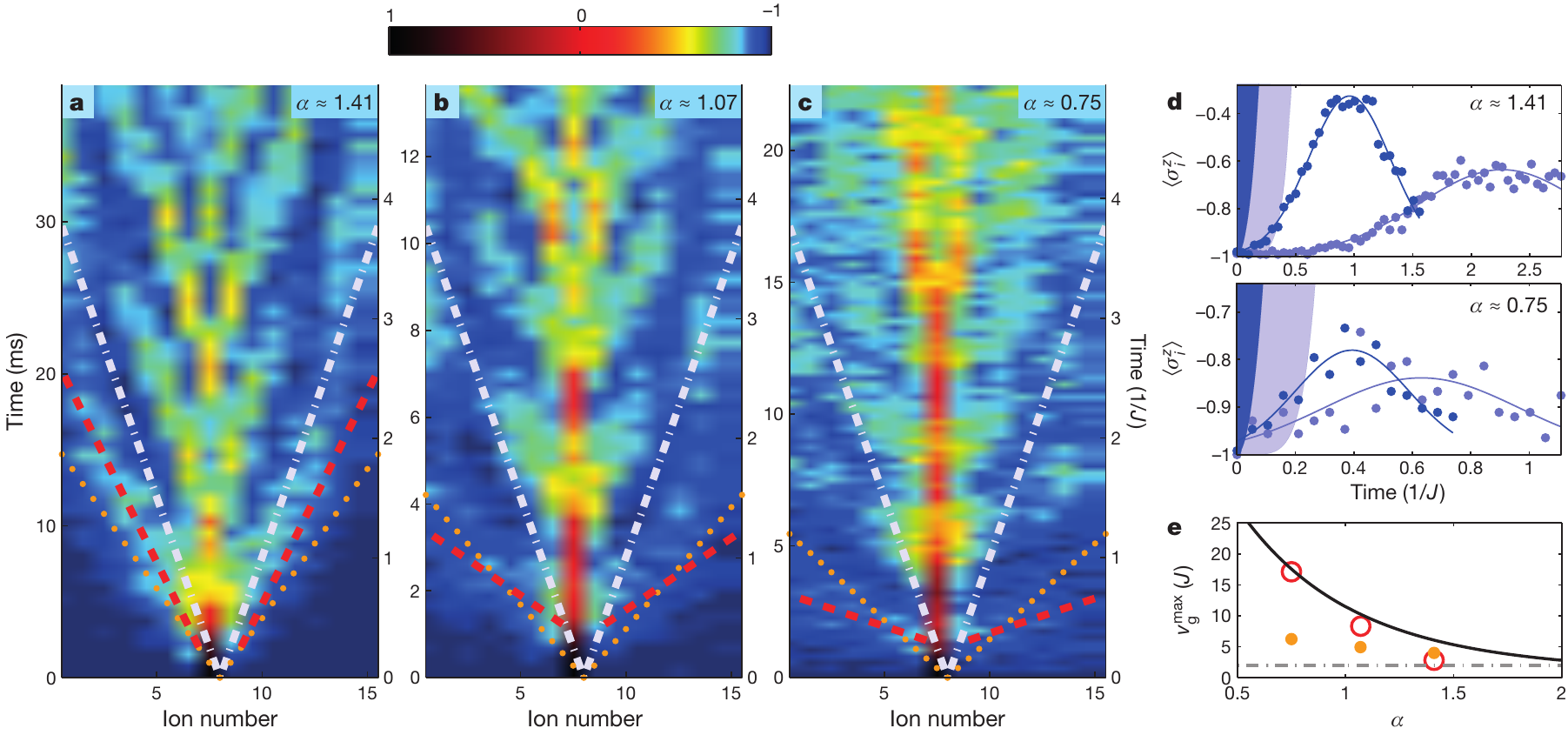}
\caption{\label{fig:local} (a-c) Measured magnetization $\langle \sigma^z_i(t)\rangle$ (color coded) following a local quench. From (a) to (c), the interaction ranges are $\alpha \approx 1.41, 1.07, 0.75$. In (a), an effective light cone is evident and the dynamics are approximately described by nearest-neighbor interactions only. Red lines, fits to the observed magnon arrival times [examples in (d)]; white lines, light cone for averaged nearest-neighbor interactions; orange dots, after renormalization by the algebraic tail. As the interaction range is increased (b,c) the light cone disappears and nearest-neighbor models fail to capture the dynamics. (d), Magnetization of spins (ions) 6 and 13, from (a) (top) and (c) (bottom). Solid lines, Gaussian fits to measured magnon arrival. Top: for $\alpha = 1.41$, a nearest-neighbor Lieb-Robinson bound captures most of the signal (shaded region). Bottom: for $\alpha = 0.75$, it does not. (e) Maximum group velocity. With increasing $\alpha$, the measured magnon arrival velocities (red circles) approach the group velocity of the non-renormalized nearest-neighbor model (grey dash-dotted line). If renormalized by the algebraic tail, the nearest-neighbor group velocity increases at small $\alpha$ (orange dots), but much less than the increase of the observed magnon velocity. For small $\alpha$, the measured arrival times are consistent with the divergent behaviour predicted for full power-law interactions (black line).  Adapted from \citealp{jurcevic2014quasiparticle}.}
\end{figure}

Thanks to Eq.\ (\ref{eq:local}), Lieb-Robinson bounds on $||[A(t),B]||$ directly constrain information propagation after local quenches.  A local quench experiment on a trapped-ion chain has realized the XY model of hopping hard-core bosons corresponding to $h_i = 0$ and $h_{ij} = J_{ij}/|i-j|^\alpha (\sigma^+_i \sigma^-_j + \sigma^-_i \sigma^+_j)$ in Eq.\ (\ref{eq:hlr}) \cite{jurcevic2014quasiparticle}. Figure\ \ref{fig:local} shows results for a local quench on $15$ spins from the experiment. After flipping up the middle spin (corresponding to $B = \sigma^x_8$) in a chain of down spins, the experiment measures $A = \sigma^z_i$ for various $i$. Since the number of flipped spins is conserved during the time evolution, the evolution in this restricted Hilbert space of spins is well-described using the language of single-flip eigenstates called magnons. 

Information propagation can also be studied in global quench experiments, where connected correlations are measured as a simple initial state (such as a product state) evolves under a given Hamiltonian. Linear light-cone-like spreading of such correlations due to a nearest-neighbor Hamiltonian was first measured in neutral atoms confined in an optical lattice \cite{cheneau12}.  For long-range interacting systems such as trapped-ion spins, the spread of correlations may no longer be confined within a linear light cone.
In particular, suppose the system starts in an initial product state $\ket{\psi}$ and evolves under the Hamiltonian in Eqs.\ (\ref{eq:hlr})-(\ref{eq:power}). At time $t = 0$, the connected correlation function $C_{ij}(t) = \langle O_i(t) O_j(t) \rangle - \langle O_i(t) \rangle \langle O_j(t) \rangle$ (where operator $O_i$ acts on site $i$) vanishes since the first expectation value factorizes. As time goes on, the supports of $O_i(t)$ and $O_j(t)$ grow, making $C_{ij}(t)$ grow in return. For the case of short-range interactions, $C_{ij}(t)$ is bounded in the $r$-$t$ plane (where $r = |i-j|$) by a linear light cone similar to the corresponding light cone for the unequal time commutator $||[A(t),B]||$ \cite{bravyi06}. For general $\alpha$, a bound on $||[A(t),B]||$ can also be used to derive a bound on $C_{ij}(t)$ \cite{gong14v1,tran20}, but the relationship between the two light cones is not as trivial as in the nearest-neighbor ($\alpha = \infty$) case. However, the physical picture is similar: the Lieb-Robinson bound constraints the spread of operators $O_i(t)$ and $O_j(t)$, so if the support of $O_i(t)$ ($O_j(t)$) hasn't spread significantly outside of a ball of radius $r/2$ around $i$ ($j$), then $O_i(t)$ and $O_j(t)$ have approximately disjoint supports, leading to small $C_{ij}(t)$.  

\begin{figure}[t!]
\includegraphics*[width=\columnwidth]{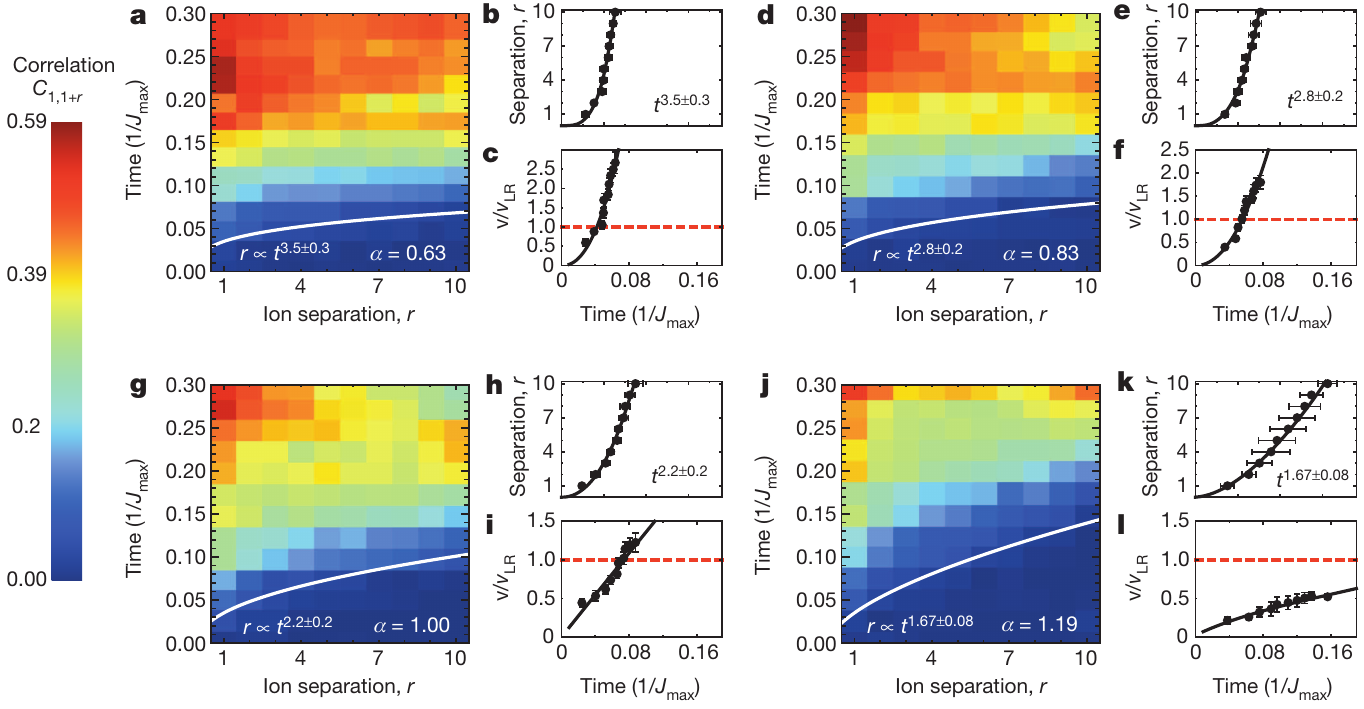}
\caption{\label{fig:global} (a-c) Spatial and time-dependent correlations (a), extracted light-cone boundary (chosen as the contour $C_{i,j} = 0.04$) (b) and correlation propagation velocity (c) following a global quench of a long-range XY model with $\alpha = 0.63$. The curvature of the boundary shows an increasing propagation velocity (b), quickly exceeding the short-range Lieb-Robinson velocity bound, $v_\textrm{LR}$ (c). Solid lines give a power-law fit to the data, which slightly depends on the choice of  contour $C_{i,j}$. (d-l), Complementary plots for $\alpha = 0.83$ (d-f), $\alpha = 1.00$ (g-i) and $\alpha = 1.19$ (j-l). As the range of the interactions decreases, correlations do not propagate as quickly through the chain. For the shortest-range interaction (j-l), the experiment demonstrates a faster-than-linear growth of the light-cone boundary, despite having $\alpha > 1$. Error bars, 1 s.d.  Adapted from \citealp{richerme2014non}.}
\end{figure}

Connected correlations following a global quench have been measured in a chain of trapped ions \cite{richerme2014non}. This experiment also studied the XY model corresponding to $h_i = 0$ and $h_{ij} = J_{ij}/|i-j|^\alpha(\sigma^x_i \sigma^x_j + \sigma^z_i \sigma^z_j)$ in Eq.\ (\ref{eq:hlr}). Figure\ \ref{fig:global} shows data on 10 ions following a global quench. Starting with an initial state of all spins pointing down (in the $z$ basis), the experiment measures the time evolution of connected correlations $C_{ij}(t) = \langle \sigma^z_i(t) \sigma^z_j(t) \rangle - \langle \sigma^z_i(t) \rangle \langle \sigma^z_j(t) \rangle$. The growth of connected correlations following a global quench may also be accompanied by the growth of entanglement, as was observed experimentally in \cite{friis18}.

Experiments on ultracold polar molecules \cite{yan13} and defect centers in solid state \cite{choi17} do not yet have the single-spin resolution necessary for studying the shape of the causal region after local or global quenches in long-range-interacting systems. On the other hand, experiments on ultracold neutral atoms interacting via Rydberg-Rydberg interactions \cite{bernien17,lienhard18,guardado-sanchez18,zeiher17} should be able to access the particularly interesting parameter regime of $\alpha = 3$ (dipolar interactions) \cite{leseleuc18b} in one, two, and three spatial dimensions. 

\subsection{Disorder Induced Localization \label{sec:mbl}}

Many-body localization (MBL) has become one of the most studied nonequilibrium phases of matter, receiving considerable scrutiny in both experiment and theory in the past decade \cite{oganesyan2007localization,nandkishore2015many-body,abanin2019colloquium,schreiber2015observation,choi2016exploring}. The localization effect is a generalization of single-particle Anderson localization, which is characterized by a cessation of quasiparticle transport in non-interacting systems subject to a random potential landscape~\cite{anderson1958absence}. Surprisingly, in the case of MBL similar insulator-like properties are observed even when particles are strongly interacting~\cite{oganesyan2007localization,nandkishore2015many-body,abanin2019colloquium}. When prepared with a quench, the quantum states become highly entangled many-body superpositions of excited eigenstates spanning the entire energy spectrum of the disordered system Hamiltonian. MBL can be distinguished from Anderson localization by the logarithmic growth of entanglement entropy at long-times \cite{bardarson2012unbounded}. The distribution of eigenstates occupied in an MBL phase is decidedly non-thermal and a number of observables have been identified to characterize phase transitions between MBL and thermal states when varying the interaction strength or disorder in the Hamiltonian~\cite{nandkishore2015many-body,abanin2019colloquium,luitz2015many-body,serbyn2014interferometric}. 

Signatures of MBL have been observed in a trapped ion quantum simulator in both \citealp{smith2016many} and \citealp{brydges2019probing} by engineering a locally disordered but programmable potential ($H_{D}$), which is applied simultaneously with an effective long range interacting transverse field Hamiltonian (see section~\ref{sec:Eff_B_and_J})
\begin{equation}
H_{\textrm{MBL}} = \sum_{i<j} J_{ij} \sigma_i^x \sigma_j^x + \frac{B}{2} \sum_i \sigma_i^z + H_{D}.
\label{eq:MBLHAM}
\end{equation}
The disordered potential is implemented with single site resolution across the ion chain such that
\begin{equation}
H_{D} = \sum_{i} D_i \sigma_i^z,
\label{eq:HDisorder}
\end{equation}
where $D_i$ is sampled from a uniform distribution, $D_i \in [-W/2 , W/2]$, with width $W$. For finite system sizes, this Hamiltonian exhibits  features consistent with many-body localization and demonstrates a disorder-induced, long-lived memory of the system's initial conditions~\cite{hauke2015many-body,maksymov2017comment}. Understanding the thermodynamic stability of localization with power-law interactions remains an intriguing open question \cite{burin2006energy,yao2014many,burin2015localization,nandkishore2017many}. 

In both experiments, the MBL state is created by initially preparing the 10 spin N\'{e}el state with staggered order $(\ket{\psi_0} = \ket{\downarrow \uparrow \downarrow \uparrow \downarrow \uparrow \downarrow \uparrow \downarrow \uparrow}_z)$ which is highly excited with respect to the disordered Ising Hamiltonian of Eq.~\ref{eq:MBLHAM}. This Hamiltonian is rapidly quenched on and the resulting single spin magnetization dynamics $\expval{\sigma_i^z(t)}$ are measured for times up to $t = 10/J_0$. The experiment is repeated under multiple instances of disorder with Stark-shifts $(D_i)$ applied programmatically to each ion using a rastered individual addressing laser~\cite{lee2016engineering}. This individual addressing laser is also used create the initial N\'{e}el state using a sequence of controlled spin-flips.

\begin{figure}[t!]
\includegraphics*[width=\columnwidth]{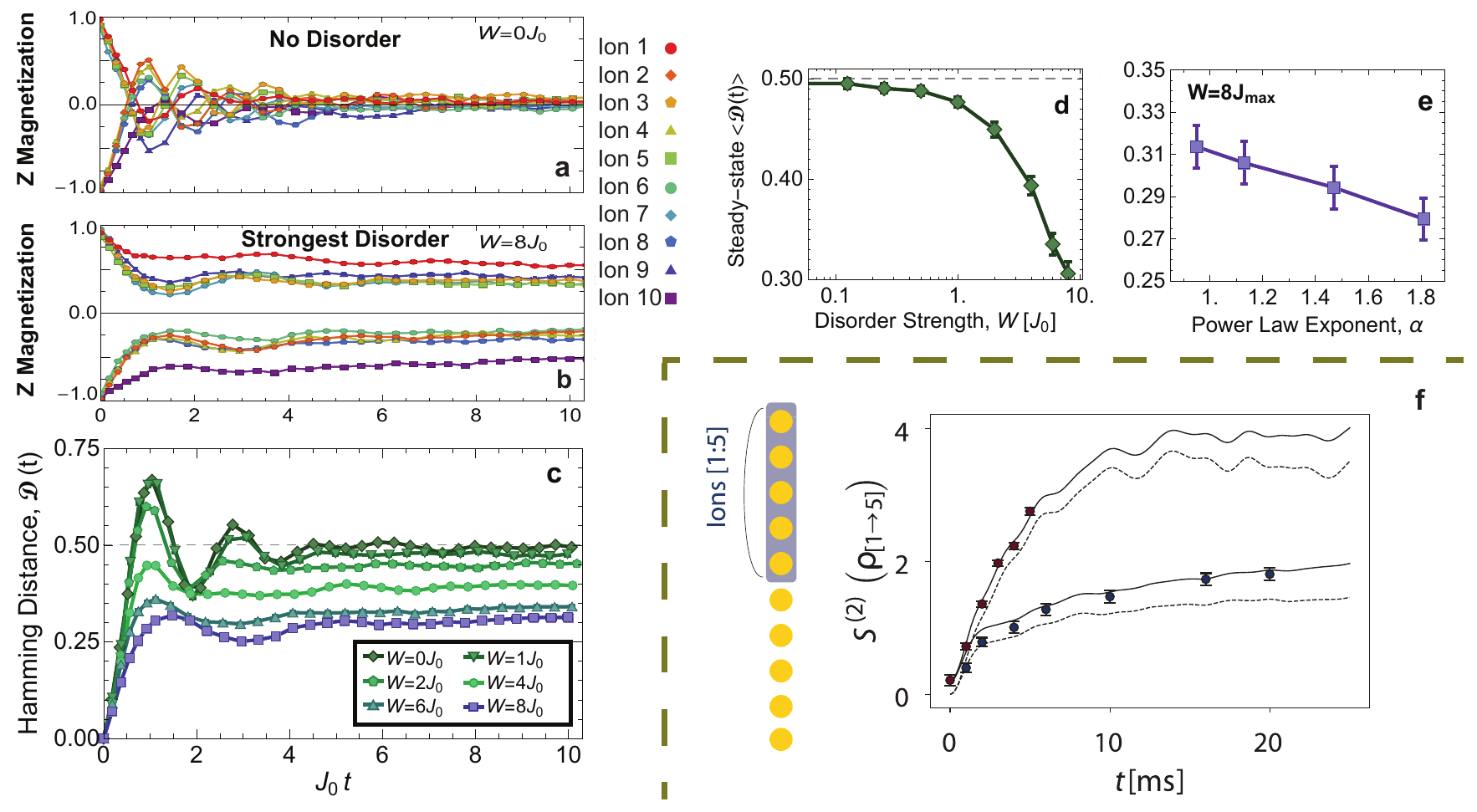}
\caption{Many-Body Localization: (a)-(b) Temporal dynamics for each of the 10 ion's Z magnetization $\expval{\sigma_i^z(t)}$ in the case of zero disorder and strong disorder for panels (a) and (b) respectively. (c) The normalized Hamming distance $\mathcal{D}(t)$ has plateaued to a steady state value for $J_0 t \geq 5$ at all measured disorder strengths. (d) The time averaged steady state value $\expval{\mathcal{D}(t)}$ after the plateau shows the onset of a crossover between a thermalizing regime $(\expval{\mathcal{D}(t)} = 0.5)$ and localizing regime $(\expval{\mathcal{D}(t)} = 0)$ as disorder increases. (e) The steady state Hamming distance increases with longer range interactions. (f) The half-chain entropy growth in the absence of disorder (red points) and for a disordered chain with $W=6J_0$ (blue points), compared with numerical simulations of unitary dynamics (dotted lines) and including known sources of decoherence (solid lines). Panels (a)-(e) are adapted from \citealp{smith2016many}, and panel (f) is adapted from \citealp{brydges2019probing}.}
\label{fig:MBL}
\end{figure}

In the absence of disorder these initial spin states will thermalize if the uniform transverse field $B$ is sufficiently large~\cite{deutsch1991quantum, srednicki1994chaos, rigol2008thermalization}. In \citealp{smith2016many}, global rotations are used to prepare eigenstates of both $\sigma^x$ and $\sigma^z$ and measure the resulting single ion magnetization projected into those directions after evolution under $H_{\textrm{MBL}}$. In the case of a thermalizing system, memory of the initial spin configuration will be lost in all directions of the Bloch sphere, namely $\expval{\sigma_i^x} = \expval{\sigma_i^z} = 0$ at long times. Above the threshold transverse field $(B \gtrsim 4 J_0)$ the system rapidly thermalizes to zero magnetization after relatively short timescales ($t < 5/J_0$) (Fig.~\ref{fig:MBL}a).

However, with the transverse field held fixed at $B = 4 J_0$, the data clearly shows that applied disorder localizes the spin chain, retaining memory of the initial N\'{e}el state in measurements of the $z$ magnetization $\expval{\sigma_i^z}$ (Fig.~\ref{fig:MBL}b). Each measurement of magnetization dynamics for disorder width $W$ is repeated with at least 30 different realizations of disorder, which are subsequently averaged together. This is sufficient to reduce the finite depth disorder sampling error to be of the same order as other noise sources. After some initial decay and oscillations, the magnetization of each spin settles to a steady state value for $J_0 t \geq 5$. The degree of localization can be quantified using the normalized Hamming distance (HD)
\begin{equation}
	\begin{aligned}
		\mathcal{D}(t) &=  \frac{1}{2} - \frac{1}{2N} \sum_{i} \expval{\sigma_i^z(t)\sigma_i^z(0)} \\
               &=  \frac{1}{2} - \frac{1}{2N} \sum_{i} (-1)^i  \expval{\sigma_i^z(t)}.
	\end{aligned}
\end{equation}
This observable counts the number of spin flips from the initial state, normalized to the length of the spin chain. At long times, a randomly oriented thermal state shows $\mathcal{D} = 0.5$ while one that remains fully localized has $\mathcal{D} = 0$ (Fig.~\ref{fig:MBL}c). 

The average steady state value $\expval{\mathcal{D}(t)}$ for $J_0 t \geq 5$ can serve as an order parameter to display the crossover between the localizing and thermalizing regimes. The most relevant adjustable experimental control parameters for probing the MBL phase diagram are the amplitude of disorder $W$ and the interaction range $\alpha$. Increasing $W$ pins each spin closer to its initial state and pushes the entire spin chain towards a localized regime (Fig.~\ref{fig:MBL}d). Likewise, the localization strengthens as $\alpha$ is increased towards shorter range interactions (Fig.~\ref{fig:MBL}e), recovering Anderson localization via a Jordan-Wigner transformation \cite{JW} in the $\alpha \rightarrow \infty$ limit. Numerical studies have confirmed that full localization occurs within experimentally accessible disorder strengths and interaction ranges~\cite{wu2016understanding}.

The slow growth of entanglement entropy $(S)$ has long been understood as a distinguishing feature of localization~\cite{bardarson2012unbounded}. A short range interacting MBL state should exhibit a slower entanglement growth rate than interacting quantum states without disorder, where entanglement spreads ballistically. The dynamics of the entanglement entropy are also quite different for non-interacting Anderson localized systems, where the entanglement saturates at short times once the system's dynamics have reached the localization length~\cite{abanin2019colloquium}. In a trapped ion quantum simulator with algebraically decaying interactions the entanglement entropy of an MBL state should also grow algebraically, $S \sim t^q$, but with $q <1$ the dynamics are still distinct from those of non-localized or Anderson localized systems~\cite{Pino2014entanglement}.

It is generally difficult to measure entanglement entropy in quantum simulators due to the exponential system-size scaling of the number of measurements required for full state tomography. In \citealp{smith2016many} the observed slow growth in quantum-fisher information (QFI) is used as a proxy for half-chain entanglement entropy, motivated by a similar scaling with disorder and interaction strength as observed in numerical simulations. A more direct measurement is made in \citealp{brydges2019probing}, where they develop a technique to probe the second-order half-chain R\'enyi entanglement entropy in their 10 ion quantum simulator ($S^{(2)}(\rho_{[1\rightarrow5]})$) using randomized measurements. They find the entanglement growth to be significantly suppressed in the presence of strong disorder, in good agreement with numerical predictions (Fig.~\ref{fig:MBL}f).

\begin{figure}[t!]
\includegraphics*[width=0.5\columnwidth]{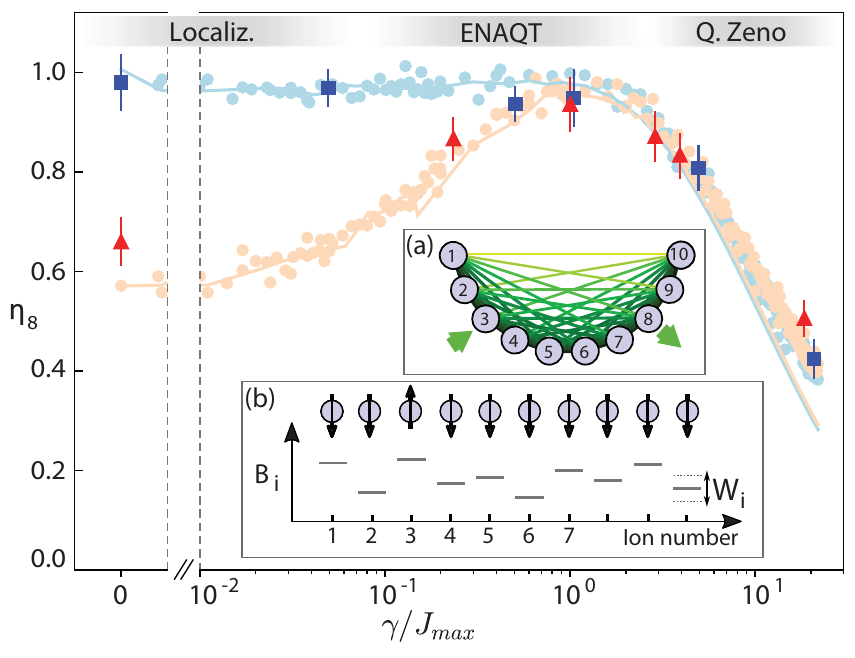}
\caption{Breakdown of Anderson Localization: Main graph: Transport efficiency $\eta_8$ to the target (ion 8)
under different strengths of static disorder (blue: $B_{\max} = 0.5J_0$; red: $B_{\max} = 2.5J_0$) and Markovian-like dephasing with rate $\gamma$. Experimental points (shown as dark squares and triangles) result from averaging over 20–40 random realizations of disorder and noise, with 25 experimental repetitions each. The regimes of localization, ENAQT \cite{Rebentrost_2009}, and the quantum Zeno effect are indicated in gray. The data agree well with theoretical simulations of the coin-tossing random process (light bullets) realized in the experiment, while simulations with ideal Markovian white noise (lines) underestimate ENAQT at large $\gamma$. {Inset (a)}: Sketch of the transport network. The ions experience a long-range coupling, with darker and thicker connections indicating higher coupling strengths. The green arrows denote the source (3) and the target (8) for the excitation in the ion network. {Inset (b)}: Sketch of the ion chain representing interacting spin-$1/2$ particles as circles, with the spin states denoted by arrows. The ions are subject to random static and dynamic on-site excitation energies, indicated by $B_i$ and $W_i(t)$. Figure and caption reproduced from \citealp{Maier2019environment}. }
\label{fig:ENAQT}
\end{figure}

Interestingly, Anderson localization can be explored using the same long-range disordered Hamiltonian, even though it is not strictly non-interacting, by observing the dynamics of a single spin excitation in the ion chain. For example, in \citealp{Maier2019environment} the transport efficiency of a spin excitation from initial site $i=3$ to the target site $i=8$ is observed as a function of time in a 10-ion spin chain (Fig.~\ref{fig:ENAQT}a). The evolution of this single spin excitation can be described by an XY model Hamiltonian
\begin{equation}
H= \sum_{i \neq j} J_{i j}\left(\sigma_{i}^{+} \sigma_{j}^{-}+\sigma_{i}^{-} \sigma_{j}^{+}\right)+ \sum_{i}\left[B_{i}+W_{i}(t)\right] \sigma_{i}^{z}    
\label{eq:DynDis}
\end{equation}
where the disorder field contains single site static $(B_i)$ and time-dependant $[W_i(t)]$ components (Fig.~\ref{fig:ENAQT}b). 
In the absence of disorder, the XY interaction term in this Hamiltonian conserves the total magnetization of the system, allowing a single spin excitation to hop around the chain. 
The transport efficiency on site $i=8$ is then quantified by integrating the instantaneous probability of the excitation appearing on spin 8, 
\begin{equation}
\eta_{8} \equiv \int_{0}^{t_{\max }} \frac{\langle\sigma_{8}^{z}(t)\rangle+1}{2} dt  
\label{eq:TranEff}
\end{equation}
over the full duration of the experiment $t_{\max}$.

The transport efficiency is reduced in the presence of strong static disorder ($B_{\max} > J_0$), consistent with  Anderson localization of the initial spin excitation (Fig.~\ref{fig:ENAQT} main). However, adding temporal variations in the form of white dephasing noise in $W_i(t)$ destroys the localization, a phenomenon known as environment-assisted quantum transport (ENAQT) \cite{Rebentrost_2009}. The spectral density of the Markovian-like noise ($S(\omega) \propto W^2_{\max}$) in $W_i(t)$ determines the rate of dephasing $\gamma = S(\omega)$. This experiment can clearly access regimes where transport is inhibited by either Anderson localization $\gamma < J_0$ or the quantum Zeno effect $\gamma > J_0$. In the ENAQT regime, when $\gamma \approx J_0$, the temporal noise modifies the destructive interference necessary for Anderson localization and transport is revived.
Section~\ref{sec:dtc} further explores experiments studying trapped ion spin dynamics under the influence of both static disorder and a periodically time-varying Hamiltonian.

Many-body localization is a unique case in which a closed quantum system remains non-ergodic and localized even up to infinite times. The trapped ion quantum simulations of \citealp{smith2016many} and \citealp{brydges2019probing} are limited by finite experimental coherence times to only one decade in $J_0 t$. This makes it difficult for these experiments to quantify how long-lived the magnetization or slow the entanglement growth might be. Fortunately, other experiments that have studied MBL using cold neutral atoms can achieve several orders of magnitude longer evolution time relative to their interaction timescale. For example, MBL can be realized using cold fermions in quasi-random 1D optical lattices~\cite{schreiber2015observation,luschen2017observation, kohlert2019observation, lukin2019probing}, verifying MBL-like behavior in a variety of Hubbard Hamiltonians. This system has also been used to confirm the breakdown of MBL in open quantum systems~\cite{bordia2016coupling,luschen2017signatures}. Moreover, experiments have started to probe whether MBL can exist in systems with dimensionality $> 1$~\cite{kondov2015disorder, choi2016exploring, bordia2017probing}, where the stability of MBL is still an open question~\cite{deroeck2017many-body}. Other experimental platforms have used novel metrics to probe many-body localization, including many-body spectroscopy~\cite{roushan2017spectroscopic}, measuring out-of-time order correlators~\cite{wei2018exploring} (see Sec. \ref{sec:dtc}), and performing full state tomography to compute entanglement entropy~\cite{xu2018emulating}.
 
\subsection{Prethermalization} \label{sec:pretherm}
Hamiltonians that support MBL are believed to be non-ergodic, even after evolution times exponentially long in the system size \cite{nandkishore2015many-body}. There are also systems that are non-ergodic for a shorter amount of time (but often still much longer than the coherence time of typical quantum simulation experiments) before eventually thermalizing. Usually, these systems are not disordered and can be described by models of weakly interacting (quasi-)particles, such as 1D Bose gases~\cite{Kinoshita2006, Gring2012, Langen2015}. The generic behavior of such a system is called prethermalization, meaning that the system relaxes to a quasi-stationary state different from thermal state  before thermalizing eventually. The prethermal quasi-stationary state is usually believed to be described by a generalized Gibbs ensemble (GGE)~\cite{Rigol2007} that corresponds to the model of quasi-particles without the weak interactions. Such state will have a partial memory of the initial state, because the quasi-particle occupation numbers are conserved if interactions are ignored. At sufficiently long time, the weak interactions are expected to break integrability of system and lead to thermalization in the end. This picture of prethermalization has been well studied both in theory \cite{Berges2004,Polkovnikov2011,Manmana2007} and experiment \cite{ Gring2012, Langen2015}. 

In a programmable ion-trap quantum simulator, due to long-range spin interactions, new types of prethermalization can occur with prethermal states not described by a standard GGE. An example study was first proposed theoretically \cite{Gong2013} and later demonstrated experimentally \cite{neyenhuis2017observation}. The central idea is that with sufficiently long range interactions, a non-disordered and homogenous system can have a strong emergent inhomogeneity due to the open boundary condition of an experimental spin chain. This emergent inhomogeneity can lead to trapping of quasi-particles before the system relaxes to GGE. As both kinetic energy and weak interactions can delocalize trapped quasi-particles, the dynamics of the system can reveal a rich interplay between quantum tunneling and interaction effects, leading to new types of relaxation beyond conventional prethermalization.

The model under study in Refs.\,\cite{Gong2013,neyenhuis2017observation} is the same transverse field Ising model described by Eq. (\ref{eqn:TransversIsing}). With long-range interactions, $H_{TI}$ is generally nonintegrable, in contrast to the nearest-neighbor case where the 1D model is integrable through a Jordan-Wigner transformation \cite{sachdev2011quantum}, so thermalization is anticipated in the long-time limit according to the eigenstate thermalization hypothesis \cite{rigol2008thermalization}. To better understand the dynamics of such a pre-thermal Hamiltonian, we can map each spin excitation along the z direction into a bosonic particle to turn Eq.\,\eqref{eqn:TransversIsing} into a bosonic model with two parts: An integrable part composed of noninteracting spin-wave bosons that can be used to construct a GGE, and an integrability-breaking part consisting of interactions among the spin-wave bosons, which is responsible for the thermalization \cite{neyenhuis2017observation}. When the initial state has a low spin/ bosonic excitation density and the magnetic field is much larger than the average Ising coupling $J_{ij}$, the bosonic excitation density will remain low during the dynamics and the interactions among the bosons will remain weak.

The experiment in~\cite{neyenhuis2017observation} begins by preparing a single spin excitation on either edge of a 7-ion chain $\ket{\psi_{R}} = \ket{\downarrow \downarrow \downarrow \downarrow \downarrow \downarrow \uparrow}_z$ or $\ket{\psi_{L}} = \ket{\uparrow \downarrow \downarrow \downarrow \downarrow \downarrow \downarrow}_z$. The spins then evolve under Eq.\,\eqref{eqn:TransversIsing} and the time evolution of the spin projection in the $z$-basis is measured. The magnetic field $B$ is at least an order of magnitude larger than $J_0$ in the experiment so the number of spin excitations along the $z$ direction is approximately conserved in the short time dynamics where the system can be regarded as a single spin excitation. But in the long time dynamics, multiple spin excitations will be created and interacting with each other.

To characterize the dynamics of the spin excitations, we introduce a single observable that measures the relative location of the spin excitation in the chain
\begin{equation}
C = \sum_{i=1}^N \left( \frac{2 i-N-1}{N-1} \right) \left( \frac{\sigma_i^z+1}{2} \right),
\end{equation} 
where $N$ is the number of ions. The expectation value of $C$ varies between -1 and 1 for a spin excitation on the left and right ends, respectively. The choice of initial states ensures that the initial value of $\langle C \rangle$ is either $1$ or $-1$. Due to the spatial inversion symmetry of the underlying Hamiltonian in Eq.\,\eqref{eqn:TransversIsing}, both the GGE and thermal values of $\langle C\rangle$ should be zero. 

\begin{figure}[t!]
\includegraphics*[width=\columnwidth]{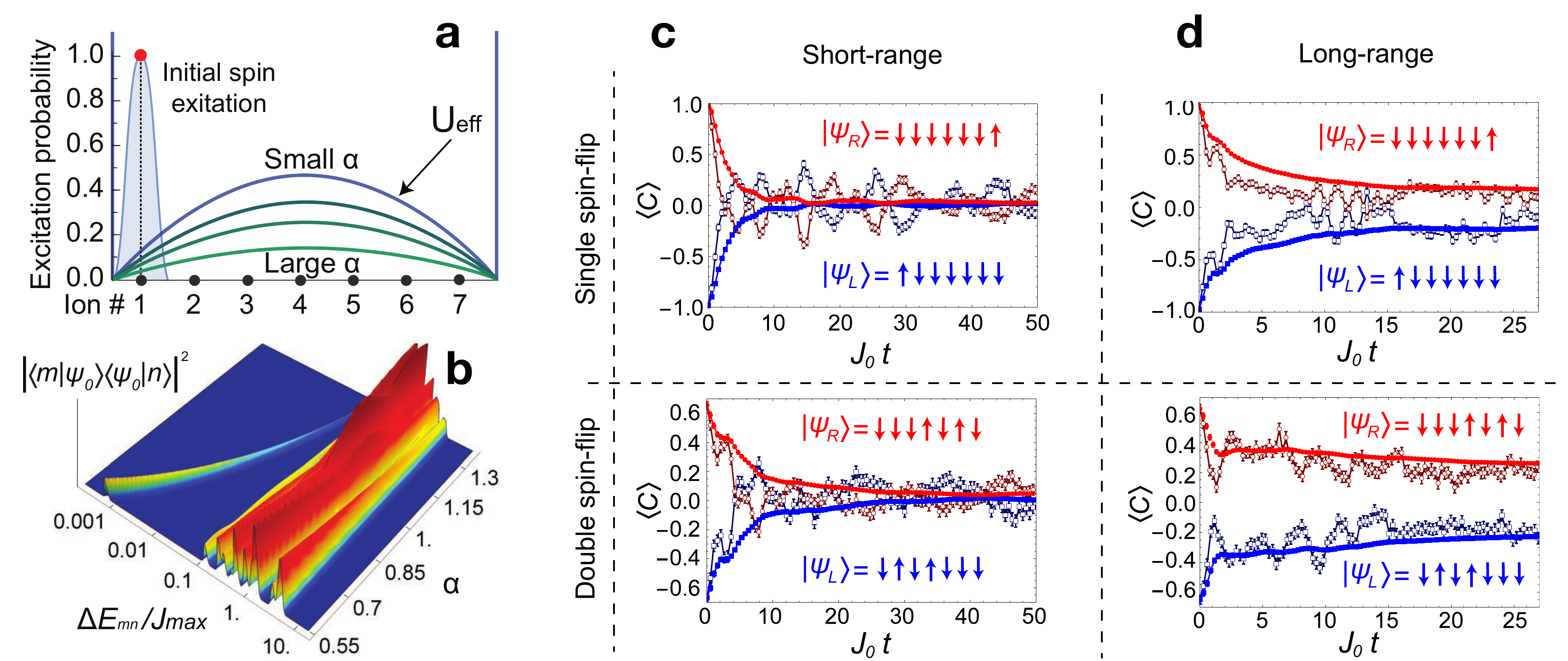}
\caption{{\q (a) An initial spin excitation is prepared on one side of a 7 ion chain subject to open boundary conditions and long-range $XY$ interactions. As the interaction range increases ($\alpha$ decreases), the effective potential energy for the single excitation changes from a square well potential to a potential resembling a double well. (b) The double-well potential gives rise to near-degenerate eigenstates when $\alpha$ is decreased, as seen in the calculated energy difference between all pairs of eigenstates versus $\alpha$. The height of the plot quantifies the off-diagonal density matrix elements of the initial state. (c) In the case of short range interactions, either one or two spin flips delocalize across the chain during time evolution, so that the quantity $\expval{C} \approx 0$ in the long time limit, consistent with the prediction of GGE. (d) For long-range interactions memory of initial conditions is preserved in a long-lived perthermal state. In both (c) and (d), the open squared/circles plot $\expval{C}$ for initial states prepared on the left/right side of the spin chain, while the filled circles/squares plot the cumulative time average $\expval{\overline{C}}$ for this data.} Adapted from \citealp{neyenhuis2017observation}.}
\label{fig:Pretherm}
\end{figure}

In Fig. \ref{fig:Pretherm}(c-d) the value of $\langle C \rangle$ along with its cumulative time average $\langle\overline{C}\rangle$ are shown for the two initial states with a single spin flips on either end of the spin chain. In the short-range interacting case ($\alpha = 1.33$), where the system rapidly evolves to a prethermal state predicted by the GGE associated (with $\langle\overline{C}\rangle=0$) with the integrals of motion corresponding to the momentum space distribution of the single particle representing the spin excitation. The memory of the initial spin excitation location is thus not preserved. However, in the long-range interacting case ($\alpha=0.55$), the position of the spin excitation reaches a quasi-stationary value that retains a memory of the initial state out to the longest experimentally achievable time of $25/J_{\textrm{max}}$. This prethermal state differs clearly from both a thermal state and the GGE prediction, which both maintain the left-right spatial symmetry of the system. 

For evolution times too short to generate more than one spin flip, the dynamics of the Hamiltonian in Eq. (\ref{eqn:TransversIsing}) for the initial states are similar to those of a free-particle in a potential, with the location of the particle representing that of the single spin excitation. For short-range spin interactions, the shape of the potential is approximated by a square well, due to the open boundary condition and no explicit spatial inhomogeneity of the interactions. However, as we increase the range of spin-spin interactions, the shape of the potential distorts from a square well to a double well shaped potential formed by the two hard walls at the ends of the spin chain and the bump at the center of the chain, as shown in Fig.\,\ref{fig:Pretherm}a. For a single particle on a lattice with a double well potential, there will be an extensive number of near-degenerate eigenstates that are symmetric and antisymmetric superpositions of wavefunctions in the left and right potential wells. For seven lattice sites, the spectrum of energy differences between all pairs of eigenstates as a function of $\alpha$ is shown in Fig.\,\ref{fig:Pretherm}b, together with the overlap of eigenstates  with the initial state. For the longest range interaction ($\alpha=0.55$), the two lowest energy states are almost degenerate, with an energy difference approximately 1000 times smaller than $J_{\text{max}}$. This stems from the tunneling rate between the two double well, which is exponentially small in the barrier height, resulting in the spin excitation remaining in its initial well until it tunnels across the potential barrier at very long times. 

\begin{figure}[t!]
\includegraphics*[width=0.75\columnwidth]{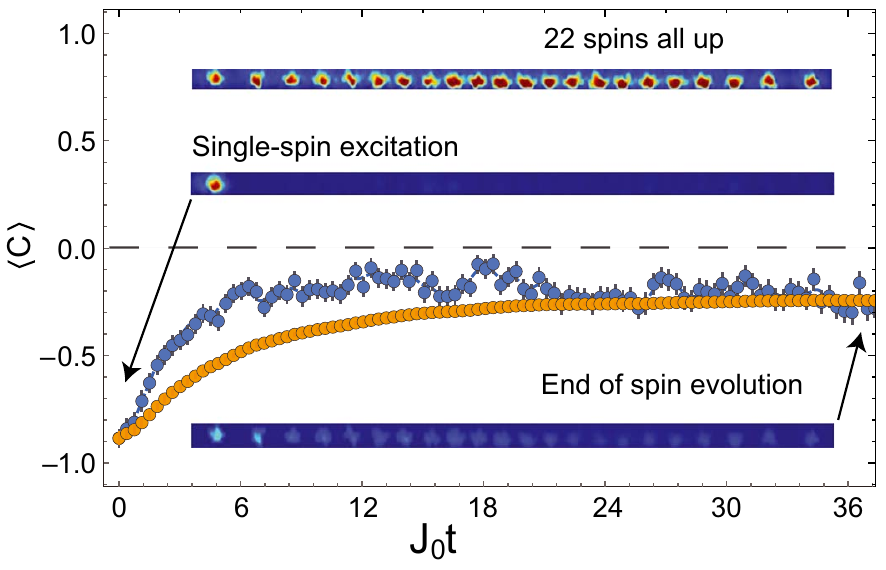}
\caption{Time evolution (light blue) and cumulative time average (orange) of the averaged center of excitation $\langle C \rangle$ in a 22-ion chain. The ion spins are initialized with a
single spin excitation on the left end (middle image). After evolving in the XY spin Hamiltonian for time $t=36/J_0$, where $J_0$ is the nearest-neighbor coupling, the spin excitation is delocalized, but its average position remains stuck on the left half of the chain (bottom image), the signature of prethermalization. Effective range of the interaction is $\alpha \approx 0.9$. Adapted from \citealp{neyenhuis2017observation}.}
\label{fig:Pretherm22}
\end{figure}

To go beyond the single particle picture above, the experiment in \cite{neyenhuis2017observation} prepares initial states with two spin excitations. In this case, there will be weak interactions between the two particles that represent the spin excitations, similar to the scenario for many-body localization \cite{smith2016many}. Despite the presence of weak interactions, similar prethermal states were found, as shown in the bottom panel of Fig.\,\ref{fig:Pretherm}(c-d): Relaxation to GGE is found for shorter-range interactions ($\alpha=1.3$) while for longer-range interactions the system clearly does not relax to GGE. Similar results were also found in a spin chain of 22 ions, as shown in Fig. \ref{fig:Pretherm22} \cite{neyenhuis2017observation}. The persistence of the same prethermalization observed with more than a single spin excitation is attributed by the existence of extensive number of nearly degenerate eigenstates for the single particle spectrum in the double well shown in Fig.\,\ref{fig:Pretherm22}a. Thus an extensive number of spin excitations near one end of the chain will still be localized by the double well before tunneling happens at a later time.

The interplay between single-particle tunneling in the effective double well and particle-particle interactions is always present in this system; even when initialized in a single spin excitation state, the finite transverse field in Eq.\,\eqref{eqn:TransversIsing} will create more spin excitations over time. The effect of interactions will thermalize the system, while the effect of tunneling will bring the system to the GGE. Thus, depending on the range of spin-spin interactions, it is possible to either observe the prethermalization to GGE after the prethermalization caused by the trapping in double well potential, or observe thermalization directly after the observed prethermalization. Determining which scenario is relevant would require a much longer coherence time than is possible in current experiments. While improving the experimental coherence time of the spin interactions may be challenging, simulating the long-time dynamics of an non-integrable, long-range interacting spin chain is equally, or even more challenging on a classical computer.

It should be emphasized that the emergence of an effective double-well potential for the spin excitations is a phenomenon unique to an open spin chain with strongly long-range interactions. A spin chain with periodic boundary condition and without spatial inhomogeneity is fully translationally invariant, and translational invariance is not expected to be broken in the long time behavior of the system. It may be surprising that changing boundary conditions of a long-range interacting system can significantly impact its bulk properties. Prethermalization in trapped ion spin crystals is thus a good example of new physics that is possible with programmable quantum simulators, reaching beyond existing condensed matter frameworks, as also demonstrated by other experiments mentioned in this section.

\subsection{Stroboscopic Dynamics and Floquet Phases of Matter \label{sec:dtc}}
Section \ref{sec:mbl} treated static (time-independent) Hamiltonians whose non-equilibrium nature arises from the presence of quenched disorder leading to many-body localization; such localization prevents the system's internal dynamics from thermalizing and leads to certain memory of local initial conditions \cite{nandkishore2015many-body,abanin2019colloquium}.
An alternate setting for exploring non-equilibrium phases is to begin with a time-dependent Hamiltonian whose equations of motion are intrinsically dynamical. 
{\q This setting is particularly ideal for trapped ion quantum spin simulators, where pulsed control of both the interactions and fields of the transverse field Ising model of Eq. (\ref{eqn:TransversIsing}) naturally allow for the realization of time-dependent Hamiltonians (e.g.~see \ref{sec:adiabatic_prep} where time-dependent magnetic fields are utilized).}

Recently, a tremendous amount of theoretical and experimental work has been devoted to exploring the mildest case of such time-dependence, where the system is governed by a periodic Hamiltonian, $H(t + T) = H(t)$. Such Floquet systems \cite{floquet1883equations,bukov2015universal,kuwahara2016floquet} are particularly ideal from the perspective of quantum simulation since they do not require the complexities of cooling to the many-body ground state in order to observe novel dynamical phenomena~\cite{bloch2008many}.
In the case of trapped ions, as we have previously discussed in section \ref{sec:MB_hamil}, there exists a natural capability to stroboscopically apply different microscopic Hamiltonians, making this platform an ideal Floquet quantum simulator. 
Here we focus on two specific examples.

First, we describe the implementation of a novel class of measurements termed out-of-time-ordered correlation (OTOC) functions \cite{larkin1969quasiclassical,maldacena2016bound}.
Such correlators have recently been proposed as powerful diagnostics of quantum chaos and their dynamical behavior remains the subject of intense interest \cite{swingle2016measuring,yoshida2019disentangling,landsman2019verified}; indeed, the possibility of defining a quantum Lyapunov exponent based on the exponential growth of OTOCs in certain systems has led to a conjectured bound on the rate of thermalization in many-body quantum systems \cite{maldacena2016bound}.

The functional form of the OTOC is typically written as the expectation
\begin{equation}
F(\tau) = \langle W^\dagger(\tau) V^\dagger(0) W(\tau) V(0) \rangle,
\end{equation}
where $W$ and $V$ are two commuting operators with $W(\tau)=e^{iH\tau}We^{-iH\tau}$ the evolved operator $W$ under the Hamiltonian $H$. The OTOC compares two quantum states obtained by either (a) applying V, waiting for a time t, and then applying W; or (b) applying W at time t, going back in time to apply V at t = 0, and then letting time resume its forward progression to t. The OTOC $F(\tau)$ thus describes how initially commuting operators $W$ and $V$ fail to commute at later times due to the interactions generated by $H$, and have a quite different form than the conventional autocorrelation function $\langle W^\dagger(\tau) V(0) W^\dagger(\tau) V(0) \rangle$.  Operationally, measuring an OTOC requires an intermediate step of time-reversal, representing a major challenge from an experimental implementation perspective \cite{li2017measuring,garttner2017measuring}. 

\begin{figure}[h!]
\includegraphics*[width=\columnwidth]{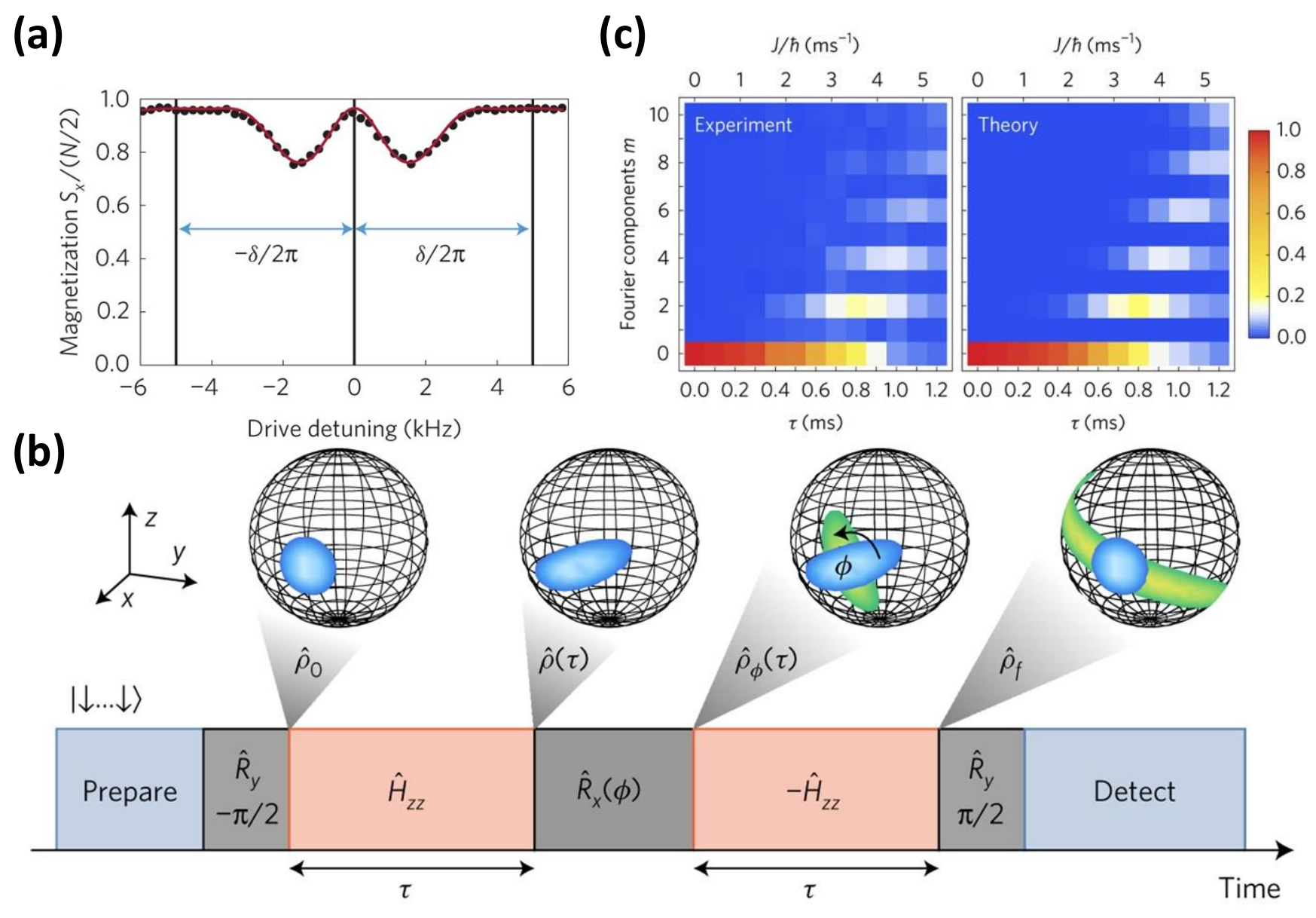}
\caption{Out-of-time-order-correlators (OTOC) in a 2D Penning trap ion simulator (see Fig. \ref{fig:traps}b for experimental schematic). 
(a) Demonstration of the ability to tune the sign of the Ising interaction through the symmetric detuning of the optical dipole force around a single mode of motion. Plotted is the residual spin-phonon couplings vs. detuning from the center of mass mode. A positive (negative) detuning gives rise to an anti-ferromagnetic (ferromagnetic) interaction. (b) Experimental sequence used to measure an out-of-time-ordered correlation function with alternation in the sign of the Ising coupling. (c) Measured Fourier amplitudes of collective magnetization dynamics under the Ising spin-spin interactions for $N=111$ spins, showing the sequential build-up of higher order spin-spin correlations. Adapted from \citealp{garttner2017measuring}.}
\label{fig:OTOC}
\end{figure}

In the case of ions, it is possible to directly measure an OTOC stroboscopically applying both an interaction Hamiltonian and its negative counterpart, thus effectively reversing time evolution. 
This was experimentally demonstrated in \cite{garttner2017measuring} by using a two-dimensional array of laser-cooled $^9$Be$^+$ ions in a Penning trap, summarized in Fig. \ref{fig:OTOC}.
Here, the spin-dependent optical dipole force couples to the axial center-of-mass motion of the 2D crystal (see Fig. \ref{fig:traps}b), resulting in a nearly uniform all-to-all Ising interaction (Eq. (\ref{Jpowerlaw}) with $\alpha \approx 0$).
Such an all-to-all interacting Ising model does not possess the complexity of interactions from a quantum chaos perspective.  However, this system is well-suited for OTOC studies, because the sign of the Hamiltonian can be controlled. Since the Ising interaction strength scales as $J \sim 1/\delta$  (Eq. (\ref{Jij}) with just a single mode contributing and $\delta_m=\delta$), this allows for the implementation of a Hamiltonian sign reversal by simply changing the sign of the detuning.
With the ability to stroboscopically apply both $H$ and $-H$, a family of different OTOCs can then be measured through the collective magnetization of the system, as shown in Fig. \ref{fig:OTOC} \cite{garttner2017measuring}. 

The second example of stroboscopic Hamiltonian simulation with trapped ion spin systems is the quantum simulation of Floquet systems, or periodically modulated Hamiltonians \cite{deng2015observation,bukov2015universal,kuwahara2016floquet}. 
In a broader context, such time-periodic manipulations have long been used for controlling quantum systems including NMR qubits and atomic ensembles \cite{goldman2014light,oka2019floquet}. 
However, recent explorations of Floquet systems have stumbled upon an intriguing question beyond the landscape of quantum control; in particular, can Floquet systems host intrinsically new phases of matter that do not have any equilibrium equivalent \cite{moessner2017equilibration,else2019discrete}? 

In the noninteracting (single-particle) case, the question has been affirmitavely answered with the discovery of a host of novel band structures that can only exist in the presence of periodic driving \cite{kitagawa2010topological,lindner2011floquet,cayssol2013floquet,titum2016anomalous,rechtsman2013photonic}. 
The many-body case is more subtle. 
On the one hand, one might naturally suspect that new phenomena can in principle arise when the driving frequency is of order the intrinsic energy scales of the system; {\q indeed, this limit is far from the Suzuki-Trotter limit (e.g.~as discussed in \ref{subsec:Trotter}) where to first order, the effective Hamiltonian describing the Floquet system is simply a sum of its stroboscopic components. }
On the other hand, it is generally expected that a driven many-body system will absorb energy from the driving field and ultimately heat to infinite temperature \cite{bukov2015universal,ponte2015periodically}.  
However, recent theoretical advances have demonstrated that it is possible to avoid such Floquet heating. 
One general scheme is to utilize many-body localization as discussed in section \ref{sec:mbl}. 
In principle, a Floquet MBL system \cite{abanin2016theory,ponte2015many} can exhibit stable dynamical phases of matter for infinitely long times \cite{khemani2016phase,else2016floquet,potirniche2017floquet}. 
Interestingly, recent studies suggest an alternative disorder-free approach can also be used to combat Floquet heating, albeit not to infinite times.
In particular, for large enough driving frequencies, the system can enter a regime of Floquet-prethermalization \cite{kuwahara2016floquet,mori2016rigorous}, where exotic non-equilibrium phases can be observed for exponentially long time-scales \cite{else2017prethermal,machado2017exponentially,machado2019prethermal}.
The underlying essence of Floquet-prethermalization is analogous to the discussions of prethermalization in section \ref{sec:pretherm}. The key difference is that here the lifetime of the quasi-stationary state is controlled by the Floquet drive frequency. 

\begin{figure}[h!]
\includegraphics*[width=\columnwidth]{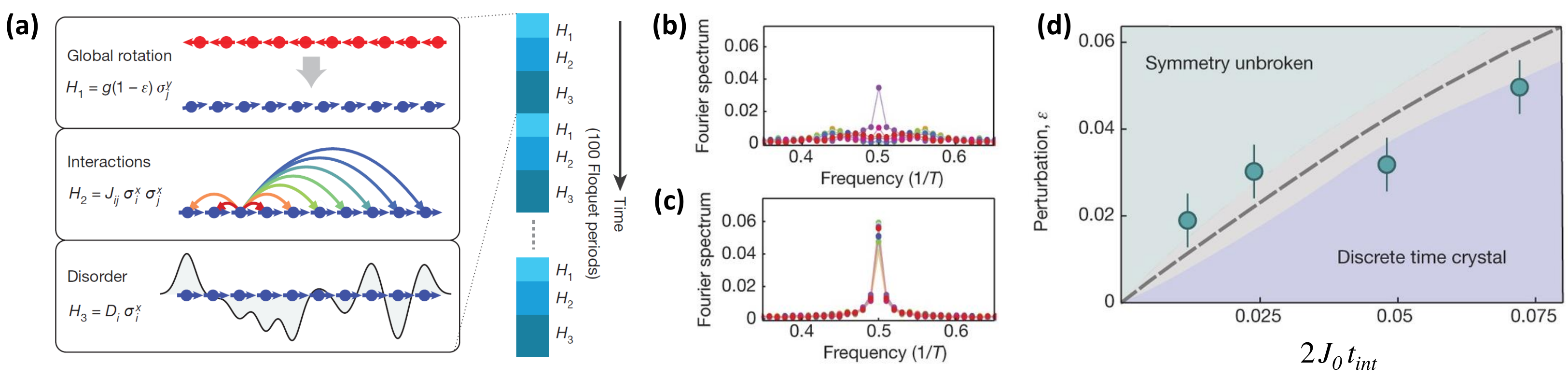}
\caption{Floquet quantum simulation of a discrete time crystal (DTC). (a) Schematic depiction of the Floquet evolution of a trapped ion spin chain. Three Hamiltonians are applied sequentially in time: (i) a global nominal $\pi$-pulse $(2gt=\pi)$ around the spin y-axis with fractional perturbation $\varepsilon$, (ii) long-range Ising interactions (see section \ref{sec:MB_hamil}), and  (iii) site-dependent disorder along the spin x axes. (b) The x-magnetization of each spin is measured after each Floquet sequence, up to 100 periods.  The Fourier transform of the oscillations show a clear peak observed at half the driving frequency, even with a programmed perturbation on the global $\pi$ pulses of $\varepsilon=0.03$,
signaling the discrete breaking of time translation symmetry and the ``rigidity" of the time crystal. (c) When the perturbation is too strong ($\varepsilon=0.11$), we cross the boundary from the DTC into a symmetry unbroken phase. (d) For stronger interactions parameterized by the nearest-neighbor Ising coupling $J_0$, the DTC can tolerate larger imperfections to the global rotation pulse, leading to a qualitative phase diagram \cite{yao2017discrete}. Adapted from \citealp{zhang2017observation}.}
\label{fig:DTC}
\end{figure}

We now turn to recent experiments that demonstrated a Floquet quantum simulation using a one dimensional trapped ion spin chain, depicted in Fig.~\ref{fig:DTC} \cite{zhang2017observation}.
In these experiments, a combination of high-precision spatial and temporal control allowed for the implementation of three distinct types of time evolution applied repeatedly in sequence: 1) global spin rotations, 2) long-range Ising interactions, and 3) disordered on-site fields (Fig.~\ref{fig:DTC}a). 

The stroboscopic combination of these evolutions is the basis for realizing a discrete time crystal (DTC) \cite{khemani2016phase,else2016floquet,von2016absolute,yao2017discrete, else2019discrete}, where a system exhibits a spontaneous breaking of the time-translation symmetry generated by the Floquet evolution. 
{\q The characteristic signature of a DTC, which is consistent with experimental observations in chains of up to $L=14$ ions \cite{zhang2017observation}, is the robust  synchronization of oscillations at sub-harmonic frequencies compared to that of the drive, as shown in Fig.~\ref{fig:DTC}b-c.} 
Within the decoherence time-scale of the experiments, the observed signatures of DTC order are independent of the initial state.
Crucially, the robustness of the sub-harmonic oscillations depends on the presence of strong interactions in the system; in the absence of interactions, even small perturbations immediately destroy signatures of a time crystal. 

In addition to implementations in trapped ion systems, a number of other experimental platforms have also observed signatures of time crystalline order \cite{choi2017observation,rovny2018observation}. Here, we note a particular set of experiments performed using ensembles of nitrogen-vacancy (NV) color centers in diamonds \cite{choi2017observation}. 
We emphasize this particular platform because it shares a number of similar features with the trapped ion system (i.e.~long range interactions and disorder), but also has a number of crucial differences (i.e.~three dimensional system with time-dependent disorder).
{\q Interestingly, both platforms exhibit similar signatures of interaction stabilized time-translation symmetry breaking, although these signatures appear to be limited to time-scales before local thermalization has fully completed \cite{else2019discrete}.}
To this end, such cross-platform verifications are especially valuable once controlled quantum simulators reach a regime where classical computers cannot calculate \cite{calarco2018quantum, leibfried2010could}. In this regard, comparing the results from analog quantum simulators to those from digital quantum computers would also be helpful in cross-checking and assessing validity \cite{mahadev2018classical}. 

\subsection{Dynamical Phase Transitions \label{sec:dpt}}
Having discussed the simulation of non-equilibrium phases in both disordered and periodically driven trapped ion experiments, we now turn to the question of understanding phase transitions in such out-of-equilibrium systems. 
Novel dynamical phases can emerge after a quantum quench, {\q and the transition between them can be observed experimentally by measuring non-analytic changes in the dynamical response of the many-body spin system} \cite{Zhang2017b, Jurcevic2017}. {\q As also described in sections \ref{sec:mbl} and \ref{sec:pretherm}}, out-of-equilibrium systems do not necessarily behave thermodynamically, so it is a fundamental question how to properly establish analogies and differences among thermodynamic equilibrium phases and their dynamical counterparts \cite{Heyl2018, Titum2019}, in terms of order parameters \cite{Ajisaka_2014, Zunkovic2018}, scaling and universality \cite{Heyl2015}, and discrete or continuous symmetry breaking \cite{Zunkovic2016,Weidinger2017,Huang2019}. 

Dynamical phases can be separated by dynamical quantum phase transitions (DQPT), characterized by a non-analytic response of the physical system as a function of quench parameters. Two types of DQPT signatures have been defined for an interacting spin-$1/2$ chain \cite{Zunkovic2018} governed by the Hamiltonian of Eq. (\ref{eqn:TransversIsing}) with the field along the z-axis:
\begin{equation}
H=\sum_{i<j} J_{ij} \sigma_x^{i} \sigma_x^{j} + B \sum_{i} \sigma_z^{i},
\label{eq:H_DPT}
\end{equation}
{\q where the spin-spin interactions $J_{ij}$ and the transverse field $B_z$ are generated using the technique explained in section \ref{subsec:Ising}}.
Both {\q types of DQPT} have been experimentally observed in a trapped-ion quantum simulator \cite{Jurcevic2017,Zhang2017b}. The first type of DQPT {\q (type I)} is based on the formal analogy between the non-analytic behaviour of the return probability to the initial state $\ket{\psi_0}$ after a quantum quench under the Hamiltonian $H$, defined as $\mathcal{G}(t)=\bra{\psi_0} e^{-i H t}\ket{\psi_0}$, and the partition function of the corresponding equilibrium system $Z={\rm Tr}(e^{-H/k_B T})$ \cite{Heyl2013}. It is possible to define the complex counterpart of the thermodynamic free energy density $f=-N^{-1}k_B T \log(Z)$ using the rate function $\lambda(t)=-N^{-1} \log[\mathcal{G}(t)]$. This quantity, in the thermodynamic limit, exhibits dynamical real-time nonanalyticities that play an analogous role as the non-analytic behaviour of the free energy density of a thermodynamic system at equilibrium. It is possible to observe experimentally these nonanalyticities in an interacting spin chain after a quantum quench evolving under the long-range Transverse Field Ising Hamiltonian of Eq. (\ref{eq:H_DPT}).

\begin{figure}[t!]
\includegraphics*[width=\columnwidth]{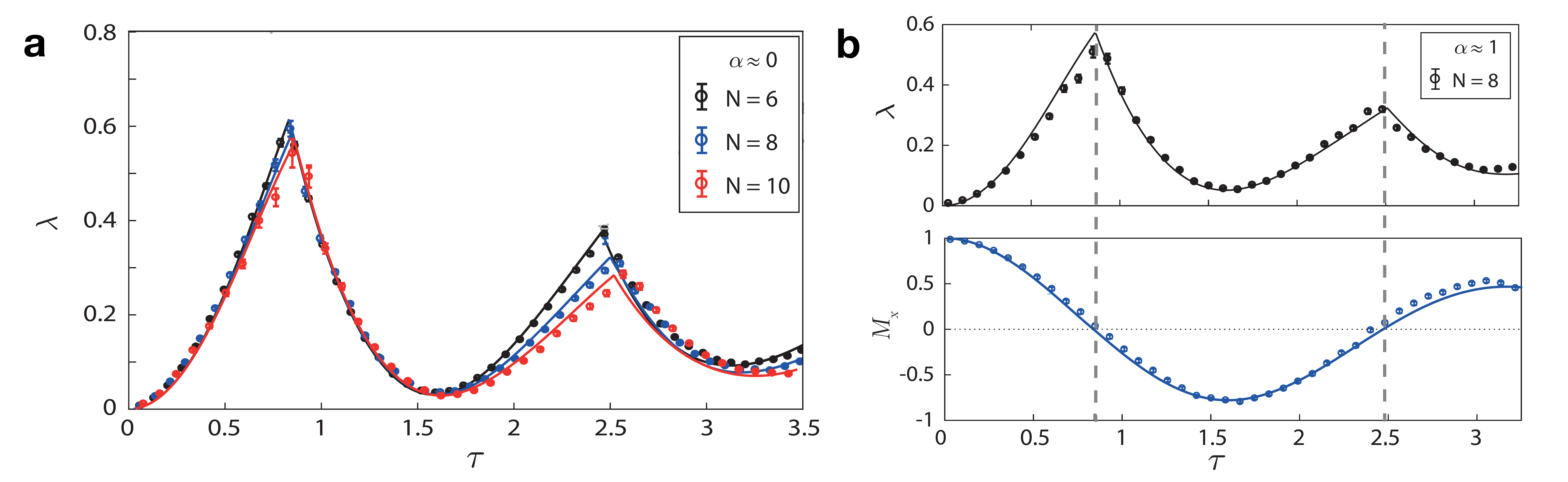}
\caption{{\q Trapped-ion} quantum simulation of DQPT {\q type I}. (a) Measured rate function $\lambda$ for three different system sizes at $B/J_0\approx2.38$, with $\tau=t B$ being the dimensionless time. The kinks in the evolution become sharper for larger $N$. In order to take into account the $\mathbb{Z}_2$ degeneracy of the ground state of $H_0$, here the rate function is defined based on the return probability to the ground state manifold, namely $\lambda(t) = N^{-1}\log(P_{\ket{\psi_0}}+P_{\ket{-\psi_0}})$, where $\ket{-\psi_0}=\ket{\uparrow\uparrow\uparrow\cdots\uparrow}_x$. {(b)} Comparison between rate function $\lambda(t)$ and magnetization evolution $M_x(t)$. The inversion of the magnetization sign corresponds to the nonanalyticity of the rate function $\lambda(t)$. Solid lines are exact numerical predictions based on experimental parameters $(B/J_0=2)$. Adapted from \citealp{Jurcevic2017}.}
\label{fig:DQPT-LO}
\end{figure}

This type of DQPT has been observed experimentally with a linear chain of trapped $^{40}$Ca$^+$ ion spins \cite{Jurcevic2017}.  The spins are initialized in the ground state of the field part of the transverse Ising model, namely $\ket{\psi_0}=\ket{\downarrow\downarrow\downarrow ...\downarrow}_z$.
Then the transverse field Hamiltonian [Eq. (\ref{eq:H_DPT})] is suddenly switched on (quenched) with $B>J_0$, with $J_0$ being the average nearest-neighbour spin-spin coupling. As shown in Fig. \ref{fig:DQPT-LO}a, in this regime the rate function $\lambda$ exhibits pronounced nonanalyticities at the critical times $t_c$. This behaviour can be related to other observables, such as the global average magnetization $M_x=N^{-1}\sum_i \sigma_i^x$. Since the initial state breaks the $\mathbb{Z}_2$ symmetry of the Hamiltonian (\ref{eqn:TransversIsing}), the system restores this symmetry during the evolution at the times where the magnetization changes sign, which also corresponds to the critical times in the Loschmidt echo observable, as shown in Fig. \ref{fig:DQPT-LO}b. 

The second type of DQPT{\q-(type II)} has an order parameter defined in terms of long time averaged observables, such as asymptotic late-time steady states of local observables:
\begin{equation}
\bar{A}=\lim_{T\rightarrow\infty}\frac{1}{T}\int_0^T A(t)dt,
\end{equation}
where the operator $A$ is the magnetization or higher order correlators between the spins. Here, the DQPT occurs as the ratio $B/J_0$ is varied and the order parameter changes abruptly from ferromagnetic ($B<J_0$) to paramagnetic order ($B>J_0$). The onset of this non-analytic behaviour can be observed by measuring the late time average values of the two-body correlator
\begin{equation}
    C_2=\frac{1}{N^2}\sum_{ij}\langle \sigma_i^x \sigma_j^x\rangle,
\end{equation}
after a quantum quench with Hamiltonian (\ref{eq:H_DPT}). 

\begin{figure}[h!]
\includegraphics*[width=\columnwidth]{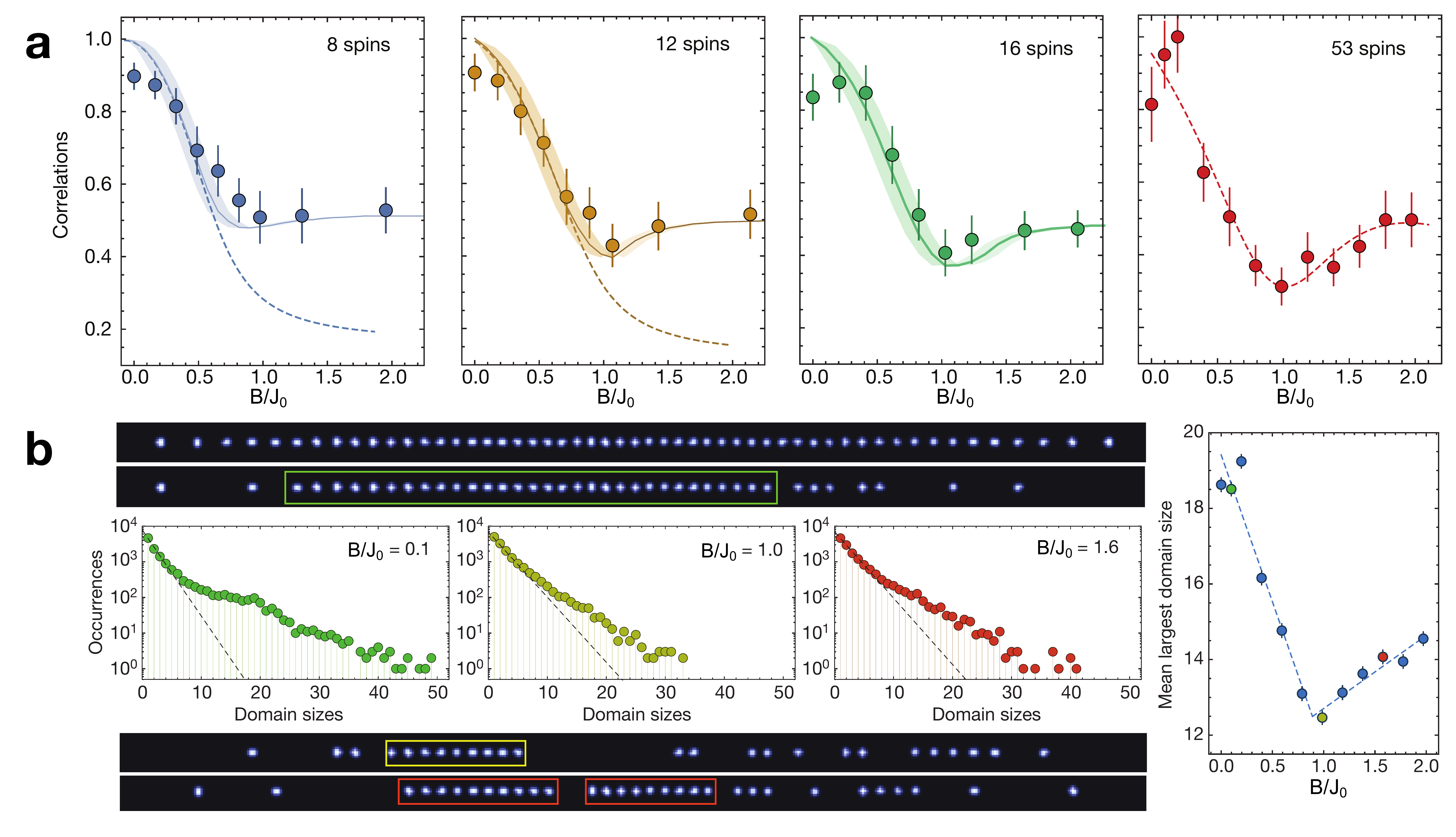}
\caption{{\q Trapped-ion} quantum simulation of a DQPT{\q-type II} {(a)} Long-time averaged values of the two-body correlations $C_2$, for different numbers of spins in the chain. Solid lines in (a)-(c) are exact numerical solutions to the Schr\"odinger equation, and the shaded regions take into account uncertainties from experimental Stark shift calibration errors. Dashed lines in (a) and (b) are calculations using a canonical (thermal) ensemble with an effective temperature corresponding to the initial energy density. {\bf (b)}
Domain statistics and reconstructed single shot images of 53 spins. (a) Top and bottom: reconstructed images based on binary detection of spin state. The top image shows a chain of 53 ions in bright spin states. The other three images show 53 ions in combinations of bright and dark spin states. Center: statistics of the sizes of domains for three different values of $B/J_0$, plotted on a logarithmic scale. Dashed lines are fits to exponential functions, which could be expected for infinite-temperature thermal state. Long tails of deviations are clearly visible, and vary depending on $B/J_0$. (b) Mean of the largest domain sizes in each single experimental shot. Dashed lines represent a piecewise linear fit, used to extract the transition point. The green, yellow, and red data points correspond to the transverse fields shown in the domain statistics data on the left. Adapted from \citealp{Zhang2017b}.}
\label{fig:DQPT-OP}
\end{figure}

This type of DQPT measurement was observed in a linear chain of trapped $^{171}$Yb$^+$ ion spins \cite{Zhang2017b}.  Here, the measured late time correlator $C_2$ exhibits a dip at the critical point that sharpens scaling up the system size $N$ up to 53 $^{171}$Yb$^+$ ions, {\q as shown in Fig. \ref{fig:DQPT-OP}a. This represents one of the largest quantum simulations ever performed on individual spins.} Further evidence of the occurrence of the phase transition can be also observed in higher-order correlations, such as the distribution of domain sizes throughout the entire chain, shown in Fig. \ref{fig:DQPT-OP}b. The occurrence of the DQPT is observed in the decreased probabilities of observing long strings {\q of aligned ions} at the critical point. This is more clearly shown measuring the mean largest domain size as a function of the transverse field strength, for late times and repeated experimental shots, which exhibits a sharp transition across the critical point of the DQPT{\q-type II}. {\q Note this measurement is in general an $N$th order spin correlation function, requiring high fidelity readout of all spins in a single shot, which is an attribute that is unique to trapped ion systems.} 

\section{Hamiltonian Sequencing and Engineering \label{sec:seq}}
\label{sec:Hsequencing}

{\q
The tailored interactions between spins available in trapped ion systems can also be used to realize a digital quantum simulator (DQS), which is a many-body quantum system with enough control to perform a universal set of quantum operations or gates. In other words, it is a quantum computer used to implement Hamiltonian evolution rather than other quantum algorithms \cite{feynman1982simulating}. The universality means that these simulations are not limited to the inherent interactions, but any local Hamiltonian can be designed and implemented as a circuit \cite{lloyd1996universal}. Additionally, error bounds and error correction protocols using fault-tolerant gate sets will be applicable to large digital simulations \cite{Steane1999}. 
Arbitrary Hamiltonians are changed into the spin or qubit basis via the Jordan-Wigner \cite{JW} or the Bravyi-Kitaev \cite{BK} transformation. First-quantization mapping can also been used \cite{Babbush2017,Bravyi2017}. There are several ways a time evolution or adiabatic ramp has been approximated by discrete operations in a DQS, Trotterized evolution \cite{trotter1959, suzuki1985} or variational approximations \cite{Yuan2019theoryofvariational}, in particular the Variational Quantum Eigensolver (VQE) \cite{Peruzzo2014} and the Quantum Approximate Optimization Algorithm (QAOA) \cite{farhi2000quantum}. There are other alternatives such as quantum walks or Taylor series expansion. We note that the latter has potential for fault-tolerant devices due to its optimal asymptotic scaling, but has not been demonstrated experimentally \cite{Berry2015}.

\subsection{Trotter Hamiltonian Expansion}
\label{subsec:Trotter}

In order to decompose the unitary evolution under a given Hamiltonian into discrete operations or gates, Suzuki-Trotter \cite{trotter1959, suzuki1985} formulas are commonly used. These provide an approximate factorization of the unitary time-evolution operator $\hat{U}$. For a time-independent Hamiltonian $H$ containing a sum of terms $H_k$, the first-order Trotter approximation $\hat{U}=e^{-i\hat{H}t/\hbar}=\textrm{lim}_{n\rightarrow \infty}(\Pi_k e^{-i\hat{H_k}t/n\hbar})^n$ for finite $n$ leads to a sequence of small time steps $\delta t = t/n$ with local evolution operators $\hat{U}_k=e^{-i\hat{H}\delta t/\hbar}$ that is repeated $n$ times. These operators can then be deconstructed into the operations from the simulator's universal set.
Similarly, a time-dependent Hamiltonian can be simulated by breaking the evolution down into short segments \cite{Poulin2011}. Since the unitaries $U_k$ do not commute in general for finite $n$, the sequence deviates from the target evolution. These Trotter errors are bounded for small step sizes but undergo a dynamical phase transition \cite{Heyleaau2019}, above which the evolution exhibits many-body quantum chaos \cite{Sieberer2019,dalessio2016from}.

This technique was used by \cite{lanyon2011universal}  in a trapped-ion experiment to simulate a wide range of different interaction models and interaction graphs of two to six spin$-1/2$ particles. This spin-model problem matches the system and does not require a transformation. Using the operations available from the toolbox described in section \ref{subsec:Ising}, a universal set of operations is realized. The versatility of this approach is shown by simulating the dynamics of two-particles under an XX (Ising), XY, and XYZ-type interaction, and an adiabatic ramp of the interaction strength. Figure \ref{fig:Lanyon_multiionflop}A shows the dynamics of four spins initialized along z under a long-range (all-to-all) Ising interaction. The oscillation frequencies correspond to energy gaps in the system. Figure \ref{fig:Lanyon_multiionflop}B presents the collective oscillation of six spins under a $\sigma_y^1 \sigma_x^2 \sigma_x^3 \sigma_x^4 \sigma_x^5 \sigma_x^6$ interaction, composed digitally of Ising-type interactions and indvidual $\sigma_z$ rotations, periodically producing a six-spin GHZ state. 

Interleaving different native operations has also been proposed for creating an XXZ-interaction in a spin-1 system \cite{ cohen2015simulating}. This will allow the study of an integer-spin systems under Heisenberg-type interactions, including the Haldane phase \cite{haldane1983}. Adding optical pumping as an incoherent process to the set of operations was used in \cite{barreiro2011open} to create the ability to engineer open-system dynamics, and show the dissipative preparation of multi-qubit stabilizer states for quantum error correction. Evolution by DQS can also be combined with standard computational gates to measure quantities of interest. A digital simulation of the two-site Fermi-Hubbard model was realized in \cite{linke2018measuring}. After a trotterized adiabatic ramp composed of two-qubit XX-interactions and single qubit rotations, the second R\'{e}nyi entropy was measured. In contrast to \cite{brydges2019probing}, this is achieved by creating two copies of the state and measuring the expectation value of the SWAP operator following \cite{Johri2017,Horodecki2002}.

The DQS approach has also been proposed for quantum field theory simulations using trapped ions \cite{banuls2019simulating}. The Schwinger model, an example of a lattice gauge theory, describes fermions coupled to an electro-magnetic field in one spatial dimension. After mapping the model to qubits with the Jordan-Wigner transformation using open boundary conditions, the Hamiltonian reads (see \cite{InnsbruckLGT1,InnsbruckLGT2} for details):
\begin{equation}
\hat{H_S}=w \sum_{n=1}^{N-1}\left(\hat{\sigma}_n^+\hat{\sigma}_{n+1}^- + \hat{\sigma}_n^-\hat{\sigma}_{n+1}^+\right)+\frac{m}{2}\sum_{n=1}^{N}(-1)^n\hat{\sigma}_n^z+g\sum_{n=1}^{N-1}\hat{L}_n^2
\label{eqn:Schwinger}
\end{equation}
where $n$ labels the lattice sites. Following a Kogut-Susskind eoncoding, spin-down (spin-up) on an odd (even) lattice site indicates the presence of a positron (electron). The first term represents particle-antiparticle pair creation and annihilation with coupling $w$. The second is a mass term and the third term represents the energy of the electric field $\hat{L}_n$ with coupling $g$. The model reproduces key features of more complex high-energy physics theories such as quantum chromodynamics \cite{Gattringer2010}, which are extremely challenging for classical numerical methods. 

An experiment realizing the Schwinger model in a trapped-ion spin chain has been implemented in \cite{martinez2016real}, showing very good agreement with the ideal evolution for the real-time dynamics of a 2-site (4-qubit) model. The difficulty in mapping gauge theories to spins lies in the infinite-dimensional Hilbert space of the gauge field operators. Instead of truncating the model to the size of the simulator, the authors analytically integrate out the gauge fields using Gauss' law, allowing the electric field to be expressed in terms of Pauli oprators: $\hat{L}_n=\epsilon_0-\frac{1}{2}\Sigma_{l=n+1}^N(\hat{\sigma}_l^z+(-1)^l)$. This creates a spin Hamiltonian with asymmetric long-range couplings \cite{Hamer1997}, which can be directly implemented by a sequence of multi-ion spin-spin operations \cite{martinez2016real,InnsbruckLGT1}. By construction, the scheme constrains the evolution within the subspace allowed by Gauss' law and hence maintains the gauge invariance of the theory \cite{InnsbruckLGT1}. These features, as well as the linear scaling of the number of gate operations and qubits with lattice size, make DQS with trapped ions a promising avenue for the simulation of more complex gauge theories. The authors also realize a variational lattice gauge theory simulation \cite{InnsbruckLGT2}, which will be discussed below.

Quantum walks are alternative framework to discretize a unitary evolution for a DQSs. Quantum walks, the quantum equivalent of classical random walks \cite{Childs2003}, see a system evolve into on a discrete grid or continuum of spacial locations based on the superposition of a quantum coin. Discrete step quantum walks have been mapped to represent simulations of different physical systems \cite{Chandrashekar2010,Molfetta2016}, including the evolution of a particle under the Dirac equation \cite{Mallick2019}. The latter was realized in trapped ions by mapping the position space to multi-qubit spin states, where different particle masses were chosen by varying the weight of the quantum coin \cite{HuertaAlderete2020}. Trapped ion experiments have previously realized quantum walks \cite{schmitz2009quantum,zahringer2010realization} and separately a Dirac equation \cite{Gerritsma2010} simulation, using the harmonic oscillator degrees of freedom in the trap rather than mapping to spins.
A non-variational digital simulation of spin models was also performed with superconducting qubits \cite{Salathe2015}.

\begin{figure}[t]
\includegraphics[width=0.8\linewidth]{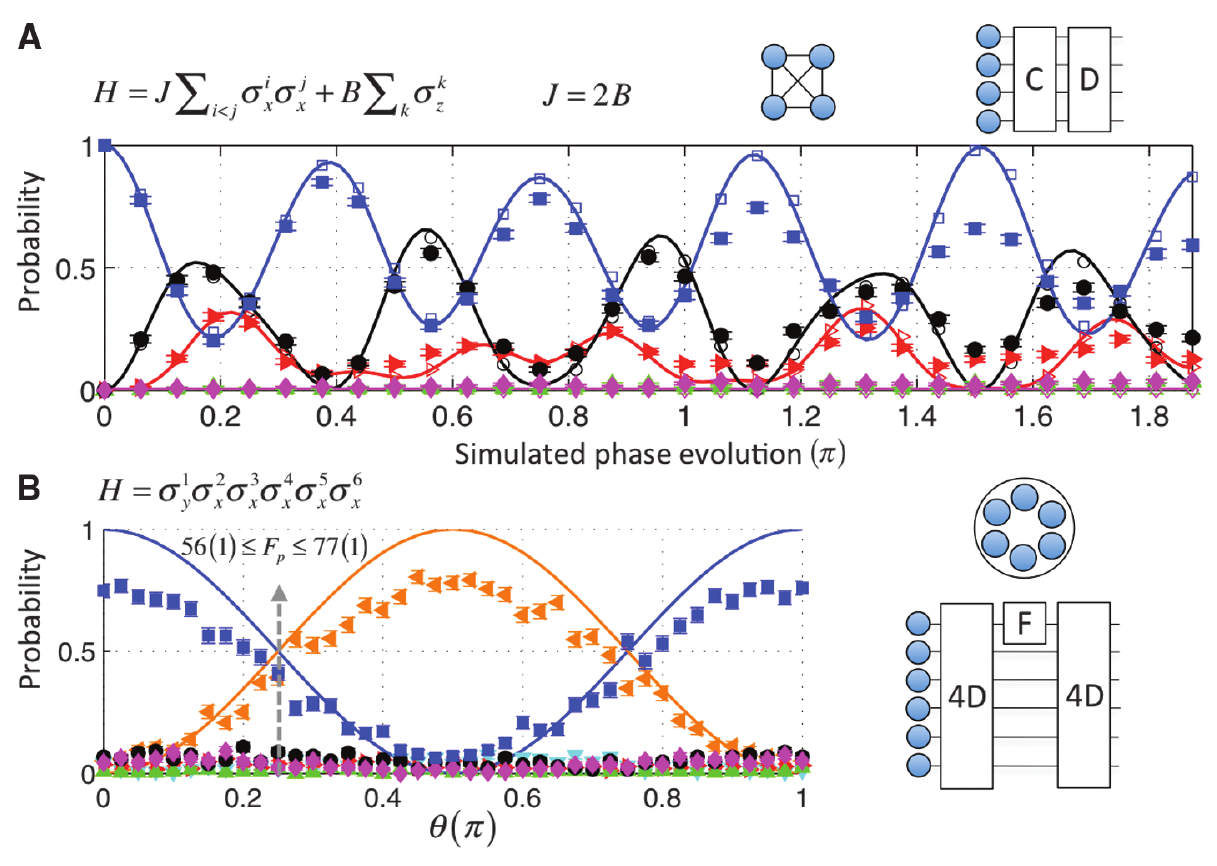}
\caption{Digital simulations of dynamics in four- and six-spin systems. Initially, all spins point up. ({\bf A}) Four spin long-range Ising system. Each digital step is given by operators $C=\textrm{exp}(-i\theta_C\Sigma_{i}\sigma^i_z)$ with $\theta_C=\pi/32$, and $D = \textrm{exp}(-i\theta_D\Sigma_{i<j}\sigma^i_{\phi_D}\sigma^j_{\phi_D})$ with $\theta_D=\pi/16$ and $\phi_D=0$. ({\bf B}) Six-body interaction with six spins. $F = \textrm{exp}(-i\theta\sigma_z)$ with $\theta$ variable, and $4D = \textrm{exp}(-i\theta_0\Sigma_{i<j}\sigma^i_{\phi_0}\sigma^j_{\phi_0})$ with $\theta_0=\pi/4$ and $\phi_0=0$. Bounds for the quantum process fidelity $F_p$ are given at $\theta = 0.25\pi$. Lines: exact dynamics; unfilled shapes: ideal digitized;
filled shapes: data (squares: $P_0$; diamonds: $P_1$; circles: $P_2$; triangles pointing up: $P_3$, right: $P_4$, down: $P_5$, left: $P_6$, where $P_i$ is the total probability of finding i spins pointing down). Adapted from \citealp{lanyon2011universal}.}
\label{fig:Lanyon_multiionflop}
\end{figure}

\subsection{Variational quantum simulation}

Variational quantum simulation (VQS) implements the versatile classical method of variational simulation \cite{Balian1988,Szabo2012,Shi2018} on a quantum computer by employing a quantum-classical hybrid approach. Hybrid quantum algorithms are a promising way to solve potentially difficult problems on non-fault-tolerant, near-term quantum systems \cite{Preskill2018NISQ} by combining classical and quantum resources. They distribute the task between a quantum and a classical computer. The quantum system is running a parameterized sequence of operations to generate a classically intractable state of interest, e.g. following an approximate evolution under a Hamiltonian or a quantum gate sequence. The classical system varies the parameters to minimize a cost function evaluated by measuring the quantum state, a task that is considered relatively easy based on its complexity class \cite{Yuan2019theoryofvariational}. Employing trapped ion spin simulators, this concept has been used to train a generative model, a routine from classical machine learning, using an ion trap-based digital quantum circuit with up to 26 parameters \cite{zhu2019training}. The authors showed that the choice of optimizer is crucial as the parameter space grows. One of the challenges of this approach lies in the cost function landscapes, which can exhibit so-called “barren plateaus” \cite{McClean2018}, making optimization hard. Finding the classical optimization strategy and making efficient use of quantum resources is an active area of research \cite{Wecker2015,Yang2017,Cirstoiu2019,Sundar2019}. 

VQS employs such a hybrid approach to simulating physical models. For solving static problems, the Rayleigh-Ritz variational method is generalized to the quantum regime. The ground state energy of a Hamiltonian $\hat{H}=\lambda_i\hat{h}_i$, given as a linear combination of tensor products of local operators $\hat{h}_i$, can be estimated by considering trial wave functions $|\phi(\vec{\theta})\rangle$ with parameters $\vec{\theta}=(\theta_1,\theta_2,...)$. and approximating the ground state energy by an upper bound \cite{Yuan2019theoryofvariational}:
\begin{equation}
    E_0\leq E_0^{est}=\min_{\vec{a}}\langle\phi(\vec{a})|\hat{H}|\phi(\vec{a})\rangle.
\end{equation}
$E_0^{est}$ is obtained by preparing $|\phi(\vec{\theta})\rangle$ through a sequence of parameterized quantum operations. Different circuit parameters implement different instances of $\vec{\theta}$, and $\langle\phi(\vec{a})|\hat{H}|\phi(\vec{a})\rangle$ is obtained by measuring each term $\langle\phi(\vec{a})|\hat{h}_i|\phi(\vec{a})\rangle$, and calculating the linear combination given by $\{\lambda_i\}$. A classical optimization routine, e.g. a gradient descent algorithm, uses this value as a cost function to converge to $E_0^{est}$.

A clear illustration of this method was presented in \cite{InnsbruckLGT2}, where the authors perform a VQS of the Schwinger model described in section \ref{subsec:Trotter} above. Starting from $|\Psi_0\rangle=|\uparrow\downarrow..\uparrow\downarrow\rangle$, which corresponds to the bare vacuum state, they apply finite-range Ising interactions (see section \ref{subsec:Ising}) to all ions, and alternate with individual $\hat{\sigma}_z$ shifts realized with a Stark-shift beam. The set of discrete operations used in this ansatz enforces the symmetries of the problem, namely Gauss' law and gauge invariance. Each operation is parameterized and the cost function is given by $\langle\Psi(\vec{\theta}) |\hat{H}_S|\Psi(\vec{\theta}) \rangle$, where $\hat{H}_S$ is the Schwinger Hamiltonian from equation \ref{eqn:Schwinger}, and $|\Psi(\vec{\theta}) \rangle$ is the trial wave function. The cost function is evaluated by expressing $\hat{H}_S$ as a string of Pauli operators and measuring their expectation values by projecting the experimental output state in the appropriate basis. An adaptive classical optimization process based on the Dividing Rectangles (DIRECT) algorithm \cite{Liu2015} is employed in a quantum-classical feedback loop find the minimum. At the minimum, the classical parameters $\vec{\theta_0}$ describe a sequence to prepare the approximate ground state wave function, and the cost function gives an upper bound estimate of the ground state energy. The authors use this method with 20 (16) ions to optimize over an up to 15-dimensional parameter space and achieve $85\%$ ($90\%$) ground state fidelity and an estimate of the ground state energy within 2$\sigma$ uncertainty. One of the challenges in variational quantum algorithms is the verification of convergence since classical simulation cannot provide a measure of success. For eight ions, the authors use the variance of the cost function as a termination criterion by measuring the Hamiltonian in additional sets of bases to determine $\langle\hat{H}_S^2 \rangle$, making the optimization "self-verifying". In a final 8-ion measurement, a phase transition \cite{Byrnes2002density} is observed by varying the mass in the Schwinger Hamiltonian and observing an order parameter \cite{InnsbruckLGT2}. The authors also show that the R\'{e}nyi entropy peaks at the critical point by the method described in \cite{brydges2019probing}.
To overcome the deep circuits required for digital LGT simulations, an analog approach with tailored spin-spin interactions in an ion chain has recently been proposed \cite{Davoudi2020}.

Different ways to construct circuits or sequences of quantum operations for preparing the trial wave function have been used. On the one hand, there is the so-called hardware-efficient ansatz as used in \cite{InnsbruckLGT2}, where the sequence is constructed from available operations without trying to reproduce the precise problem Hamiltonian. It is straightforward to implement but more susceptible to optimization problems like barren plateaus \cite{McClean2018}. The the other hand, the physically-inspired ansatz prescribes a more problem-specific construction for the trial states, for example based on Unitary Coupled Cluster (UCC) theory as mentioned below. The cost function tends to better match the underlying problem, but it can quickly lead to deep circuits as the complexity of the physical system increases \cite{shehab2019noise}.

For the simulation of physical models two flavors of variational algorithms are distinguished, which are both based on the VQS method but used in different domains of application, the variational quantum eigensolver (VQE) \cite{Peruzzo2014} for the determination of Hamiltonian eigenvalues \cite{Yuan2019theoryofvariational}, e.g. in the determination of ground and excited state energies quantum chemistry, and the quantum approximate optimization algorithm (QAOA) \cite{Farhi2014quantum} for combinatorial optimization such as graph problems and approximate state preparation. The following sections will review results on both from recent work with trapped ions. We note that a  probabilistic variant has also been proposed for implementation with trapped-ions \cite{zhang20}.

\subsubsection{Variational Quantum Eigensolvers}

The Variational Quantum Eigensolver (VQE) is a DQS algorithm for the calculation of operator eigenvalues, which has been employed to tackle molecular electronic structure and molecular dynamics problems in quantum chemistry \cite{Peruzzo2014,Yung2014,McClean2016} as an efficient alternative to quantum phase estimation \cite{PEA}. Under the Born-Oppenheimer approximation, which fixes the inter-nuclear distances, the quantum chemistry Hamiltonian can be written in a second quantization formulation using a basis of molecular orbitals. These are formed following the Hartree-Fock method by a linear combination of atomic orbitals \cite{Helgaker2000}. The electronic part of the Hamiltonian then reads:
\begin{equation}
\hat{H}=\sum^M_{pq} h_{pq}\hat{a}^\dagger_p\hat{a}_q+\frac{1}{2}\sum^M_{pqrs}h_{pqrs}\hat{a}^\dagger_p\hat{a}^\dagger_q\hat{a}_r\hat{a}_s,
\label{eq:QChem}
\end{equation} 
where the summation is over all of the $M$ molecular basis states.
The factors $h_{pq}$ and $h_{pqrs}$ are related to one-electron and two-electron transitions respectively and are calculated numerically \cite{Yung2014}.
The fermionic creation and annihilation operators fulfill $\{\hat{a},\hat{a}^\dagger\}=1$, which enforces the antisymmetry of the wave function. The Hamiltoninan is mapped to spins \cite{JW,BK} and the expectation values are calculated for trial wave functions of the ground state, generated via an ansatz circuit as described previously. A references state, which represents a classical approximation to the ground state such as a Hartree-Fock solution, usually a product state, can be chosen as the initial state. A physically-inspired ansatz circuit for this type of problem is the Unitary Coupled Cluster (UCC) \cite{Helgaker2000}. The scheme prescribes a series of fermionic operators called cluster operators $T=T_1+T_2+...+T_N$ for an $N$ electron system \cite{Peruzzo2014}. These are essentially electron-hopping operators of increasing order that form a unitary to evolve the system into the ground state through the exploration of Hilbert space by going to increasing Hamming distances from the initial state \cite{McClean2016}, $\hat{U}_{UCC}=e^{\theta (T-T^\dagger)}$. Using a pseudo-time evolution of this unitary by first-order Trotterzation is a good approximation of this evolution \cite{Yung2014,InnsbruckMolecule2018}: $\hat{U}_{UCC}\simeq \Pi_i e^{\theta_i(T_i-T^\dagger_i)}$. In the spin basis, this creates circuits of higher-order spin-spin interactions, which can be realized on a quantum computer, particularly in trapped ions via multi-ion Ising-operations, or MS-gates. Classical coupled-cluster calculations can be employed to find reasonable starting values for $\theta$ \cite{Romero2018}. The computational complexity of the UCC method scales polynomially with the number of molecular orbitals $M$. The procedure can be repeated for different nuclear configurations to sample out the energy surface. Alternatively, a quantum phase estimation algorithm can be employed to this end \cite{Yuan2019theoryofvariational}.

\begin{figure}[t]
\includegraphics[width=0.8\linewidth]{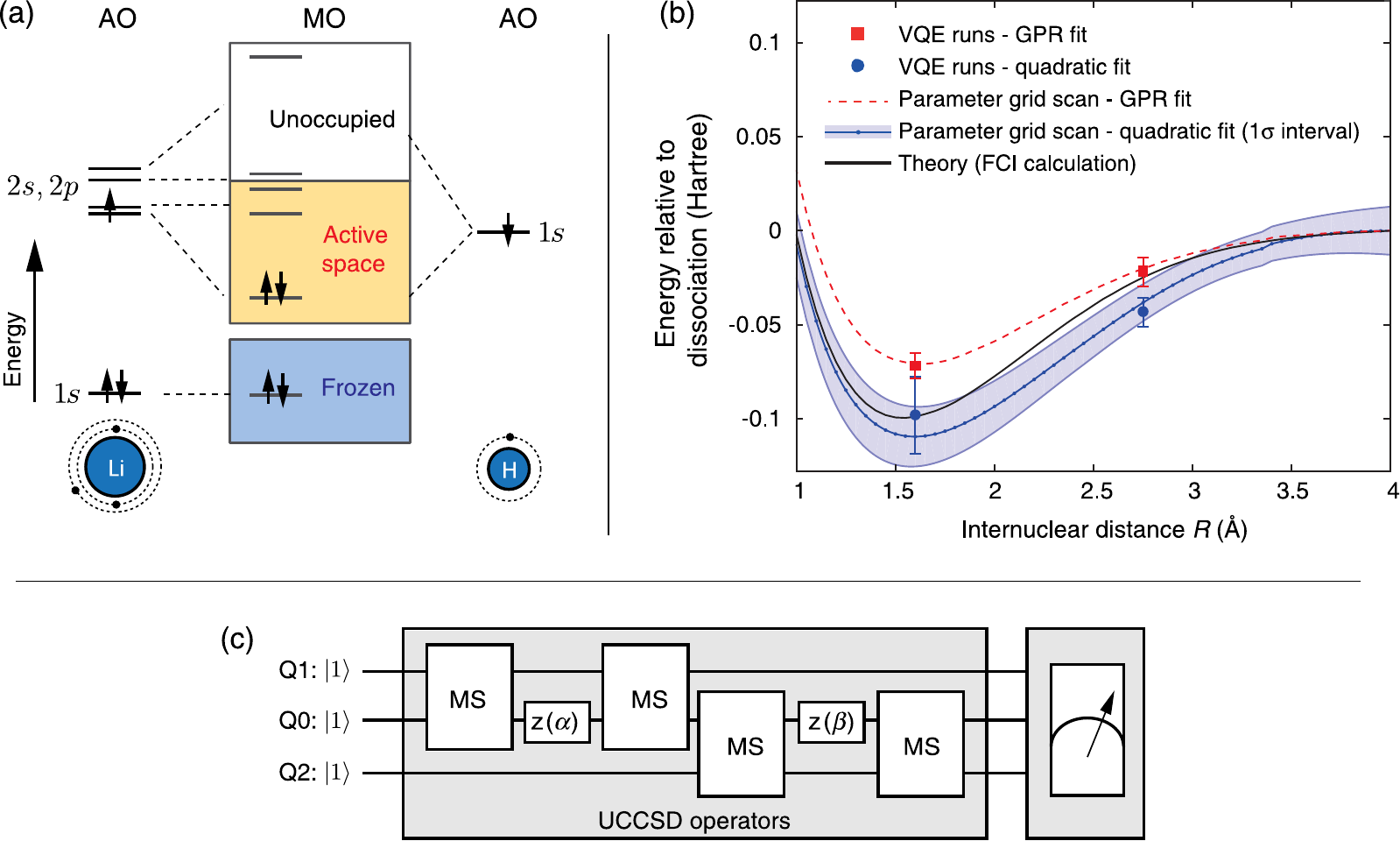}
\caption{Variational Quantum Eigensolver simulation of the Lithium Hydride (LiH) molecule with three trapped-ion qubits. (a) LiH molecular orbitals (MO) formed out of atomic orbitals (AO) contributed by each element. The active space in which the implemented UCC excitation operators act is highlighted in yellow. (b) The theoretical LiH potential energy surface calculated for the minimal basis set (black) is shown in comparison with experimentally obtained results, offset to overlap at maximum distance R to better illustrate the well-depth differences in the grid-scan reconstructions. The data points result from sampling the energy landscape using the VQE algorithm or a parameter grid scan and fitting the explored space with a GPR-based machine learning algorithm (red dashed line) or a 2D quadratic fit (blue solid line). The error bars are obtained from the fits with the underlying data weighted by quantum projection noise. (c) Abstract quantum circuit implementing each of the two target UCCSD unitary operators on the qubits indexed Q0, Q1, and Q2. A fully entangling gate (MS) acts locally between the enclosed qubits and surrounds a local qubit rotation z quantified by parameters $\alpha$ and $\beta$, respectively. Adapted from \citealp{InnsbruckMolecule2018}.}
\label{fig:IBK_molecule}
\end{figure}

Below we document recent implementations of VQE with trapped ions, although VQE has also been implemented with other experimental platforms  \cite{Peruzzo2014, OMalley2016, Kandala2017, Colless2018, Dumitrescu2018, GoogleVQE2020}.
VQE was implemented in trapped $^{171}$Yb$^+$ ions \cite{shen2017quantum} to simulate the electronic structure of the molecular cation HeH$^+$.  Instead of Ising operations, the authors used the four available states in the ground level of a $^{171}$Yb$^+$ ion coupled by microwave- and radiofrequency-driven transitions. The experiment matches the ground state energy from exact diagonalization. The authors also look for excited states by measuring the expectation value of a changed target Hamiltonian to $\langle(\hat{H}-\lambda \mathbb{1})^2\rangle$. The measurement involved scanning $\lambda$ and looking for zero-valued expectation values and resolved three out of four excited states.

In \cite{InnsbruckMolecule2018}, the authors mapped two molecular problems to spins realized in chains of trapped $^{40}$Ca$^+$ ions, the Hydrogen molecule, H$_2$, and Lithium Hydride, LiH. For H$_2$, the molecular orbitals are formed by linear combinations of the two 1s orbitals of each atom. The Bravyi-Kitaev \cite{BK} and Jordan-Wigner \cite{JW} encoding are used to map the problem onto two and four qubits, respectively. The former is more efficient since both the Hamiltonian and the UCC ansatz can be reduced to two qubits by appropriate choice of reference state \cite{OMalley2016}. In both mappings, the UCC operator leads to a single-parameter ansatz circuit, which is implemented using a pair of MS gates based on a technique developed in \cite{M_ller_2011}. The authors do a full sweep of the parameter as well as a VQE optimization to map out the ground state potential energy using a Nelder-Mead search algorithm, which was found to not converge in some instances due to the presence of noise. For the simulation of LiH, a minimal basis set for the molecular orbitals was chosen. The four electrons are filled into the energy-ordered orbitals, as determined by a Hartree-Fock calculation. The UCC ansatz is truncated to single and double excitations (UCCSD) and an active space of two electrons and three orbitals identified \cite{Kandala2017}. The two core electrons can be considered to not contribute to molecular bonding \cite{Roos1980}. Using the classical theory of configuration interaction of singles and doubles (CISD) \cite{Shavitt1984}, the active space is further reduced to two singlet-excitations. This active space is illustrated in Fig. \ref{fig:IBK_molecule}a. The resulting UCC unitary was mapped to spins via Bravyi-Kitaev encoding, which results in a three-qubit operator with two circuit parameters, $\hat{U}_{UCCSD}(\alpha,\beta)=e^{-i\alpha \sigma_0^x\sigma_1^y}e^{-i\beta \sigma_0^x\sigma_2^y}$, which was realized in the circuit shown in Fig. \ref{fig:IBK_molecule}c. The authors performed a grid scan of the parameters at different inter-nuclear distances and reconstructed the potential energy curve for the electronic ground state by fitting the minimum in two ways, by Gaussian process regression (GPR) and to 2D-quadratic surface, where the latter yields very good results, reproducing the well depth within statistical uncertainties, see line plots in Fig. \ref{fig:IBK_molecule}b. The authors also performed a feedback VQE using a classical optimizer that incorporates elements of simulated annealing and matches the results of the grid scan, see data points in Fig. \ref{fig:IBK_molecule}b.

In \cite{Nam2020Water} the authors estimated the ground state energy of the water molecule (H$_2$O). The UCC ansatz circuits are optimized by taking advantage of an active subspace and by sorting the UCC operators into bosonic and non-bosonic excitatios from the Hartree-Fock ground state. Bosonic terms refer to excitations of two electrons into a molecular orbital such that their spins pair up. These excitations can be represented by a single-qubit operation. Non-bosonic excitations are implemented via the Jordan-Wigner transformation. The resulting circuits are further optimized by term ordering \cite{Hastings2014ImprovingQA}. The authors present the cost of adding an increasing number of ansatz terms to the circuit in terms of qubits and MS gates, see fig.\ref{fig:ionqwater}a. Their results show that chemical accuracy can be reached with about $15$ terms, which corresponds to $11$ qubits and $140$ MS-gates. Experimentally, the authors realize circuits with up to three terms and three qubits. The measurement results agree within errors with classical calculations, see fig.\ref{fig:ionqwater}b. Finally, the authors proposed the implementation of this simulation problem with a growing number of terms as a benchmark for quantum computing platforms \cite{Nam2020Water}. 

\begin{figure}[t]
\includegraphics[width=0.95\linewidth]{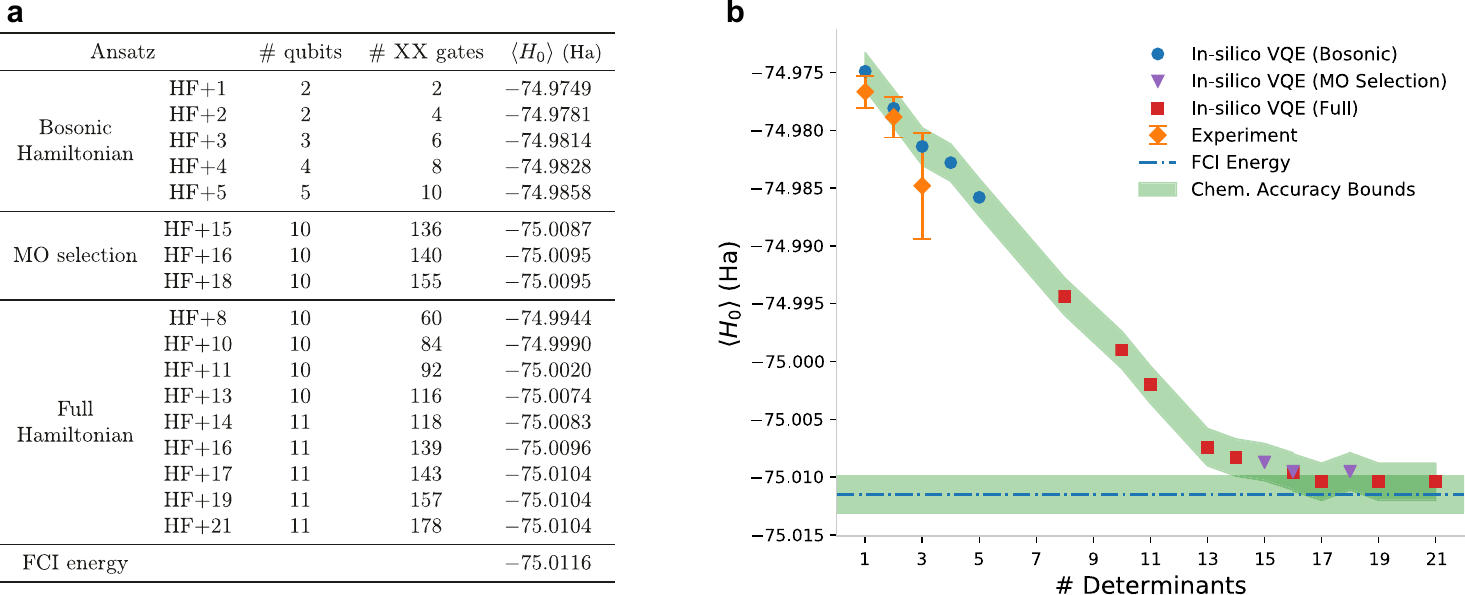}
\caption{VQE simulation of the ground state energy of the water molecule, H$_2$O. a: Metrics for each circuit are labeled HF+N, as up to N of the most significant interaction terms are added to the ansatz state. The bosonic terms through HF+5 can be represented as pair excitations to reduce the qubit resource requirements, while the molecular orbital (MO) selection strategy prunes the two least significant molecular states to reduce the qubit count slightly at the expense of accuracy at the mHa level. Energies should
be compared to the full configuration interaction (FCI) ground-state energy, which is the exact result from diagonalizing the complete Hamiltonian in the minimal chemical basis. b: Comparison of ground-state energy estimates as additional interactions are included in the UCC ansatz state (labeled HF+N, for N significant determinants). The orange diamonds indicate experimental results from implementations with up to three qubits, with $1\sigma$ error bars from the bootstrap distribution.
The remaining points are from the in silico VQE simulation and show how the ansatz states converge to the full configuration-interaction ground state, indicated by the dot-dashed blue line. Computational error equivalent to the bound for chemical accuracy ($1.6\:$mHa) is indicated by the shaded green region. Figure and table adapted from \citealp{Nam2020Water}}
\label{fig:ionqwater}
\end{figure}

VQE was used in \cite{shehab2019toward} to solve a nuclear physics problem, finding the binding energy of the Deuteron nucleus. The nuclear interaction is modeled in pionless effective field theory (EFT) following \cite{Dumitrescu2018}. The Hamiltonian is expressed in the $N$-oscillator basis, 
\begin{equation}
    \hat{H}_D=\sum_{n,n'=0}^{N-1} \langle n'|(\hat{T}+\hat{V})|n\rangle \hat{a}^\dagger_{n'} \hat{a}_n,
\end{equation}
where the operators $\hat{a}^\dagger_{n}$ and $\hat{a}_n$ create and annihilate, respectively, a deuteron in the harmonic-oscillator $s$-wave state. Using finite harmonic oscillator spaces introduces errors by imposing sharp cut-offs in both position and momentum space, called infrared (IR) and ultraviolet (UV) errors. These depend on the basis size and the potential \cite{Koenig2014}. Using the Jordan-Wigner transformation \cite{shehab2019toward} find the qubit Hamiltonians for $N=2$,$3$, and $4$. The value of the largest coefficient in the Hamiltonian grows with basis size $N$ and its uncertainty bounds the overall uncertainty of the Hamiltonian expectation value. This means that increasing the basis, which already comes with more gates and qubits, also requires an increase in the number of measurements that need to be taken to reproduce the gain in model accuracy experimentally. The UCC ansatz used is based on single-excitations of the different $s$-wave states \cite{Lu2019Simulations}, and leads to an iterative scheme for circuit construction with $N-1$ variational parameters. The authors perform a line sweep of the parameters for $N=4$, and measure the expectation values at theoretically optimal parameters for $N=2$, $3$, and $4$.  In order to improve the results, the authors employ an error mitigation scheme based on Richardson extrapolation \cite{Temme2017, Mcardle2019}, by which each circuit is implemented with an increasing number of pairwise-cancelling MS-gates, i.e. 3, 5, 7,... MS-gates per actual circuit gate. These additional measurements are used to scale the noise and extrapolate the value of each Hamiltonian term to zero noise. These extrapolated values are then combined to give the final ground state energy estimate. For the 3- and 4-qubit ansatz circuits the theoretical predictions lie within the error bars of the experimental results \cite{shehab2019toward}. The authors also compared their results directly with the superconducting-circuit implementation in \cite{Dumitrescu2018}.

The VQE implementation of the Deuteron nuclear binding energy problem was also improved upon by some of the same authors in \cite{shehab2019noise} by applying a so-called past-light-cone method to the VQE circuits. This breaks them down into minimal circuits for each term in the Hamiltonian at the cost of a larger number of measurements. The authors showed a significant reduction of $80\%$ in the error of the estimated ground state energy. This made it consistent with the theoretical value without using error mitigation.

VQE is also applicable to condensed matter problems. \cite{Rungger2019dynamical} designed and implemented a variational algorithm to solve dynamical mean field theory (DMFT) problems. DMFT is a correction to density functional theory (DFT) and describes materials with strongly correlated electrons \cite{Georges1996}. The model problem consists of an interacting impurity coupled to a bath and is hard to solve using classical means. After the Jordan-Wigner transformation the Hamiltonian can be expressed in terms of orbital electron creation and destruction operators, which in turn are linear combinations of Pauli operators. The number of qubits required is $2(N_{\textrm{imp}}+N_\textrm{bath})$, where $N_{\textrm{imp}}$ and $N_\textrm{bath}$ are the number of impurity and bath spin orbitals included, respectively. The authors use a hardware-efficient ansatz. The procedure consists of two nested optimization loops. The inner loop is a quantum-classical hybrid VQE optimization that solves for the eigenvalues based on a set of impurity Hamiltonian parameters. The outer loop is purely classical and calculates local retarded Greens functions for the original lattice model and the impurity problem. It updates these parameters until a set of self-consistent impurity parameters is achieved \cite{Liebsch2011}. From these, the electronic structure of the system can be calculated.

\subsubsection{Quantum Approximate Optimization Algorithms}
\label{subsec:qaoa}

The Quantum Approximate Optimization Algorithm (QAOA) \cite{Farhi2014quantum} is a framework for using a quantum simulator to perform tasks such as combinatorial optimization or solving satisfiability and graph problems. QAOA can also be employed for producing highly entangled target states or finding the ground state and energy spectrum of critical Hamiltonians \cite{Ho2019}. QAOA encodes the objective function of the optimization in a spin Hamiltonian $H_A$, which is applied in a "bang-bang" protocol followed by a non-commuting mixing operator $H_B$ and can provide resource-efficient approximate answers, making it attractive for applications on current systems with limited gate depth or coherence time (see Fig. \ref{fig:qaoapagano}a). The sequence is repeated $p$ times where each layer $i$ is characterized by the variational parameters or angles $(\gamma_i, \beta_i)$, giving rise to the following evolution:

\begin{equation}
    \ket{\vec{\beta},\vec{\gamma}}=\prod_{i}^{p\leftarrow1}e^{-i \beta_i H_B}e^{-i \gamma_i H_A}\ket{\psi_0},
    \label{eq_QAOA_evolution}
\end{equation}
where $\ket{\psi_0}$ is the initial state.

For certain classes of problems QAOA has been shown to provide results that cannot be generated efficiently on a classical device \cite{Farhi2016qaoasupr}, but in practice its performance is highly problem- and problem-instance dependent \cite{Willsch2019}. Below we highlight recent QAOA implementations in trapped ions, although there have been implementations of QAOA in other quantum computing platforms \cite{Otterbach2017,Qiang2018,Bengtsson2019,Willsch2019,GoogleQAOA2020}.

A seven-qubit digital processor based on $^{171}$Yb$^+$ trapped ion spins was used to execute a QAOA-inspired protocol \cite{zhu2019qaoa} in order to generate a ``Thermofield-Double" (TFD) state \cite{Wu2019}. The TFD state is a pure two-mode squeezed state that behaves as a thermal state for one mode when traced over the other mode, and is of great interest in a number of areas of physics. These states provide a way to prepare thermal (Gibbs) states of a many-body Hamiltonian in a quantum simulator, which underpin phenomena like high temperature superconductivity \cite{lee2006doping} and quark confinement in quantum chromodynamics \cite{gross1981}. TFD states also play a key role in the holographic correspondence where they represent so-called traversable wormholes \cite{traversable1, traversable2} and a number of approaches for their preparation have been proposed \cite{Wu2019, Martyn2019, lokhande, qimaldacena}.
A TFD state corresponding to inverse temperature $\beta$ is defined on a joint system of two identical Hilbert spaces $A$ and $B$:
\begin{equation}\label{eq:TFD}
    \ket{TFD(\beta)}=\frac{1}{Z(\beta)}\sum_{n}e^{-\beta E_n/2}\ket{n}_A\ket{n}_B
\end{equation} 
where $Z(\beta)$ is a normalization factor and $\ket{n}_A$ and $\ket{n}_B$ are the eigenstates in space $A$ and $B$ with corresponding energy $E_n$. 

\begin{figure}[b]
\includegraphics[width=0.75\linewidth]{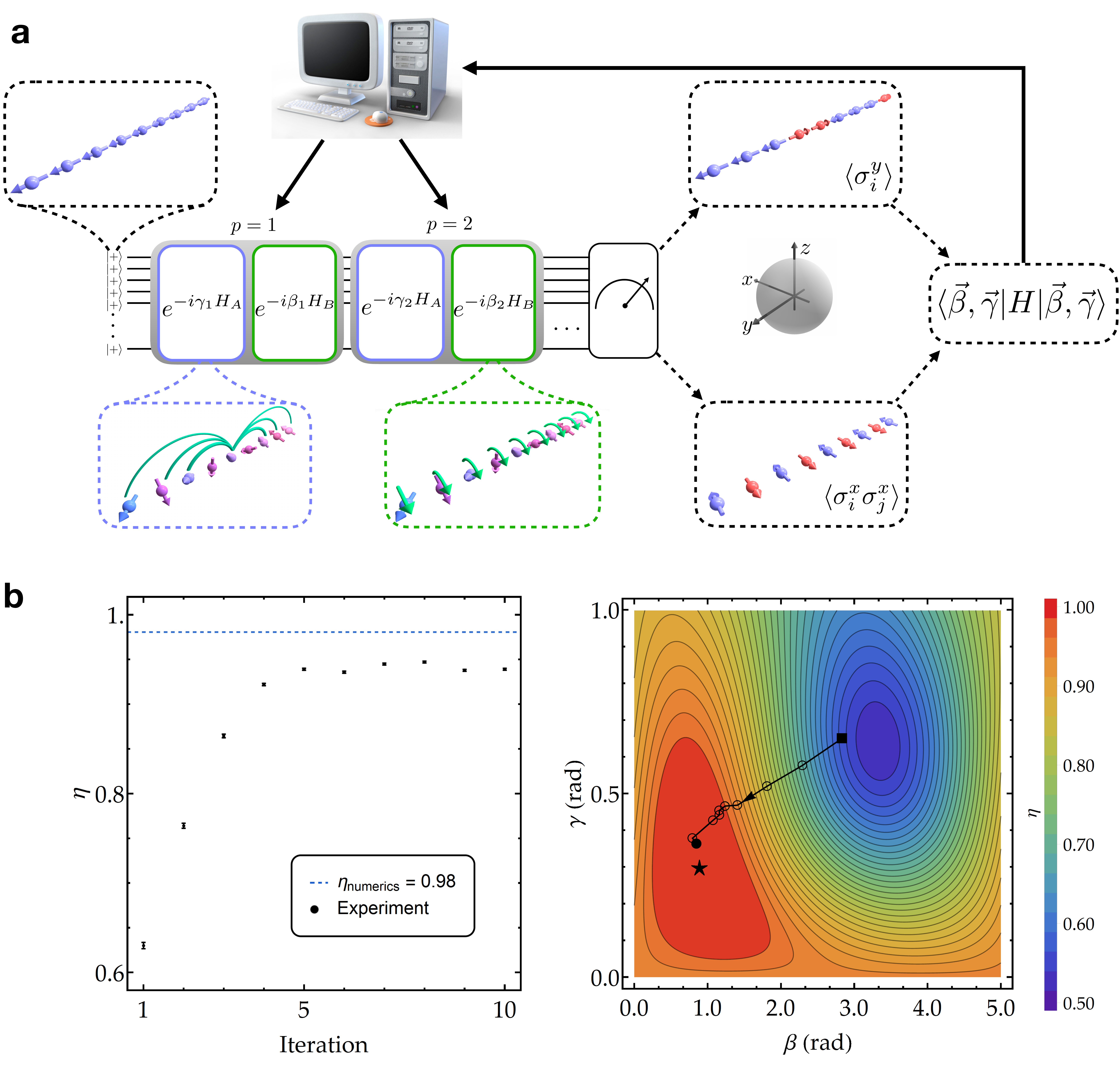}
\caption{QAOA applied to transverse-field long range Ising model.
{\bf (a)} QAOA protocol: The system is initialized along the $y$ direction on the Bloch sphere in the ground state of the mixing Hamiltonian $H_B$, namely $\ket{+}^{\otimes N}$. The unitary evolution under $H_{A(B)}$ is implemented for angles $\gamma_i (\beta_i$) and repeated $p$ times. At the end of the algorithm global measurements in the $x$ and the $y$ basis are performed to compute the average energy $E(\vec{\beta},\vec{\gamma})=\bra{\vec{\beta},\vec{\gamma}}  H \ket{\vec{\beta},\vec{\gamma}}$. The measurement results are then processed by a classical optimization algorithm to update the variational parameters. {\bf (b)} Closed-loop optimization for $p=1$ and $N=20$ qubits. Starting from an initial guess, the local gradient is approximated by performing the energy measurements along two orthogonal directions in  parameter space. To quantify the performance of the QAOA, the dimensionless quantity $\eta\equiv \frac{E(\vec{\beta},\vec{\gamma})-E_{max}}{E_{gs}-E_{max}}$ is used, where  $E_{max}(E_{gs})$ are the energies of the highest (lowest) excited state, therefore mapping the entire many-body spectrum to the $[0,1]$ interval. The algorithm converges after about 7 iterations. Adapted from \citealp{pagano2019qaoa}.}
\label{fig:qaoapagano}
\end{figure}

In \cite{zhu2019qaoa}, the authors generated a TFD state similar in form to Eq. (\ref{eq:TFD}) under a nearest-neighbor transverse-field Ising Hamiltonian at various temperatures, with three qubits per subsystem. Preparation starts by initializing a product of Bell-pair singlets $\dn_A\up_B-\up_A\dn_B$ between pairs of $A$ and $B$ qubits. This is an infinite temperature TFD state since $\rho_A$ is maximally mixed. The evolution alternates between  inter-system coupling $H_{AB}=\sum_i X_{i,A} X_{i,B} + Z_{i,A} Z_{i,B}$ and the intra-system Hamiltonians $H_A+H_B$, where $H_A$ and $H_B$ are identical transverse-field Ising Hamiltonians of the subsystems, $H_{XX} + g H_Z \equiv \sum_{i=1}^{L}X_iX_{i+1} + g \sum_{i=1}^{L}Z_i$. This evolution was approximated by variational application of these Hamiltonians. The sequence was executed with theoretically optimal parameters for $p=1$, and found to reproduce the expected inter- and intra- system correlators well. The authors also use a related approach \cite{Ho2019ultrafast} to directly prepare the zero-temperature ground state of the quantum critical transverse field Ising model with seven trapped ion spins using quantum-classical feedback, starting from a random point for $p=1$ and from the ideal parameters for $p=2$. 

The QAOA approach was used in an ion trap system to approximately generate the critical ground state energy of the long-range anti-ferromagnetic transverse field Ising model (Eq. (\ref{eqn:TransversIsing})) \cite{pagano2019qaoa},
where the spin-spin interaction $J_{ij}$ is defined in Eqs. (\ref{Jij},\ref{Jpowerlaw}) and $B$ is the transverse field. Here we set $H_A$ to be the Ising coupling term and $H_B$ to be the field term in Eq. (\ref{eqn:TransversIsing}).
QAOA was employed with up to 40 qubits in this system, using both grid search and closed loop approaches. In the grid search approach, the whole parameter space $(\vec{\beta},\vec{\gamma})$ was explored experimentally to find the optimum; in the closed-loop approach the analog trapped-ion quantum simulator was interfaced with a greedy gradient-descent algorithm to optimize the measured energy (Fig. \ref{fig:qaoapagano}a) starting from an initial guess. In the $p=1$ QAOA, the optimization trajectory in the experiment can be visualized on the theoretical performance surface as shown in Fig. \ref{fig:qaoapagano}b. 

In \cite{shehab2019noise} the authors mapped the MaxCUT problem for a specific 5-node graph to a five-qubit trapped ion processor and solved it approximately using an optimized QAOA protocol, that reduces circuit width and depth by splitting it into circuits for each Hamiltonian term or correlator at the cost of a larger number of measurements. For the five-node Dragon graph, the authors reported a $36\%$ improvement in the accuracy of the solution over the standard circuit in their system. 

Finally, we note that VQS can also be used to study dynamics by employing generalizations of time-dependent variational principles \cite{Broeckhove1988} to simulate real \cite{Ying2017efficient,Heya2019} and imaginary \cite{McArdle2019img,Tyson2019,Chen2019} time evolution of pure quantum systems, as well as mixed states \cite{Yuan2019theoryofvariational}. 
}
\section{Outlook and Future Challenges}
Quantum interacting spin models are among the simplest many-body quantum systems with nontrivial features that can elude classical computational approaches. Trapped atomic ion spins offer the ability to implement and control quantum spin models with tunability of the interaction form and range. Much of the research in this field has concentrated on studies of the long-range transverse Ising model, which can feature frustrated ground states with associated degeneracies and entanglement in the ground state.  The many types of phase transitions and dynamical processes in this system form a fruitful test-bed for studying quantum non-equilibrium processes, in many cases challenging classical computational power even for small numbers of spins.  There are many extensions in this physical system to simulating more complex spin models with trapped ions such as Heisenberg couplings \cite{grass2014trapped}, higher-dimensional spin models, and interactions involving three or more spins. These future directions may allow the quantum simulation of more exotic spin phases such as spin liquids \cite{balents2010spin}, or topological orders in spin systems such as the Haldane chain \cite{haldane1983} or the Kitaev lattice \cite{kitaev2006anyons}.

{\q 
There are a number of technical challenges to overcome in the path to scaling up trapped-ion systems to $\sim100$ qubits and beyond. However none of these obstacles are fundamental and many groups have shown effective mitigation techniques in this growing area of research. With a single linear chain of ions, it should be possible to extend the number of ions to hundreds using anharmonic trapping potentials that generate uniform spacing and thus avoid structural phase transitions for larger ion crystals \cite{lin2009large,Pagano2018}. Two technical challenges include the increased sensitivity of equally-spaced ion crystals to background fluctuating electric fields \cite{cetina2020quantum} and the shorter lifetime of the ion chain due to collisions with the background gas. Solutions to these challenges include the use of cryogenic trapped-ion systems, that significantly lower the vacuum pressure so that storage times of hours or longer can be achieved with large ion crystals \cite{Poitzsch1996,Pagano2018}, and the use of sympathetic cooling with multiple species ions in the system \cite{Larson1986sympathetic, lin2009large, Chou2010frequency, Honeywell}, allowing the background heating to be quenched without disturbing the spins in the simulation.

With a 1D ion chain, it is possible to simulate any frustrated spin graph with multiple engineered laser pulses \cite{korenblit2012quantum,Davoudi2020}. However, it is possible to further increase the number of ions in a quantum simulator by using higher-dimensional ion crystals \cite{Wang2015}. The Penning trap geometry is amenable to 2D and 3D crystal structures, as discussed in this review.
In a rf trap, 2D or 3D ion crystals are necessarily accompanied with significant rf micro-motion for ions not positioned at a rf null \cite{dehmelt1967radiofrequency}, which is limited to a point or a line in space. Although micro-motion will introduce a number of subtleties for laser cooling, gate operations, and ion detection, it is a coherent and well-controlled motion whose effects can be mitigated through proper gate design and the engineering of spin interactions \cite{Shen2014scalable,Wang2015}. Alternatively, an array of rf traps can be engineered to create multidimensional designer ion crystals \cite{Cirac2000,Kumph2016}, although this may require very small distances between the ions and the electrodes to generate sufficiently strong spin interactions.
Apart from offering an avenue to increase the ion number, 2D and 3D ion crystals provide a natural platform to realize rich frustrated spin Hamiltonians in a more complicated geometry that can match the native dimension of the underlying target spin Hamiltonian.

Finally, we note that well-known modular approaches to scaling trapped ion systems can be directly applied to quantum spin simulations. For instance, large numbers of ions can be grouped in spatially-separated modules, where each module contains many trapped ion spins with Coulomb-mediated spin-spin couplings discussed in section \ref{sec:intro}.  Here, the system is scaled by shuttling subsets of ions between modules to extend the size of the system \cite{Kielpinski:2002, Honeywell}.  Because the idle spin states are nearly perfectly decoupled from ion motion between modules, the scaling procedure is largely a systems engineering task. Ultimately, separated modules of trapped ion crystals can also be connected via heralded photonic interconnects, even between separated trap structures, for a remote modular scalable architecture \cite{Duan2010,Li2012,Monroe:2014}. 


Useful quantum simulations should have unambiguous benchmarks to indicate performance beyond what is possible with classical computational simulation. As opposed to recent simulations of random quantum circuit evolution \cite{GoogleSupremacy2019}, we desire a benchmark that also indicates the usefulness of the underlying problem. When we use a quantum simulator to probe the ground-state of a quantum many-body Hamiltonian for instance, a possible heuristic benchmark could be the ground-state variational energy of the Hamiltonian. Although in general there is no quantum algorithm to guarantee that we can successfully find the genuine ground-state of a many-body Hamiltonian, the associated variationtal energy of the approximate ground state realized by a quantum simulator can always be efficiently measured by the experimental setup. If such a measured energy is lower than the energy obtained by any classical simulation even with the most powerful computers, it is an indication that the quantum simulator finds a useful solution that better approximates the ground state of the target Hamiltonian compared with any known classical method.} 

{\q Of course,} there is a close relationship between spin simulations and quantum computations with qubits, and the underlying mechanism behind the Ising couplings in trapped ion spin simulations is exactly that used for discrete quantum gates between trapped ion qubits \cite{sorensen2000entanglement}, which are sometimes called Ising gates. Quantum simulations in this sense can be considered as a special case of a quantum computation, and it should be expected that as trapped ion quantum computers scale in the future \cite{monroe2013scaling, brown2016codesigning, Wright:2019, Honeywell}, so will the reach of trapped ion quantum spin simulators.

\section*{Acknowledgments}
We acknowledge key collaborations with Patrick Becker, Howard Carmichael, Ming-Shien Chang, Itzak Cohen, Katherine Collins, Arinjoy De, Emily Edwards, Lei Feng, Michael Foss-Feig, James Freericks, Mohammad Hafezi, Phillip Hauke, Markus Heyl, David Huse, Dvir Kafri, Harvey Kaplan, Antonis Kyprianidis, Simcha Korenblit, Aaron Lee, Guin-Dar Lin, Mohammad Maghrebi, Brian Neyenhuis, Nhung H. Nguyen, Changsuk Noh, Andrew Potter, Alex Retzker, Jacob Smith, Wen-Lin Tan, Ashvin Vishwanath, C.-C. Joseph Wang, and Jiehang Zhang.  We thank R. Blatt, J. Bollinger, T. Sch\"atz, and C. Wunderlich for discussions.

%

\end{document}